\documentclass[reprint,
superscriptaddress,
amsmath,
amssymb,
aps,
prb,
floatfix,
english
]{revtex4-2}
 
\usepackage{bibunits}
\usepackage[unicode=true, colorlinks=true, citecolor={blue!80!black}, urlcolor={blue!50!black}, linkcolor = {blue!80!black}]{hyperref}
\defaultbibliography{references}
\defaultbibliographystyle{apsrev4-2}
\usepackage{graphicx}
\usepackage{svg}
\usepackage{dcolumn}
\usepackage{siunitx}
\usepackage[T1]{fontenc}
\usepackage[english]{babel}
\usepackage[utf8]{inputenc}
\usepackage{bbm}
\usepackage[normalem]{ulem}
\usepackage[dvipsnames]{xcolor}

\newcommand{\wq}{\omega_\mathrm{q}}

\newcommand{\wres}{\omega_\mathrm{r}}

\newcommand{\nbar}{\bar{n}}

\newcommand{\ghz}{\:\mathrm{GHz}}
\newcommand{\mhz}{\:\mathrm{MHz}}

\makeatletter
\global\let\@extrafn\@empty
\def\doauthor#1#2#3{%
  \ignorespaces#1\unskip\@listcomma
  \begingroup
   #3%
  \@if@empty{#2}{\endgroup{}{}}{\endgroup{\comma@space}{}\frontmatter@footnote{#2}}%
  \@extrafn\global\let\@extrafn\@empty
  \space \@listand
}%
\newcommand\extrathanks[1]{\gdef\@extrafn{{\comma@space}\frontmatter@footnote{#1}}}
\makeatother
 
\begin{document}
\widetext

\title{High-frequency dispersive readout of a superconducting qubit}
\author{Pavel~D.~Kurilovich}
\thanks{These two authors contributed equally.\\
pkurilovich@fas.harvard.edu, tom.connolly@yale.edu}
\extrathanks{Present address: Department of Physics, Harvard University, Cambridge, Massachusetts 02138, USA}
\affiliation{Departments of Applied Physics and Physics, Yale University, New Haven, Connecticut 06520, USA}
\author{Thomas~Connolly}
\thanks{These two authors contributed equally.\\
pkurilovich@fas.harvard.edu,
tom.connolly@yale.edu}
\affiliation{Departments of Applied Physics and Physics, Yale University, New Haven, Connecticut 06520, USA}
\author{Charlotte~G.~L.~B\o ttcher}
\thanks{{Present address: Department of Applied Physics, Stanford University, Stanford, California 94305, USA}}
\affiliation{Departments of Applied Physics and Physics, Yale University, New Haven, Connecticut 06520, USA}
 
\author{Daniel~K.~Weiss}
\thanks{{Present address: Quantum Circuits, Inc., New Haven, CT, USA}}
\affiliation{Departments of Applied Physics and Physics, Yale University, New Haven, Connecticut 06520, USA}
\affiliation{Yale Quantum Institute, Yale University, New Haven, Connecticut 06511, USA}
 
\author{Sumeru~Hazra}
\affiliation{Departments of Applied Physics and Physics, Yale University, New Haven, Connecticut 06520, USA}
\author{Vidul~R.~Joshi}
\thanks{{Present address: Microsoft Quantum}}
\affiliation{Departments of Applied Physics and Physics, Yale University, New Haven, Connecticut 06520, USA}
 
\author{Andy~Z.~Ding}
\affiliation{Departments of Applied Physics and Physics, Yale University, New Haven, Connecticut 06520, USA}
\author{Heekun~Nho}
\affiliation{Departments of Applied Physics and Physics, Yale University, New Haven, Connecticut 06520, USA}
\author{Spencer~Diamond}
\affiliation{Departments of Applied Physics and Physics, Yale University, New Haven, Connecticut 06520, USA}
\author{Vladislav~D.~Kurilovich}\thanks{{Present address: Google Quantum AI, 301 Mentor Dr, Goleta, CA93111, USA}}
\affiliation{Departments of Applied Physics and Physics, Yale University, New Haven, Connecticut 06520, USA}
 
\author{Mykola~Kishmar}
\affiliation{Physics Department, Columbia University, New York, New York 10027, USA}
 
\author{Wei~Dai}
\affiliation{Departments of Applied Physics and Physics, Yale University, New Haven, Connecticut 06520, USA}
\author{Valla~Fatemi}
\affiliation{Departments of Applied Physics and Physics, Yale University, New Haven, Connecticut 06520, USA}
\affiliation{School of Applied and Engineering Physics, Cornell University, Ithaca, New York 14853, USA}
\author{Luigi Frunzio}
\affiliation{Departments of Applied Physics and Physics, Yale University, New Haven, Connecticut 06520, USA}
\author{Leonid~I.~Glazman}
\affiliation{Departments of Applied Physics and Physics, Yale University, New Haven, Connecticut 06520, USA}
\affiliation{Yale Quantum Institute, Yale University, New Haven, Connecticut 06511, USA}
\author{Michel~H.~Devoret}\thanks{michel.devoret@yale.edu\\
{Present address: Physics Dept., U.C. Santa Barbara, Santa Barbara, California 93106, USA and Google Quantum AI, 301 Mentor Dr, Goleta, California 93111, USA}}
\affiliation{Departments of Applied Physics and Physics, Yale University, New Haven, Connecticut 06520, USA}
 
\begin{abstract}
Driving a {low-dissipation, non-linear} system periodically can destabilize it by exciting {indirect} resonances, which occur whenever an integer multiple of the drive frequency matches an integer combination of the system's mode frequencies. Avoiding such resonances is a central design constraint across physics and engineering, from radio-frequency ion traps to particle accelerators. Remarkably, the measurement of quantum bits based on superconducting circuits has so far remained limited by non-linear resonances. Measurements of qubit states are customarily performed in a regime where the frequency of the microwave readout tone is comparable to that of the circuit. In that case, the drive excites resonances that eject the qubit from its computational subspace. This limits the readout power and hence the achievable measurement fidelity. We show that a readout {tone} with frequency much larger than that of the circuit circumvents this problem by exponentially suppressing the resonances. Using a readout tone with frequency 12 times larger than that of our transmon qubit, we achieve a quantum non-demolition measurement fidelity of $99.931\pm 0.003\%$, with an error an order of magnitude smaller than the previous record. {Under these conditions, the} probability of leakage to non-computational states per readout is $0.018\pm0.002\%$.
\end{abstract}
 
\maketitle
\section*{Introduction}
 
\begin{figure*}[t]
  \begin{center}
    \includegraphics[scale = 1]{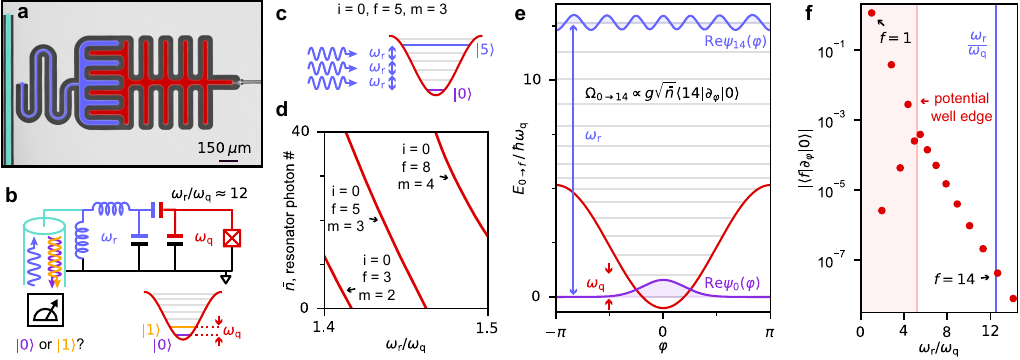}
\caption{Overcoming the deficiencies of dispersive readout in a superconducting qubit by using a high-frequency probe. \textbf{a}, False-colored device image. A transmon superconducting qubit with frequency $\wq / 2\pi = 0.757\ghz$ (red) is coupled to a readout resonator with frequency $\wres/2\pi = 9.227\ghz$ (blue). The resonator is coupled to a feed-line (turquoise). \textbf{b}, Readout is achieved by monitoring the probe signal reflected from the resonator. The reflected phase encodes the qubit state. \textbf{c}, Unwanted resonances in the qubit spectrum degrade measurement performance by exciting the qubit to non-computational states. Resonances occur when $m \hbar \wres = E_{i\rightarrow f}$ (where $|i\rangle$ is the initial computational state, $|f\rangle$ is a non-computational state, $E_{i\rightarrow f}$ is the energy difference between the two states, and $m$ is an integer). \textbf{d}, Due to the AC Stark shift, the frequencies at which the resonances occur {depend on the drive power and thus the photon number $\bar{n}$ in the resonator}. This makes resonances hard to avoid in the standard frequency regime $\wres \sim \wq$. The slopes of the lines depend on the dispersive shift $\chi$ (fixed to the value used in our experiment). The offset charge is fixed to $n_g=1/4$. Only processes with $m,f\leq 10$ are shown. \textbf{e}, In our new regime, $\wres \gg \wq$, unwanted resonances can only happen between distant states. The corresponding transition amplitude $\Omega_{i\rightarrow f}$ is proportional to the matrix element of $\partial_\varphi$ operator ($\varphi$ is the superconducting phase difference). $g$ is the strength of the qubit-resonator coupling. \textbf{f}, The transition amplitude decreases exponentially as $\wres$ is increased. Unwanted resonances become irrelevant for readout performance. For each $\wres$, the final state $|f\rangle$ is the one satisfying $\hbar\wres = E_f - E_i$ with $|i\rangle = |0\rangle$. The horizontal axis is therefore discrete: each point corresponds to a different final state and to the frequency at which this state is resonant.}
    \label{fig:1}
  \end{center}
\end{figure*}
 
In a variety of physical contexts, {a} periodic drive can destabilize a system by resonantly exciting its modes. {In non-linear, low-dissipation systems, the drive frequency does not have to match the frequency of the mode to excite it.} {Such a} non-linear resonance is possible whenever an integer multiple of the excitation frequency matches an integer combination of the mode frequencies. The effect is famously visible in celestial mechanics, where the asteroid belt is depleted in narrow bands known as Kirkwood gaps. These gaps occur where the orbital period of asteroids is commensurate with that of Jupiter, most prominently at ratios of 3:1, 5:2 and 2:1. Avoiding unwanted resonances of this type is a key consideration for any application relying on periodic drive. In Paul traps, where ions are confined by an oscillating electric potential, non-linear resonances cause ion loss in regions of the stability diagram that are nominally stable \cite{alheit_higher_1996}. The same effect limits particle accelerators, where charged particles are guided by periodically placed magnets: resonances between the betatron oscillations and the revolution frequency drive the beam loss \cite{guignard_general_1978}.

Remarkably, to date, measurement of qubits based on superconducting circuits has remained limited by the problem of unwanted non-linear resonances. There, measurements rely on dispersive probing \cite{blais_cavity_2004, wallraff_strong_2004}, where a microwave drive is employed to address the qubit state \cite{mallet_single-shot_2009, manucharyan_fluxonium_2009, reed_high-fidelity_2010, jeffrey_fast_2014, walter_rapid_2017, dassonneville_fast_2020}. Conventionally, the drive frequency is detuned from but comparable to a difference between the circuit energy levels. Therefore, even though the detuning prevents direct resonant excitation by the drive, the latter can still excite non-linear resonances involving the non-computational states. Such excitation ejects the qubit from the computational subspace \cite{sank_measurement-induced_2016, shillito_dynamics_2022, xiao_diagrammatic_2025, khezri_measurement-induced_2023, dumas_measurement-induced_2024} (Fig.~\ref{fig:1}c), so that the measurement outcome no longer reflects the qubit state. Because the excitation probability grows with drive strength, this limits the measurement power and therefore the achievable measurement speed and fidelity. This problem is especially severe for applications requiring a mid-circuit measurement such as quantum error correction \cite{google_quantum_ai_quantum_2025}. Leakage out of the computational manifold creates a correlated error that persists for multiple correction rounds, violating the core assumption of quantum error correction. Similarly, leakage is detrimental for quantum sensing applications \cite{dixit_searching_2021, albertinale_detecting_2021}. 

The spurious non-linear resonances cannot be eliminated in the standard regime where the probe frequency is close to that of the circuit (Fig.~\ref{fig:1}d). Even worse, the resonance condition is not fixed: the AC Stark shift moves the circuit's transition frequencies in proportion to the drive power, sweeping the working point across resonances during every measurement pulse. Therefore, no choice of drive frequency avoids the non-linear resonances. Is there a way to avoid the unwanted non-linear resonances in superconducting circuits, lifting the restriction on the readout power and thereby improving the measurement fidelity?

{Our work introduces a new high-frequency paradigm for the dispersive readout of a superconducting qubit. We show that when the probe frequency exceeds that of the circuit {by an order of magnitude}, the unwanted non-linear resonances become exponentially weak. This opens the way for improving the measurement fidelity by increasing the power of the measurement, all while retaining the desired quantum non-demolition (QND) measurement character.}

{The origin of the enhanced performance lies in the exponential suppression of the transition matrix elements for the unwanted resonances. The large separation between the transition frequency of the circuit and the readout frequency guarantees that the resonances may occur only between distant quantum states of the circuit. A general property of the wave-functions for highly-excited states is that they are quickly oscillating, in contrast to those of the computational states (Fig.~\ref{fig:1}e). Hence the exponential smallness for the transition matrix elements (Fig.~\ref{fig:1}f). By measuring a transmon superconducting qubit \cite{koch_charge-insensitive_2007, schreier_suppressing_2008} with a frequency of 0.757 GHz through a 9.227 GHz readout resonator, a frequency ratio of 12, we achieve quantum non-demolition measurement fidelity of 99.93\%, record-breaking among superconducting qubits, with a leakage probability of only 0.02\%. The same frequency separation also suppresses radiative decay of the qubit through the readout channel by more than two orders of magnitude, removing the need for a Purcell filter in our setup. Our measurement technique can be readily generalized beyond the transmon qubits \cite{noauthor_see_nodate}.}

\begin{figure*}[t]
  \begin{center}
    \includegraphics[scale = 1]{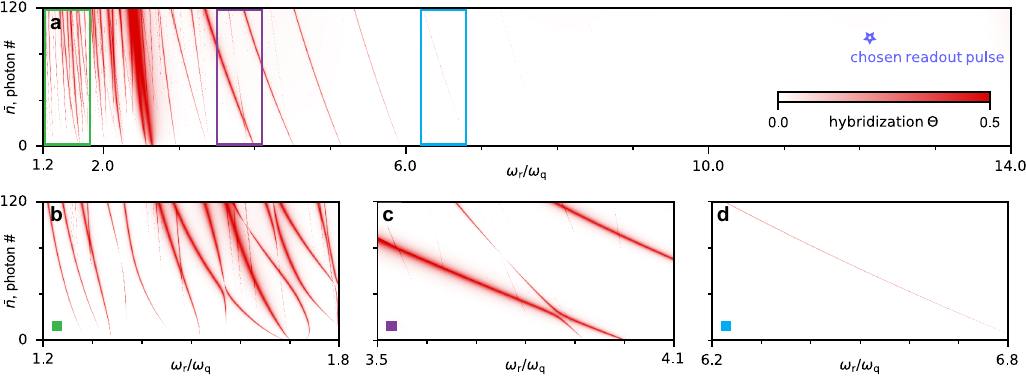}
\caption{Numerically simulated landscape of unwanted resonances for transmon in state $|1\rangle$ ($n_g = 1/4$, $\chi$ is the same as in our experiment). The horizontal axis shows the normalized readout frequency and the vertical axis shows the number of photons in the readout resonator. The color density shows the strength of hybridization $\Theta$ between the computational state $|1\rangle$ and the non-computational states; unwanted resonances correspond to the regions where this hybridization is enhanced.
Exactly on resonance, the highest value, $\Theta = 1/2$, is reached.
\textbf{a}, The resonances in a broad frequency range. The blue star shows the frequency and the maximal photon number of our chosen readout pulse.
\textbf{b}-\textbf{d}, Zoom in on three frequency windows highlighted with colored boxes in panel \textbf{a}. The resonances become progressively less significant as the readout frequency is increased.}
    \label{fig:floquet}
  \end{center}
\end{figure*}

\section*{Concept description}
 
In dispersive readout, a superconducting qubit is coupled to a readout resonator detuned from the qubit transition (Fig.~\ref{fig:1}a,b). Due to this coupling, the resonator frequency shifts by $\chi$ when the qubit is excited from its ground state $|0\rangle$ to the first excited state $|1\rangle$. By measuring the phase of the signal reflected from the resonator, it is then possible to infer the qubit state \cite{blais_cavity_2004, wallraff_strong_2004}. The rate at which we obtain information about the qubit state during readout is proportional to the dispersive shift and the number of photons in the resonator, $\Gamma_\mathrm{meas} \propto \nbar \chi$ \footnote{This result is simplified by assuming the optimal resonator linewidth.
}. 
The measurement speed can thus be enhanced by using higher-power pulses with increased $\bar{n}$. This gives a promise of achieving a fast QND qubit measurement.
 
However, as the drive power is increased, the measurement starts to excite the superconducting circuit to its non-computational states, setting a limit on the achievable QND measurement fidelity. The unwanted excitation happens due to the activation of spurious non-linear resonances in the circuit spectrum \cite{sank_measurement-induced_2016, shillito_dynamics_2022, xiao_diagrammatic_2025, khezri_measurement-induced_2023, dumas_measurement-induced_2024}. It occurs whenever condition $m \hbar \wres = E_{i\rightarrow f} = E_f - E_i$ is satisfied for some integer $m$ and for some final non-computational state $|f\rangle$ (here, $|i\rangle$ is one of the two computational states; $E_i$ and $E_f$ are the energies of the states $|i\rangle$ and $|f\rangle$, respectively). For the standard dispersive readout, the resonator frequency is of the same order as the energy level spacing of the superconducting circuit. In that regime, a spurious resonance is likely to be activated for some $m$ and $f$, regardless of the exact choice of circuit parameters. The issue is further complicated by the fact that during the measurement the energy difference $E_{i\rightarrow f}$ sweeps a certain range due to the AC Stark shift (Fig.~\ref{fig:1}d).

We put forward a new dispersive readout paradigm free of this problem. Within this paradigm, the unwanted resonances are sparse and the corresponding transition matrix elements are exponentially small, as we now explain. Specifically, we consider dispersive readout in the regime where the readout resonator frequency vastly exceeds the typical frequency spacing of the superconducting qubit circuit. In this regime, the strongest unwanted resonance employs a single readout photon, $m=1$, which would excite the circuit to a highly excited final state $|f\rangle$ with $f\gg 1$ (Fig.~\ref{fig:1}e).

The transition amplitude $\Omega_{i\rightarrow f}$ for this resonance is proportional to the matrix element $|\langle f|\partial_\varphi|i\rangle|$ (here $\varphi$ is the phase at the qubit Josephson junction). This matrix element becomes exponentially small as the readout frequency -- and the final state index $f$ -- is increased (Fig.~\ref{fig:1}f). An analytical estimate of this matrix element is given in \cite{noauthor_see_nodate}. There we also show, using a Landau--Zener estimate, that the leakage probability per readout caused by unwanted resonances is expected to be below $\sim 10^{-7}$ for our parameters, and that the suppression of the matrix elements does not require the final state to lie above the edge of the Josephson well. We note that the increase of the readout frequency reduces the dispersive shift $\chi$. This reduction, however, can be compensated by increasing the coupling strength $g$ between the qubit and the readout resonator. 
 
To further demonstrate the suppression of unwanted resonances in the high-frequency regime, we numerically simulate the response of a transmon superconducting qubit to microwave driving (Fig.~\ref{fig:floquet}). In this simulation, we evaluate by how much the computational states hybridize with the non-computational ones due to the driving as a function of drive frequency and power. The drive amplitude is converted into the readout photon number via the AC Stark shift of the $|0\rangle\rightarrow|1\rangle$ transition, $\delta\omega = \chi\nbar$; see Methods. The hybridization is the strongest ($\Theta = 0.5$) precisely at the resonant frequencies. The frequency range around each resonance in which the hybridization is substantial increases with the power. This limits the measurement performance in the standard regime of readout frequencies, $\wres \sim \wq$.

In contrast, when $\wres \gg \wq$ both the density and the strength of the resonances are strongly suppressed. This makes spurious resonances irrelevant for the readout performance. The decrease in the number of spurious resonances with the increase of drive frequency is noted in some of the numeric simulations presented in Ref.~\cite{cohen_reminiscence_2023}.
 
\section*{Experimental setup}
 
\begin{figure*}[t]
    \begin{center}
        \includegraphics[scale = 1]{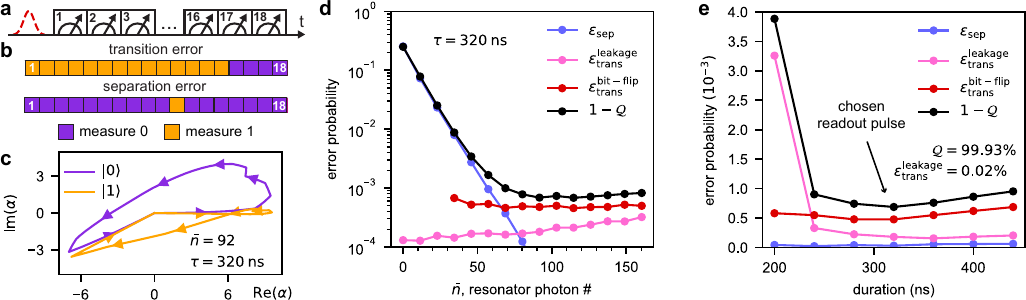}
        \caption{Achieving superconducting qubit measurement with a $99.93\%$ QND measurement fidelity. \textbf{a}, Experimental sequence used to estimate QND fidelity. A $\pi$-pulse (red) is played before the beginning of the sequence half the time. After that, 18 back-to-back qubit measurements (black) are performed. To quantify leakage, we perform additional leakage check measurements at the beginning and end of the sequence \cite{noauthor_see_nodate}. \textbf{b}, A transition error (bit flip) is identified by a sustained switch in state assignment upon repeated measurements. A {separation} error is identified as an isolated change in the state assignment. \textbf{c}, Numerically simulated resonator coherent state trajectories for our pulse (\textit{maximum} photon number is $\bar{n}$). The pulse is optimized to maximize the signal-to-noise ratio for a fixed \textit{mean} photon number. The two trajectories are not mirror images of each other because of the residual Kerr nonlinearity of the resonator. \textbf{d}, Error probability as a function of \textit{maximum} photon number in the resonator during a $320 \: \mathrm{ns}$ readout pulse. $1-\mathcal{Q}$ is the total error probability; $\varepsilon_\mathrm{sep}$ is the separation error probability; $\varepsilon_\mathrm{trans}^\mathrm{bit-flip}$ is the $|0\rangle\leftrightarrow |1\rangle$ transition probability; $\varepsilon_\mathrm{trans}^\mathrm{leakage}$ is the leakage transition probability. {Error bars from Poisson statistics are smaller than the markers. \textbf{e}, Error probability as a function of pulse duration at a fixed signal-to-noise ratio.} The optimal pulse has a duration of $320 \: \mathrm{ns}$, $\mathcal{Q} = 99.931\pm0.003\%$, and leakage probability of $0.018\pm0.002\%$.}  
    \label{fig:QND}
  \end{center}
\end{figure*}  
 
To experimentally demonstrate the advantage of high-frequency readout, we realized a device with a readout resonator with frequency $\wres/2\pi = 9.227\ghz$ and a transmon superconducting qubit with frequency $\wq/2\pi = 0.757\ghz$, such that $\wres/\wq \approx 12$ (Fig.~\ref{fig:1}b). We chose to reduce the transmon frequency from its typical value of several GHz and keep the readout in the standard range so that we can use our existing microwave components and electronics in the readout chain. The measured excited state probability $p_1 = 3\%$ corresponds to an effective temperature of $T_\mathrm{q} = 11\:\mathrm{mK}$.

The large detuning between the transmon and the resonator forces us to couple them strongly to obtain an appreciable resonator dispersive shift $\chi$, such that $g \sim \wq$. A large  interdigitated coupling capacitor in a 2D geometry results in $g / 2\pi = 0.515\ghz$ and $\chi/2\pi = -0.90\mhz$ \footnote{When designing the capacitor structures, we make sure that the frequencies of the associated spurious modes exceed 15 GHz.}. For an efficient readout, the coupling between the resonator and the transmission line $\kappa$ should be comparable to the dispersive shift $\chi$. For our inductively-coupled resonator, $\kappa/2\pi = 1.80\mhz$.
 
An important consideration in dispersive readout is the possibility of radiative decay of the qubit through the readout resonator.
Conventionally, a Purcell filter \cite{reed_fast_2010} is employed to suppress this decay channel. In the high-frequency dispersive readout, however, the radiative decay is suppressed by a large factor $\sim (\wres/\wq)^3$, even without the Purcell filter \cite{noauthor_see_nodate}. Physically, the suppression originates in the frequency dependence of the impedance of the coupling elements: at the qubit frequency, the capacitor coupling the transmon to the resonator acts as an open, while the inductor coupling the resonator to the transmission line acts as a short. The transmon is therefore weakly connected to the dissipative environment at its own frequency. This suppression -- which exceeds two orders of magnitude for our device -- allows us to achieve qubit relaxation time $T_1 = 388\:\mu\mathrm{s}$ limited by material properties. Eliminating the Purcell filter significantly reduces the hardware overhead associated with qubit measurement in quantum processors. The frequency dependence of radiative decay was recently discussed in Ref.~\cite{yen_interferometric_2025}.

\section*{Readout performance}
 
We read out the transmon state by sending a microwave pulse to the readout resonator. Upon reflecting from the resonator, the pulse acquires information about the qubit state. The amplified signal is then demodulated, digitized, and integrated with an envelope that maximizes the signal-to-noise ratio \cite{gambetta_protocols_2007, touzard_gated_2019} by the room-temperature measurement electronics. We assign a state $|0\rangle$ or $|1\rangle$ to the outcome of the measurement based on whether it is closer to the average outcome for $|0\rangle$ or $|1\rangle$, respectively.
 
The readout operation is considered to be complete when the resonator is sufficiently empty of readout photons such that the qubit no longer experiences significant dephasing. To speed up the measurement, we employ a shaped readout pulse which is designed to leave the readout resonator empty of photons at the end of the measurement \cite{mcclure_rapid_2016, hazra_benchmarking_2025}.
The pulses are numerically optimized to maximize the discernibility of $|0\rangle$ and $|1\rangle$ while keeping the time-integrated number of photons in the resonator fixed. Simulated trajectories of the resonator field amplitude under the action of our readout pulse are shown in Fig.~\ref{fig:QND}c.

Ideally, the dispersive measurement described above results in a QND readout of the qubit state. In reality, the performance of the readout is limited by two imperfections. First, noise in the readout signal can lead to a state separation error, where the measurement outcome for a qubit in state $|1\rangle$ falls on the side of the threshold corresponding to state $|0\rangle$ (or vice versa). Second, the transmon can experience a transition error where its state changes during the readout. For example, state $|1\rangle$ can decay to $|0\rangle$ during the readout pulse. The QND fidelity $\mathcal{Q}$ is correspondingly given by
\begin{equation}
    1 - \mathcal{Q} = \varepsilon_{\rm sep} + \varepsilon_{\rm trans}
\end{equation}
with  $\varepsilon_{\rm sep}$ being the misassignment probability and $\varepsilon_{\rm trans}$ being the transition probability \cite{noauthor_see_nodate}. Here, we average $\mathcal{Q}$ over the computational qubit states $|0\rangle$ and $|1\rangle$. Both types of errors should be minimized to obtain a near-unity QND fidelity.

However, minimization of $\varepsilon_{\rm trans}$ is in conflict with minimization of $\varepsilon_{\rm sep}$. The noise in the readout signal cannot be reduced below limits set by the Heisenberg uncertainty principle, so the separation error can only be reduced by increasing the power or duration of the readout. Increasing the duration is undesirable because the transmon would have more time to decay, which increases transition errors. One would hope to reduce both $\varepsilon_{\rm sep}$ and $\varepsilon_{\rm trans}$ by using a short pulse with a high power. However, at higher powers, the drives themselves start to induce state transitions. Such drive-induced transitions bound the achievable QND fidelity. Our high-frequency readout technique suppresses the drive-induced transitions and thus allows us to improve $\mathcal{Q}$.

To optimize the readout performance, we measure the error probabilities as a function of pulse duration and power. We extract the error probabilities from a sequence of back-to-back measurements (Fig.~\ref{fig:QND}a): a transition error is identified by a change of the outcome persisting for at least two readouts, whereas a separation error appears as an isolated change of the outcome (Fig.~\ref{fig:QND}b). A separation error mimics a transition error only if it occurs twice in a row, so the two are distinguishable as long as $\varepsilon_{\rm sep}^2\ll\varepsilon_{\rm trans}$; we report the two errors separately only where this condition holds.

{As a result of optimization of pulse duration and power, we obtain a maximum QND fidelity of $99.931\pm0.003\%$ using a $320\:\mathrm{ns}$-long pulse with a maximum number of photons in the readout resonator of $\bar{n} = 92$ (Fig.~\ref{fig:QND}d,e). To the best of our knowledge, the error in the QND fidelity that we measure is about an order of magnitude smaller than the previous state-of-the-art result for superconducting qubits \cite{swiadek_enhancing_2024, hazra_benchmarking_2025, spring_fast_2025}. This improvement, obtained with a drive power far exceeding that of past experiments, results from suppression of unwanted resonances.}

{The remaining readout error largely stems from two channels: qubit bit flips and leakage from the computational manifold. Bit flips account for $0.048\pm 0.003\%$ of the error; they largely originate from conventional qubit relaxation. The probability of leakage per readout that we observe is about $0.018\pm0.002\%$; the leakage predominantly happens to states $|2\rangle$ and $|3\rangle$ \cite{noauthor_see_nodate}. This leakage is unrelated to unwanted transmon resonances, which would excite the transmon to state $|14\rangle$ or higher in our device. Instead, the residual leakage error is set by a combination of thermal heating as well as inelastic scattering of readout photons \cite{connolly_full_2025}. As discussed in Ref.~\cite{connolly_full_2025}, microwave engineering can be used to suppress errors caused by inelastic scattering. Thermal heating can be suppressed by either elevating the qubit frequency or lowering the mixing chamber temperature.}
 
\section*{Outlook}
We expect the novel high-frequency readout technique that we demonstrated to substantially improve the performance of quantum error correction codes \cite{kitaev_fault-tolerant_2003}. For such codes, achieving low measurement-induced qubit leakage rates is of a paramount importance, as leakage may create uncorrectable correlated errors \cite{aliferis_fault-tolerant_2007, mcewen_et_al_removing_2021, miao_overcoming_2023}. Our work can remove this roadblock leading to a more favorable scaling of error rate with code distance while also simplifying the hardware \cite{noauthor_see_nodate}. High-frequency dispersive readout technique can also benefit sensing applications where precise non-demolition measurement of a superconducting qubit is desirable.
These include search for dark matter \cite{dixit_searching_2021}, detection of single spins \cite{albertinale_detecting_2021}, and probing of exotic Josephson junctions \cite{hays_coherent_2021}.
 
An additional benefit of high-frequency readout is the suppression of frequency crowding in a quantum processor. In our approach, the qubits and the readout resonators occupy separate frequency bands by construction, in contrast to the standard regime where both are placed within the same $4$--$9\ghz$ window. This should allow one to avoid unwanted spectral collisions between processor components, and leaves the full readout band available for multiplexing. The elimination of the Purcell filter removes an additional set of frequencies from the layout.
 
In our work, the transmon superconducting qubit has an unusually small frequency of $\wq / 2\pi = 0.757\ghz$. This allowed us to use a conventional readout resonator frequency $\wres / 2\pi = 9.227\ghz$ while keeping the ratio $\wres/\wq$ high. At face value, such low-frequency transmons seem to encounter two challenges: significant thermal population ($p_1 \sim 3\%$ in our experiment) and slow gate speed limited by reduced transmon anharmonicity. We believe that neither of these factors poses a significant limitation. The thermal population can be straightforwardly constrained using either measurement-based feedback or an unconditional reset \cite{magnard_fast_2018, sunada_fast_2022}.
While the small anharmonicity does limit the gate speed, it need not limit the gate fidelity. At fixed $E_J/E_C$, the anharmonicity scales linearly with the qubit frequency, as does the dielectric-loss-limited decay rate at fixed material quality factor; the coherence-limited gate error is thus expected to be independent of the qubit frequency, and the only drawback is a slower processor clock.

Alternatively, one could use a transmon in a more typical frequency range of several GHz and an unusually high-frequency readout resonator.
 
The readout frequency, however, cannot be raised indefinitely. The readout photons of sufficiently high frequency may break Cooper pairs, either directly or through a multi-photon process, causing a qubit transition error in the same event \cite{houzet_photon-assisted_2019, kishmar_quasiparticle-induced_2026, chowdhury_theory_2026}. At the current error levels, staying below the threshold of the three-photon process is required to make this mechanism irrelevant for readout performance; the inequality $3\wres + \wq < 2\Delta/\hbar$ should thus hold, where $\Delta$ is the superconducting gap. For standard aluminum Josephson junctions, this limits $\wres/2\pi$ to $25$--$30\ghz$. Together with the requirement on high $\wres/\wq$, this points to a transmon at $2$ to $2.5\ghz$ measured with a tone at $\wres/2\pi\sim25\ghz$ as an optimum. We note that higher-gap junctions \cite{anferov_improved_2024} can be used to further increase the readout frequency.
 
{Finally, we note that} a high-frequency drive could be employed to avoid undesired resonances in applications other than measurement. Among them are two-qubit gates \cite{paik_experimental_2016} as well as quantum control of linear oscillators \cite{eickbusch_fast_2022}. The latter can be used for realizing novel bosonic quantum error-correction codes \cite{campagne-ibarcq_quantum_2020, sivak_real-time_2023}.
 
\section*{Acknowledgments}

We thank Christian K. Andersen, Arno Bargerbos, Marta Pita-Vidal, Lukas Splitthoff, Jaap Wesdorp, and Daniel Sank for discussions. We thank Alessandro Miano and Elifnaz {\"O}nder for help with the measurement setup. We thank Jayameenakshi Venkatraman, Xu Xiao, and Rodrigo G. Corti\~{n}as for discussions of Floquet simulations. We thank Daniil Antonenko for discussions of pulse optimization. Finally, we thank Y. Sun, K. Woods, L. McCabe, and M. Rooks for their assistance and guidance in the device fabrication processes. Fabrication facilities use was supported by the Yale Institute for Nanoscience and Quantum Engineering (YINQE) and the Yale University Cleanroom.

\section*{Funding}

This research was sponsored by the Army Research Office (ARO) under grants no.~W911NF-22-1-0053 and W911NF-23-1-0051, by DARPA under grant no.~HR0011-24-2-0346, by the U.S. Department of Energy (DoE), Office of Science, National Quantum Information Science Research Centers, Co-design Center for Quantum Advantage (C2QA) under contract number DE-SC0012704, and by the Air Force Office of Scientific Research (AFOSR) under award number FA9550-21-1-0209. The views and conclusions contained in this document are those of the authors and should not be interpreted as representing the official policies, either expressed or implied, of the ARO, DARPA, DoE, AFOSR or the US Government. The US Government is authorized to reproduce and distribute reprints for Government purposes notwithstanding any copyright notation herein.
 
\section*{Contributions}
T.C. and P.D.K. conceived of the experiment, in equal part. P.D.K., T.C., H.N., and S.D. designed the device. P.D.K., A.Z.D., C.G.L.B., H.N., S.D., and L.F. fabricated the
device. V.R.J. designed the parametric amplifier. A.Z.D. and V.R.J. fabricated the parametric amplifier. T.C., S.H., P.D.K., and W.D. calculated the zeroing pulses. D.W. and T.C. performed
Floquet simulations with feedback from P.D.K.. T.C., P.D.K., C.G.L.B., A.Z.D., V.R.J. designed the packaging and mounted the device. T.C., P.D.K., C.G.L.B., and H.N. collected and analyzed the data. P.D.K. described the device theoretically with input from L.I.G., V.D.K., and T.C.. 
M.K., V.D.K., and P.D.K. generalized the proposed technique beyond transmon.
P.D.K., T.C., L.I.G., and M.H.D. wrote the manuscript with the feedback from other authors.
M.H.D, L.I.G, and V.F. supervised the project.
 
\section*{Competing interests}
L.F. is a consultant and equity holder of D-Wave Quantum. The authors are inventors in a pending patent related to the subject of this work.
 
\section*{Methods}
 
\subsection*{Experimental setup}
 
Our transmon is fabricated on an annealed EFG Sapphire substrate. The transmon Josephson junction is an Al/AlOx/Al trilayer made using the bridge-free technique. The transmon pads, readout resonator, and ground plane are made of tantalum. This was shown to improve the quality factor compared to an all-aluminum transmon fabrication \cite{place_new_2021}. The chip hosting the transmon is placed in a copper package. The package is then embedded in an additional light-tight shield which protects it from stray infrared radiation \cite{connolly_coexistence_2024}; this suppresses the qubit heating and reduces quasiparticle-related decoherence. Inside the shield, we used a low-pass filter to suppress the infrared radiation coming from the readout line. Further details of the setup and fabrication are in the Supplementary Information.
 
In our measurement, we use a near-quantum-limited SNAIL parametric amplifier \cite{frattini_optimizing_2018} as a first stage of amplification. This allows us to make the measurement robust to thermal noise that is inevitably added to the signal as it leaves the dilution refrigerator. For our readout pulses, we use phase sensitive amplification mode to enhance the measurement efficiency, see Supplementary Information for details. The particular SNAIL amplifier that we use \cite{joshi_lumped-element_2025} has a bandwidth of about 250 MHz at 20 dB of gain, where it adds 0.61 photons of noise in the phase-preserving mode. We operate it, however, at an elevated value of gain of $26\:\mathrm{dB}$ with a corresponding reduction in the bandwidth, which remains in excess of that required for our readout pulses. The measurement efficiency that we achieve is $\eta_m = 0.45$, consistent with the loss budget of the output chain \cite{noauthor_see_nodate}.
 
\subsection*{Transmon parameters}
For our transmon, the Josephson energy is $E_J/h = 2.195\ghz$ and the charging energy is $E_C/h = 0.036\ghz$. This results in the qubit frequency $\wq/2\pi = 0.757\ghz$. The Josephson and charging energies are determined from the measured transmon frequency and anharmonicity by comparing these parameters with a numerical simulation.
 
\subsection*{Details of readout pulses}
 
Our readout pulses are numerically optimized to maximize the signal-to-noise ratio for a fixed average photon number in the resonator~\cite{mcclure_rapid_2016, hazra_benchmarking_2025}. When performing the optimization, we add a constraint that the pulses should leave no photons in the readout resonator when the measurement is completed. We choose the frequency of our pulses to be in the middle between resonator frequencies for transmon states $|0\rangle$ and $|1\rangle$. We choose the amplitude of the drive acting on the resonator, $\mathcal{E}(t)$, to be piecewise constant. In contrast to previous works, we use a larger number of segments in our pulses; the segment duration that we choose is 20 ns. In our pulse optimization it is important to account for the residual non-linearity of the readout resonator. For our device, the nonlinearity can be estimated as $K/h = -0.0056\:\mathrm{MHz}$.
 
To verify that our readout pulse empties the resonator after the measurement, we perform a ring-down experiment. In this experiment, we play a single readout pulse and continue to acquire the demodulated signal after the readout pulse is over.
We can determine the number of photons remaining in the resonator based on the amplitude of the acquired signal after the pulse ends.
We calibrate the conversion between the photon number and the recorded amplitude via the qubit AC Stark shift measurement. We find that the residual photon number after the measurement is only $\sim 0.02$ for our optimal readout pulse.
 
\subsection*{Leakage checks}
 
Our protocol for measuring the QND fidelity is complicated by the possibility of leakage to non-computational transmon states. Our readout is only effective at distinguishing states $|0\rangle$ and $|1\rangle$ from each other; non-computational states are mistakenly assigned to either $|0\rangle$ or $|1\rangle$. To remedy this problem, we interleave repeated qubit measurement sequences with leakage checks performed in the phase-preserving amplification mode at a different readout frequency. These checks can distinguish the non-computational states from the computational ones, but are not effective at distinguishing $|0\rangle$ from $|1\rangle$. The leakage checks allow us to both quantify the transition probability to non-computational states and to discard the sequences where the transmon starts from a non-computational state when computing the QND fidelity. Further details of the leakage check measurements are presented in the Supplementary Information.
 
\subsection*{Floquet simulations}
In our Floquet simulations of unwanted resonances, we model the readout signal as a classical monochromatic drive applied to the transmon~\cite{shillito_dynamics_2022, xiao_diagrammatic_2025, dumas_measurement-induced_2024}. The resonances can be identified as regions in the parameter space where the considered computational state appreciably hybridizes with non-computational states due to the presence of the drive. The information about this hybridization is contained in the Floquet modes of the model. To quantify the hybridization, we introduce the quantity
\begin{equation}
\label{eq:hybridization}
    \Theta = 1 - |\langle c | \psi\rangle|^2,
\end{equation}
where $|c\rangle$ is the Stark-shifted computational state and $|\psi\rangle$ is the Floquet mode which has the highest overlap with $|c\rangle$. Away from unwanted resonances $|\psi\rangle\approx |c\rangle$ and $\Theta$ is close to zero. In contrast, exactly on resonance, the Floquet mode becomes an equal superposition between the computational state and a non-computational state leading to $\Theta = 1/2$.
 
To obtain Floquet modes, we numerically diagonalize the propagator over one drive period,
\begin{equation}
\label{eq:propagator}
U(0,T) = \mathcal{T}\exp\left(-\frac{i}{\hbar} \int_0^T H(t)dt\right).
\end{equation}
Here $H(t)$ is the Hamiltonian of the driven transmon and symbol $\mathcal{T}$ denotes time ordering. The propagator is numerically evaluated by trotterizing the time-ordered product with a sufficiently small step. Further details of our numerical simulations are laid out in the Supplementary Information.
 
The drive in this semi-classical model enters through its amplitude $\zeta$, whereas the readout is characterized by the photon number $\nbar$ of the quantum model. The two are connected through the AC Stark shift $\delta\omega$ of the $|0\rangle\rightarrow|1\rangle$ transition: $\delta\omega$ is a known function of $\zeta$ and $\wres$ \cite{noauthor_see_nodate}, and $\delta\omega = \chi\nbar$. We use this correspondence to label the vertical axis of Fig.~\ref{fig:floquet} by $\nbar$.

\section*{Data availability}

The data that support the findings of this study are available from the corresponding authors upon reasonable request.

\section*{Code availability}

The code used for the Floquet simulations is available as the open-source Python package \cite{weiss_floquet_nodate}.
 
\bibliography{references}

\begin{thebibliography}{53}%
\makeatletter
\providecommand \@ifxundefined [1]{%
 \@ifx{#1\undefined}
}%
\providecommand \@ifnum [1]{%
 \ifnum #1\expandafter \@firstoftwo
 \else \expandafter \@secondoftwo
 \fi
}%
\providecommand \@ifx [1]{%
 \ifx #1\expandafter \@firstoftwo
 \else \expandafter \@secondoftwo
 \fi
}%
\providecommand \natexlab [1]{#1}%
\providecommand \enquote  [1]{``#1''}%
\providecommand \bibnamefont  [1]{#1}%
\providecommand \bibfnamefont [1]{#1}%
\providecommand \citenamefont [1]{#1}%
\providecommand \href@noop [0]{\@secondoftwo}%
\providecommand \href [0]{\begingroup \@sanitize@url \@href}%
\providecommand \@href[1]{\@@startlink{#1}\@@href}%
\providecommand \@@href[1]{\endgroup#1\@@endlink}%
\providecommand \@sanitize@url [0]{\catcode `\\12\catcode `\$12\catcode `\&12\catcode `\#12\catcode `\^12\catcode `\_12\catcode `\%12\relax}%
\providecommand \@@startlink[1]{}%
\providecommand \@@endlink[0]{}%
\providecommand \url  [0]{\begingroup\@sanitize@url \@url }%
\providecommand \@url [1]{\endgroup\@href {#1}{\urlprefix }}%
\providecommand \urlprefix  [0]{URL }%
\providecommand \Eprint [0]{\href }%
\providecommand \doibase [0]{https://doi.org/}%
\providecommand \selectlanguage [0]{\@gobble}%
\providecommand \bibinfo  [0]{\@secondoftwo}%
\providecommand \bibfield  [0]{\@secondoftwo}%
\providecommand \translation [1]{[#1]}%
\providecommand \BibitemOpen [0]{}%
\providecommand \bibitemStop [0]{}%
\providecommand \bibitemNoStop [0]{.\EOS\space}%
\providecommand \EOS [0]{\spacefactor3000\relax}%
\providecommand \BibitemShut  [1]{\csname bibitem#1\endcsname}%
\let\auto@bib@innerbib\@empty
\bibitem [{\citenamefont {Alheit}\ \emph {et~al.}(1996)\citenamefont {Alheit}, \citenamefont {Kleineidam}, \citenamefont {Vedel}, \citenamefont {Vedel},\ and\ \citenamefont {Werth}}]{alheit_higher_1996}%
  \BibitemOpen
  \bibfield  {author} {\bibinfo {author} {\bibfnamefont {R.}~\bibnamefont {Alheit}}, \bibinfo {author} {\bibfnamefont {S.}~\bibnamefont {Kleineidam}}, \bibinfo {author} {\bibfnamefont {F.}~\bibnamefont {Vedel}}, \bibinfo {author} {\bibfnamefont {M.}~\bibnamefont {Vedel}},\ and\ \bibinfo {author} {\bibfnamefont {G.}~\bibnamefont {Werth}},\ }\bibfield  {title} {\bibinfo {title} {Higher order non-linear resonances in a {Paul} trap},\ }\href {https://doi.org/10.1016/0168-1176(96)04380-7} {\bibfield  {journal} {\bibinfo  {journal} {International Journal of Mass Spectrometry and Ion Processes}\ }\textbf {\bibinfo {volume} {154}},\ \bibinfo {pages} {155} (\bibinfo {year} {1996})}\BibitemShut {NoStop}%
\bibitem [{\citenamefont {Guignard}(1978)}]{guignard_general_1978}%
  \BibitemOpen
  \bibfield  {author} {\bibinfo {author} {\bibfnamefont {G.}~\bibnamefont {Guignard}},\ }\bibfield  {title} {\bibinfo {title} {A general treatment of resonances in accelerators},\ }\bibfield  {journal} {\bibinfo  {journal} {CERN Academic Training Lecture}\ }\href {https://doi.org/10.5170/CERN-1978-011} {10.5170/CERN-1978-011} (\bibinfo {year} {1978})\BibitemShut {NoStop}%
\bibitem [{\citenamefont {Blais}\ \emph {et~al.}(2004)\citenamefont {Blais}, \citenamefont {Huang}, \citenamefont {Wallraff}, \citenamefont {Girvin},\ and\ \citenamefont {Schoelkopf}}]{blais_cavity_2004}%
  \BibitemOpen
  \bibfield  {author} {\bibinfo {author} {\bibfnamefont {A.}~\bibnamefont {Blais}}, \bibinfo {author} {\bibfnamefont {R.-S.}\ \bibnamefont {Huang}}, \bibinfo {author} {\bibfnamefont {A.}~\bibnamefont {Wallraff}}, \bibinfo {author} {\bibfnamefont {S.~M.}\ \bibnamefont {Girvin}},\ and\ \bibinfo {author} {\bibfnamefont {R.~J.}\ \bibnamefont {Schoelkopf}},\ }\bibfield  {title} {\bibinfo {title} {Cavity quantum electrodynamics for superconducting electrical circuits: {An} architecture for quantum computation},\ }\href {https://doi.org/10.1103/PhysRevA.69.062320} {\bibfield  {journal} {\bibinfo  {journal} {Physical Review A}\ }\textbf {\bibinfo {volume} {69}},\ \bibinfo {pages} {062320} (\bibinfo {year} {2004})}\BibitemShut {NoStop}%
\bibitem [{\citenamefont {Wallraff}\ \emph {et~al.}(2004)\citenamefont {Wallraff}, \citenamefont {Schuster}, \citenamefont {Blais}, \citenamefont {Frunzio}, \citenamefont {Huang}, \citenamefont {Majer}, \citenamefont {Kumar}, \citenamefont {Girvin},\ and\ \citenamefont {Schoelkopf}}]{wallraff_strong_2004}%
  \BibitemOpen
  \bibfield  {author} {\bibinfo {author} {\bibfnamefont {A.}~\bibnamefont {Wallraff}}, \bibinfo {author} {\bibfnamefont {D.~I.}\ \bibnamefont {Schuster}}, \bibinfo {author} {\bibfnamefont {A.}~\bibnamefont {Blais}}, \bibinfo {author} {\bibfnamefont {L.}~\bibnamefont {Frunzio}}, \bibinfo {author} {\bibfnamefont {R.-S.}\ \bibnamefont {Huang}}, \bibinfo {author} {\bibfnamefont {J.}~\bibnamefont {Majer}}, \bibinfo {author} {\bibfnamefont {S.}~\bibnamefont {Kumar}}, \bibinfo {author} {\bibfnamefont {S.~M.}\ \bibnamefont {Girvin}},\ and\ \bibinfo {author} {\bibfnamefont {R.~J.}\ \bibnamefont {Schoelkopf}},\ }\bibfield  {title} {\bibinfo {title} {Strong coupling of a single photon to a superconducting qubit using circuit quantum electrodynamics},\ }\href {https://doi.org/10.1038/nature02851} {\bibfield  {journal} {\bibinfo  {journal} {Nature}\ }\textbf {\bibinfo {volume} {431}},\ \bibinfo {pages} {162} (\bibinfo {year} {2004})}\BibitemShut {NoStop}%
\bibitem [{\citenamefont {Mallet}\ \emph {et~al.}(2009)\citenamefont {Mallet}, \citenamefont {Ong}, \citenamefont {Palacios-Laloy}, \citenamefont {Nguyen}, \citenamefont {Bertet}, \citenamefont {Vion},\ and\ \citenamefont {Esteve}}]{mallet_single-shot_2009}%
  \BibitemOpen
  \bibfield  {author} {\bibinfo {author} {\bibfnamefont {F.}~\bibnamefont {Mallet}}, \bibinfo {author} {\bibfnamefont {F.~R.}\ \bibnamefont {Ong}}, \bibinfo {author} {\bibfnamefont {A.}~\bibnamefont {Palacios-Laloy}}, \bibinfo {author} {\bibfnamefont {F.}~\bibnamefont {Nguyen}}, \bibinfo {author} {\bibfnamefont {P.}~\bibnamefont {Bertet}}, \bibinfo {author} {\bibfnamefont {D.}~\bibnamefont {Vion}},\ and\ \bibinfo {author} {\bibfnamefont {D.}~\bibnamefont {Esteve}},\ }\bibfield  {title} {\bibinfo {title} {Single-shot qubit readout in circuit quantum electrodynamics},\ }\href {https://doi.org/10.1038/nphys1400} {\bibfield  {journal} {\bibinfo  {journal} {Nature Physics}\ }\textbf {\bibinfo {volume} {5}},\ \bibinfo {pages} {791} (\bibinfo {year} {2009})}\BibitemShut {NoStop}%
\bibitem [{\citenamefont {Manucharyan}\ \emph {et~al.}(2009)\citenamefont {Manucharyan}, \citenamefont {Koch}, \citenamefont {Glazman},\ and\ \citenamefont {Devoret}}]{manucharyan_fluxonium_2009}%
  \BibitemOpen
  \bibfield  {author} {\bibinfo {author} {\bibfnamefont {V.~E.}\ \bibnamefont {Manucharyan}}, \bibinfo {author} {\bibfnamefont {J.}~\bibnamefont {Koch}}, \bibinfo {author} {\bibfnamefont {L.~I.}\ \bibnamefont {Glazman}},\ and\ \bibinfo {author} {\bibfnamefont {M.~H.}\ \bibnamefont {Devoret}},\ }\bibfield  {title} {\bibinfo {title} {Fluxonium: {Single} {Cooper}-{Pair} {Circuit} {Free} of {Charge} {Offsets}},\ }\href {https://doi.org/10.1126/science.1175552} {\bibfield  {journal} {\bibinfo  {journal} {Science}\ }\textbf {\bibinfo {volume} {326}},\ \bibinfo {pages} {113} (\bibinfo {year} {2009})}\BibitemShut {NoStop}%
\bibitem [{\citenamefont {Reed}\ \emph {et~al.}(2010{\natexlab{a}})\citenamefont {Reed}, \citenamefont {DiCarlo}, \citenamefont {Johnson}, \citenamefont {Sun}, \citenamefont {Schuster}, \citenamefont {Frunzio},\ and\ \citenamefont {Schoelkopf}}]{reed_high-fidelity_2010}%
  \BibitemOpen
  \bibfield  {author} {\bibinfo {author} {\bibfnamefont {M.~D.}\ \bibnamefont {Reed}}, \bibinfo {author} {\bibfnamefont {L.}~\bibnamefont {DiCarlo}}, \bibinfo {author} {\bibfnamefont {B.~R.}\ \bibnamefont {Johnson}}, \bibinfo {author} {\bibfnamefont {L.}~\bibnamefont {Sun}}, \bibinfo {author} {\bibfnamefont {D.~I.}\ \bibnamefont {Schuster}}, \bibinfo {author} {\bibfnamefont {L.}~\bibnamefont {Frunzio}},\ and\ \bibinfo {author} {\bibfnamefont {R.~J.}\ \bibnamefont {Schoelkopf}},\ }\bibfield  {title} {\bibinfo {title} {High-{Fidelity} {Readout} in {Circuit} {Quantum} {Electrodynamics} {Using} the {Jaynes}-{Cummings} {Nonlinearity}},\ }\href {https://doi.org/10.1103/PhysRevLett.105.173601} {\bibfield  {journal} {\bibinfo  {journal} {Physical Review Letters}\ }\textbf {\bibinfo {volume} {105}},\ \bibinfo {pages} {173601} (\bibinfo {year} {2010}{\natexlab{a}})}\BibitemShut {NoStop}%
\bibitem [{\citenamefont {Jeffrey}\ \emph {et~al.}(2014)\citenamefont {Jeffrey}, \citenamefont {Sank}, \citenamefont {Mutus}, \citenamefont {White}, \citenamefont {Kelly}, \citenamefont {Barends}, \citenamefont {Chen}, \citenamefont {Chen}, \citenamefont {Chiaro}, \citenamefont {Dunsworth}, \citenamefont {Megrant}, \citenamefont {O’Malley}, \citenamefont {Neill}, \citenamefont {Roushan}, \citenamefont {Vainsencher}, \citenamefont {Wenner}, \citenamefont {Cleland},\ and\ \citenamefont {Martinis}}]{jeffrey_fast_2014}%
  \BibitemOpen
  \bibfield  {author} {\bibinfo {author} {\bibfnamefont {E.}~\bibnamefont {Jeffrey}}, \bibinfo {author} {\bibfnamefont {D.}~\bibnamefont {Sank}}, \bibinfo {author} {\bibfnamefont {J.}~\bibnamefont {Mutus}}, \bibinfo {author} {\bibfnamefont {T.}~\bibnamefont {White}}, \bibinfo {author} {\bibfnamefont {J.}~\bibnamefont {Kelly}}, \bibinfo {author} {\bibfnamefont {R.}~\bibnamefont {Barends}}, \bibinfo {author} {\bibfnamefont {Y.}~\bibnamefont {Chen}}, \bibinfo {author} {\bibfnamefont {Z.}~\bibnamefont {Chen}}, \bibinfo {author} {\bibfnamefont {B.}~\bibnamefont {Chiaro}}, \bibinfo {author} {\bibfnamefont {A.}~\bibnamefont {Dunsworth}}, \bibinfo {author} {\bibfnamefont {A.}~\bibnamefont {Megrant}}, \bibinfo {author} {\bibfnamefont {P.}~\bibnamefont {O’Malley}}, \bibinfo {author} {\bibfnamefont {C.}~\bibnamefont {Neill}}, \bibinfo {author} {\bibfnamefont {P.}~\bibnamefont {Roushan}}, \bibinfo {author} {\bibfnamefont {A.}~\bibnamefont {Vainsencher}}, \bibinfo {author} {\bibfnamefont {J.}~\bibnamefont {Wenner}},
  \bibinfo {author} {\bibfnamefont {A.}~\bibnamefont {Cleland}},\ and\ \bibinfo {author} {\bibfnamefont {J.~M.}\ \bibnamefont {Martinis}},\ }\bibfield  {title} {\bibinfo {title} {Fast {Accurate} {State} {Measurement} with {Superconducting} {Qubits}},\ }\href {https://doi.org/10.1103/PhysRevLett.112.190504} {\bibfield  {journal} {\bibinfo  {journal} {Physical Review Letters}\ }\textbf {\bibinfo {volume} {112}},\ \bibinfo {pages} {190504} (\bibinfo {year} {2014})}\BibitemShut {NoStop}%
\bibitem [{\citenamefont {Walter}\ \emph {et~al.}(2017)\citenamefont {Walter}, \citenamefont {Kurpiers}, \citenamefont {Gasparinetti}, \citenamefont {Magnard}, \citenamefont {Potočnik}, \citenamefont {Salathé}, \citenamefont {Pechal}, \citenamefont {Mondal}, \citenamefont {Oppliger}, \citenamefont {Eichler},\ and\ \citenamefont {Wallraff}}]{walter_rapid_2017}%
  \BibitemOpen
  \bibfield  {author} {\bibinfo {author} {\bibfnamefont {T.}~\bibnamefont {Walter}}, \bibinfo {author} {\bibfnamefont {P.}~\bibnamefont {Kurpiers}}, \bibinfo {author} {\bibfnamefont {S.}~\bibnamefont {Gasparinetti}}, \bibinfo {author} {\bibfnamefont {P.}~\bibnamefont {Magnard}}, \bibinfo {author} {\bibfnamefont {A.}~\bibnamefont {Potočnik}}, \bibinfo {author} {\bibfnamefont {Y.}~\bibnamefont {Salathé}}, \bibinfo {author} {\bibfnamefont {M.}~\bibnamefont {Pechal}}, \bibinfo {author} {\bibfnamefont {M.}~\bibnamefont {Mondal}}, \bibinfo {author} {\bibfnamefont {M.}~\bibnamefont {Oppliger}}, \bibinfo {author} {\bibfnamefont {C.}~\bibnamefont {Eichler}},\ and\ \bibinfo {author} {\bibfnamefont {A.}~\bibnamefont {Wallraff}},\ }\bibfield  {title} {\bibinfo {title} {Rapid {High}-{Fidelity} {Single}-{Shot} {Dispersive} {Readout} of {Superconducting} {Qubits}},\ }\href {https://doi.org/10.1103/PhysRevApplied.7.054020} {\bibfield  {journal} {\bibinfo  {journal} {Physical Review Applied}\ }\textbf {\bibinfo {volume}
  {7}},\ \bibinfo {pages} {054020} (\bibinfo {year} {2017})}\BibitemShut {NoStop}%
\bibitem [{\citenamefont {Dassonneville}\ \emph {et~al.}(2020)\citenamefont {Dassonneville}, \citenamefont {Ramos}, \citenamefont {Milchakov}, \citenamefont {Planat}, \citenamefont {Dumur}, \citenamefont {Foroughi}, \citenamefont {Puertas}, \citenamefont {Leger}, \citenamefont {Bharadwaj}, \citenamefont {Delaforce}, \citenamefont {Naud}, \citenamefont {Hasch-Guichard}, \citenamefont {Garcia-Ripoll}, \citenamefont {Roch},\ and\ \citenamefont {Buisson}}]{dassonneville_fast_2020}%
  \BibitemOpen
  \bibfield  {author} {\bibinfo {author} {\bibfnamefont {R.}~\bibnamefont {Dassonneville}}, \bibinfo {author} {\bibfnamefont {T.}~\bibnamefont {Ramos}}, \bibinfo {author} {\bibfnamefont {V.}~\bibnamefont {Milchakov}}, \bibinfo {author} {\bibfnamefont {L.}~\bibnamefont {Planat}}, \bibinfo {author} {\bibfnamefont {E.}~\bibnamefont {Dumur}}, \bibinfo {author} {\bibfnamefont {F.}~\bibnamefont {Foroughi}}, \bibinfo {author} {\bibfnamefont {J.}~\bibnamefont {Puertas}}, \bibinfo {author} {\bibfnamefont {S.}~\bibnamefont {Leger}}, \bibinfo {author} {\bibfnamefont {K.}~\bibnamefont {Bharadwaj}}, \bibinfo {author} {\bibfnamefont {J.}~\bibnamefont {Delaforce}}, \bibinfo {author} {\bibfnamefont {C.}~\bibnamefont {Naud}}, \bibinfo {author} {\bibfnamefont {W.}~\bibnamefont {Hasch-Guichard}}, \bibinfo {author} {\bibfnamefont {J.}~\bibnamefont {Garcia-Ripoll}}, \bibinfo {author} {\bibfnamefont {N.}~\bibnamefont {Roch}},\ and\ \bibinfo {author} {\bibfnamefont {O.}~\bibnamefont {Buisson}},\ }\bibfield  {title} {\bibinfo
  {title} {Fast {High}-{Fidelity} {Quantum} {Nondemolition} {Qubit} {Readout} via a {Nonperturbative} {Cross}-{Kerr} {Coupling}},\ }\href {https://doi.org/10.1103/PhysRevX.10.011045} {\bibfield  {journal} {\bibinfo  {journal} {Physical Review X}\ }\textbf {\bibinfo {volume} {10}},\ \bibinfo {pages} {011045} (\bibinfo {year} {2020})}\BibitemShut {NoStop}%
\bibitem [{\citenamefont {Sank}\ \emph {et~al.}(2016)\citenamefont {Sank}, \citenamefont {Chen},\ and\ \citenamefont {Khezri~et al.}}]{sank_measurement-induced_2016}%
  \BibitemOpen
  \bibfield  {author} {\bibinfo {author} {\bibfnamefont {D.}~\bibnamefont {Sank}}, \bibinfo {author} {\bibfnamefont {Z.}~\bibnamefont {Chen}},\ and\ \bibinfo {author} {\bibfnamefont {M.}~\bibnamefont {Khezri~et al.}},\ }\bibfield  {title} {\bibinfo {title} {Measurement-{Induced} {State} {Transitions} in a {Superconducting} {Qubit}: {Beyond} the {Rotating} {Wave} {Approximation}},\ }\href {https://doi.org/10.1103/PhysRevLett.117.190503} {\bibfield  {journal} {\bibinfo  {journal} {Physical Review Letters}\ }\textbf {\bibinfo {volume} {117}},\ \bibinfo {pages} {190503} (\bibinfo {year} {2016})}\BibitemShut {NoStop}%
\bibitem [{\citenamefont {Shillito}\ \emph {et~al.}(2022)\citenamefont {Shillito}, \citenamefont {Petrescu}, \citenamefont {Cohen}, \citenamefont {Beall}, \citenamefont {Hauru}, \citenamefont {Ganahl}, \citenamefont {Lewis}, \citenamefont {Vidal},\ and\ \citenamefont {Blais}}]{shillito_dynamics_2022}%
  \BibitemOpen
  \bibfield  {author} {\bibinfo {author} {\bibfnamefont {R.}~\bibnamefont {Shillito}}, \bibinfo {author} {\bibfnamefont {A.}~\bibnamefont {Petrescu}}, \bibinfo {author} {\bibfnamefont {J.}~\bibnamefont {Cohen}}, \bibinfo {author} {\bibfnamefont {J.}~\bibnamefont {Beall}}, \bibinfo {author} {\bibfnamefont {M.}~\bibnamefont {Hauru}}, \bibinfo {author} {\bibfnamefont {M.}~\bibnamefont {Ganahl}}, \bibinfo {author} {\bibfnamefont {A.~G.}\ \bibnamefont {Lewis}}, \bibinfo {author} {\bibfnamefont {G.}~\bibnamefont {Vidal}},\ and\ \bibinfo {author} {\bibfnamefont {A.}~\bibnamefont {Blais}},\ }\bibfield  {title} {\bibinfo {title} {Dynamics of {Transmon} {Ionization}},\ }\href {https://doi.org/10.1103/PhysRevApplied.18.034031} {\bibfield  {journal} {\bibinfo  {journal} {Physical Review Applied}\ }\textbf {\bibinfo {volume} {18}},\ \bibinfo {pages} {034031} (\bibinfo {year} {2022})}\BibitemShut {NoStop}%
\bibitem [{\citenamefont {Xiao}\ \emph {et~al.}(2025)\citenamefont {Xiao}, \citenamefont {Venkatraman}, \citenamefont {Cortiñas}, \citenamefont {Chowdhury},\ and\ \citenamefont {Devoret}}]{xiao_diagrammatic_2025}%
  \BibitemOpen
  \bibfield  {author} {\bibinfo {author} {\bibfnamefont {X.}~\bibnamefont {Xiao}}, \bibinfo {author} {\bibfnamefont {J.}~\bibnamefont {Venkatraman}}, \bibinfo {author} {\bibfnamefont {R.~G.}\ \bibnamefont {Cortiñas}}, \bibinfo {author} {\bibfnamefont {S.}~\bibnamefont {Chowdhury}},\ and\ \bibinfo {author} {\bibfnamefont {M.~H.}\ \bibnamefont {Devoret}},\ }\bibfield  {title} {\bibinfo {title} {Diagrammatic method to compute the effective {Hamiltonian} of a driven nonlinear oscillator},\ }\href {https://doi.org/10.1103/d6j2-lsf7} {\bibfield  {journal} {\bibinfo  {journal} {Physical Review Applied}\ }\textbf {\bibinfo {volume} {24}},\ \bibinfo {pages} {044021} (\bibinfo {year} {2025})}\BibitemShut {NoStop}%
\bibitem [{\citenamefont {Khezri}\ and\ \citenamefont {Opremcak~et al.}(2023)}]{khezri_measurement-induced_2023}%
  \BibitemOpen
  \bibfield  {author} {\bibinfo {author} {\bibfnamefont {M.}~\bibnamefont {Khezri}}\ and\ \bibinfo {author} {\bibfnamefont {A.}~\bibnamefont {Opremcak~et al.}},\ }\bibfield  {title} {\bibinfo {title} {Measurement-induced state transitions in a superconducting qubit: {Within} the rotating-wave approximation},\ }\href {https://doi.org/10.1103/PhysRevApplied.20.054008} {\bibfield  {journal} {\bibinfo  {journal} {Physical Review Applied}\ }\textbf {\bibinfo {volume} {20}},\ \bibinfo {pages} {054008} (\bibinfo {year} {2023})}\BibitemShut {NoStop}%
\bibitem [{\citenamefont {Dumas}\ \emph {et~al.}(2024)\citenamefont {Dumas}, \citenamefont {Groleau-Paré}, \citenamefont {McDonald}, \citenamefont {Muñoz-Arias}, \citenamefont {Lledó}, \citenamefont {D’Anjou},\ and\ \citenamefont {Blais}}]{dumas_measurement-induced_2024}%
  \BibitemOpen
  \bibfield  {author} {\bibinfo {author} {\bibfnamefont {M.~F.}\ \bibnamefont {Dumas}}, \bibinfo {author} {\bibfnamefont {B.}~\bibnamefont {Groleau-Paré}}, \bibinfo {author} {\bibfnamefont {A.}~\bibnamefont {McDonald}}, \bibinfo {author} {\bibfnamefont {M.~H.}\ \bibnamefont {Muñoz-Arias}}, \bibinfo {author} {\bibfnamefont {C.}~\bibnamefont {Lledó}}, \bibinfo {author} {\bibfnamefont {B.}~\bibnamefont {D’Anjou}},\ and\ \bibinfo {author} {\bibfnamefont {A.}~\bibnamefont {Blais}},\ }\bibfield  {title} {\bibinfo {title} {Measurement-{Induced} {Transmon} {Ionization}},\ }\href {https://doi.org/10.1103/PhysRevX.14.041023} {\bibfield  {journal} {\bibinfo  {journal} {Physical Review X}\ }\textbf {\bibinfo {volume} {14}},\ \bibinfo {pages} {041023} (\bibinfo {year} {2024})}\BibitemShut {NoStop}%
\bibitem [{\citenamefont {{Google Quantum AI}}(2025)}]{google_quantum_ai_quantum_2025}%
  \BibitemOpen
  \bibfield  {author} {\bibinfo {author} {\bibnamefont {{Google Quantum AI}}},\ }\bibfield  {title} {\bibinfo {title} {Quantum error correction below the surface code threshold},\ }\href {https://doi.org/10.1038/s41586-024-08449-y} {\bibfield  {journal} {\bibinfo  {journal} {Nature}\ }\textbf {\bibinfo {volume} {638}},\ \bibinfo {pages} {920} (\bibinfo {year} {2025})}\BibitemShut {NoStop}%
\bibitem [{\citenamefont {Dixit}\ \emph {et~al.}(2021)\citenamefont {Dixit}, \citenamefont {Chakram}, \citenamefont {He}, \citenamefont {Agrawal}, \citenamefont {Naik}, \citenamefont {Schuster},\ and\ \citenamefont {Chou}}]{dixit_searching_2021}%
  \BibitemOpen
  \bibfield  {author} {\bibinfo {author} {\bibfnamefont {A.~V.}\ \bibnamefont {Dixit}}, \bibinfo {author} {\bibfnamefont {S.}~\bibnamefont {Chakram}}, \bibinfo {author} {\bibfnamefont {K.}~\bibnamefont {He}}, \bibinfo {author} {\bibfnamefont {A.}~\bibnamefont {Agrawal}}, \bibinfo {author} {\bibfnamefont {R.~K.}\ \bibnamefont {Naik}}, \bibinfo {author} {\bibfnamefont {D.~I.}\ \bibnamefont {Schuster}},\ and\ \bibinfo {author} {\bibfnamefont {A.}~\bibnamefont {Chou}},\ }\bibfield  {title} {\bibinfo {title} {Searching for {Dark} {Matter} with a {Superconducting} {Qubit}},\ }\href {https://doi.org/10.1103/PhysRevLett.126.141302} {\bibfield  {journal} {\bibinfo  {journal} {Physical Review Letters}\ }\textbf {\bibinfo {volume} {126}},\ \bibinfo {pages} {141302} (\bibinfo {year} {2021})}\BibitemShut {NoStop}%
\bibitem [{\citenamefont {Albertinale}\ \emph {et~al.}(2021)\citenamefont {Albertinale}, \citenamefont {Balembois}, \citenamefont {Billaud}, \citenamefont {Ranjan}, \citenamefont {Flanigan}, \citenamefont {Schenkel}, \citenamefont {Estève}, \citenamefont {Vion}, \citenamefont {Bertet},\ and\ \citenamefont {Flurin}}]{albertinale_detecting_2021}%
  \BibitemOpen
  \bibfield  {author} {\bibinfo {author} {\bibfnamefont {E.}~\bibnamefont {Albertinale}}, \bibinfo {author} {\bibfnamefont {L.}~\bibnamefont {Balembois}}, \bibinfo {author} {\bibfnamefont {E.}~\bibnamefont {Billaud}}, \bibinfo {author} {\bibfnamefont {V.}~\bibnamefont {Ranjan}}, \bibinfo {author} {\bibfnamefont {D.}~\bibnamefont {Flanigan}}, \bibinfo {author} {\bibfnamefont {T.}~\bibnamefont {Schenkel}}, \bibinfo {author} {\bibfnamefont {D.}~\bibnamefont {Estève}}, \bibinfo {author} {\bibfnamefont {D.}~\bibnamefont {Vion}}, \bibinfo {author} {\bibfnamefont {P.}~\bibnamefont {Bertet}},\ and\ \bibinfo {author} {\bibfnamefont {E.}~\bibnamefont {Flurin}},\ }\bibfield  {title} {\bibinfo {title} {Detecting spins by their fluorescence with a microwave photon counter},\ }\href {https://doi.org/10.1038/s41586-021-04076-z} {\bibfield  {journal} {\bibinfo  {journal} {Nature}\ }\textbf {\bibinfo {volume} {600}},\ \bibinfo {pages} {434} (\bibinfo {year} {2021})}\BibitemShut {NoStop}%
\bibitem [{\citenamefont {Koch}\ \emph {et~al.}(2007)\citenamefont {Koch}, \citenamefont {Yu}, \citenamefont {Gambetta}, \citenamefont {Houck}, \citenamefont {Schuster}, \citenamefont {Majer}, \citenamefont {Blais}, \citenamefont {Devoret}, \citenamefont {Girvin},\ and\ \citenamefont {Schoelkopf}}]{koch_charge-insensitive_2007}%
  \BibitemOpen
  \bibfield  {author} {\bibinfo {author} {\bibfnamefont {J.}~\bibnamefont {Koch}}, \bibinfo {author} {\bibfnamefont {T.~M.}\ \bibnamefont {Yu}}, \bibinfo {author} {\bibfnamefont {J.}~\bibnamefont {Gambetta}}, \bibinfo {author} {\bibfnamefont {A.~A.}\ \bibnamefont {Houck}}, \bibinfo {author} {\bibfnamefont {D.~I.}\ \bibnamefont {Schuster}}, \bibinfo {author} {\bibfnamefont {J.}~\bibnamefont {Majer}}, \bibinfo {author} {\bibfnamefont {A.}~\bibnamefont {Blais}}, \bibinfo {author} {\bibfnamefont {M.~H.}\ \bibnamefont {Devoret}}, \bibinfo {author} {\bibfnamefont {S.~M.}\ \bibnamefont {Girvin}},\ and\ \bibinfo {author} {\bibfnamefont {R.~J.}\ \bibnamefont {Schoelkopf}},\ }\bibfield  {title} {\bibinfo {title} {Charge-insensitive qubit design derived from the {Cooper} pair box},\ }\href {https://doi.org/10.1103/PhysRevA.76.042319} {\bibfield  {journal} {\bibinfo  {journal} {Physical Review A}\ }\textbf {\bibinfo {volume} {76}},\ \bibinfo {pages} {042319} (\bibinfo {year} {2007})}\BibitemShut {NoStop}%
\bibitem [{\citenamefont {Schreier}\ \emph {et~al.}(2008)\citenamefont {Schreier}, \citenamefont {Houck}, \citenamefont {Koch}, \citenamefont {Schuster}, \citenamefont {Johnson}, \citenamefont {Chow}, \citenamefont {Gambetta}, \citenamefont {Majer}, \citenamefont {Frunzio}, \citenamefont {Devoret}, \citenamefont {Girvin},\ and\ \citenamefont {Schoelkopf}}]{schreier_suppressing_2008}%
  \BibitemOpen
  \bibfield  {author} {\bibinfo {author} {\bibfnamefont {J.~A.}\ \bibnamefont {Schreier}}, \bibinfo {author} {\bibfnamefont {A.~A.}\ \bibnamefont {Houck}}, \bibinfo {author} {\bibfnamefont {J.}~\bibnamefont {Koch}}, \bibinfo {author} {\bibfnamefont {D.~I.}\ \bibnamefont {Schuster}}, \bibinfo {author} {\bibfnamefont {B.~R.}\ \bibnamefont {Johnson}}, \bibinfo {author} {\bibfnamefont {J.~M.}\ \bibnamefont {Chow}}, \bibinfo {author} {\bibfnamefont {J.~M.}\ \bibnamefont {Gambetta}}, \bibinfo {author} {\bibfnamefont {J.}~\bibnamefont {Majer}}, \bibinfo {author} {\bibfnamefont {L.}~\bibnamefont {Frunzio}}, \bibinfo {author} {\bibfnamefont {M.~H.}\ \bibnamefont {Devoret}}, \bibinfo {author} {\bibfnamefont {S.~M.}\ \bibnamefont {Girvin}},\ and\ \bibinfo {author} {\bibfnamefont {R.~J.}\ \bibnamefont {Schoelkopf}},\ }\bibfield  {title} {\bibinfo {title} {Suppressing charge noise decoherence in superconducting charge qubits},\ }\href {https://doi.org/10.1103/PhysRevB.77.180502} {\bibfield  {journal} {\bibinfo  {journal}
  {Physical Review B}\ }\textbf {\bibinfo {volume} {77}},\ \bibinfo {pages} {180502} (\bibinfo {year} {2008})}\BibitemShut {NoStop}%
\bibitem [{noa()}]{noauthor_see_nodate}%
  \BibitemOpen
  \href@noop {} {\bibinfo {title} {See supplementary materials.}}\BibitemShut {Stop}%
\bibitem [{Note1()}]{Note1}%
  \BibitemOpen
  \bibinfo {note} {This result is simplified by assuming the optimal resonator linewidth.}\BibitemShut {Stop}%
\bibitem [{\citenamefont {Cohen}\ \emph {et~al.}(2023)\citenamefont {Cohen}, \citenamefont {Petrescu}, \citenamefont {Shillito},\ and\ \citenamefont {Blais}}]{cohen_reminiscence_2023}%
  \BibitemOpen
  \bibfield  {author} {\bibinfo {author} {\bibfnamefont {J.}~\bibnamefont {Cohen}}, \bibinfo {author} {\bibfnamefont {A.}~\bibnamefont {Petrescu}}, \bibinfo {author} {\bibfnamefont {R.}~\bibnamefont {Shillito}},\ and\ \bibinfo {author} {\bibfnamefont {A.}~\bibnamefont {Blais}},\ }\bibfield  {title} {\bibinfo {title} {Reminiscence of {Classical} {Chaos} in {Driven} {Transmons}},\ }\href {https://doi.org/10.1103/PRXQuantum.4.020312} {\bibfield  {journal} {\bibinfo  {journal} {PRX Quantum}\ }\textbf {\bibinfo {volume} {4}},\ \bibinfo {pages} {020312} (\bibinfo {year} {2023})}\BibitemShut {NoStop}%
\bibitem [{Note2()}]{Note2}%
  \BibitemOpen
  \bibinfo {note} {When designing the capacitor structures, we make sure that the frequencies of the associated spurious modes exceed 15 GHz.}\BibitemShut {Stop}%
\bibitem [{\citenamefont {Reed}\ \emph {et~al.}(2010{\natexlab{b}})\citenamefont {Reed}, \citenamefont {Johnson}, \citenamefont {Houck}, \citenamefont {DiCarlo}, \citenamefont {Chow}, \citenamefont {Schuster}, \citenamefont {Frunzio},\ and\ \citenamefont {Schoelkopf}}]{reed_fast_2010}%
  \BibitemOpen
  \bibfield  {author} {\bibinfo {author} {\bibfnamefont {M.~D.}\ \bibnamefont {Reed}}, \bibinfo {author} {\bibfnamefont {B.~R.}\ \bibnamefont {Johnson}}, \bibinfo {author} {\bibfnamefont {A.~A.}\ \bibnamefont {Houck}}, \bibinfo {author} {\bibfnamefont {L.}~\bibnamefont {DiCarlo}}, \bibinfo {author} {\bibfnamefont {J.~M.}\ \bibnamefont {Chow}}, \bibinfo {author} {\bibfnamefont {D.~I.}\ \bibnamefont {Schuster}}, \bibinfo {author} {\bibfnamefont {L.}~\bibnamefont {Frunzio}},\ and\ \bibinfo {author} {\bibfnamefont {R.~J.}\ \bibnamefont {Schoelkopf}},\ }\bibfield  {title} {\bibinfo {title} {Fast reset and suppressing spontaneous emission of a superconducting qubit},\ }\href {https://doi.org/10.1063/1.3435463} {\bibfield  {journal} {\bibinfo  {journal} {Applied Physics Letters}\ }\textbf {\bibinfo {volume} {96}},\ \bibinfo {pages} {203110} (\bibinfo {year} {2010}{\natexlab{b}})}\BibitemShut {NoStop}%
\bibitem [{\citenamefont {Yen}\ \emph {et~al.}(2025)\citenamefont {Yen}, \citenamefont {Ye}, \citenamefont {Peng}, \citenamefont {Wang}, \citenamefont {Cunningham}, \citenamefont {Gingras}, \citenamefont {Niedzielski}, \citenamefont {Stickler}, \citenamefont {Serniak}, \citenamefont {Schwartz},\ and\ \citenamefont {O’Brien}}]{yen_interferometric_2025}%
  \BibitemOpen
  \bibfield  {author} {\bibinfo {author} {\bibfnamefont {A.}~\bibnamefont {Yen}}, \bibinfo {author} {\bibfnamefont {Y.}~\bibnamefont {Ye}}, \bibinfo {author} {\bibfnamefont {K.}~\bibnamefont {Peng}}, \bibinfo {author} {\bibfnamefont {J.}~\bibnamefont {Wang}}, \bibinfo {author} {\bibfnamefont {G.}~\bibnamefont {Cunningham}}, \bibinfo {author} {\bibfnamefont {M.}~\bibnamefont {Gingras}}, \bibinfo {author} {\bibfnamefont {B.~M.}\ \bibnamefont {Niedzielski}}, \bibinfo {author} {\bibfnamefont {H.}~\bibnamefont {Stickler}}, \bibinfo {author} {\bibfnamefont {K.}~\bibnamefont {Serniak}}, \bibinfo {author} {\bibfnamefont {M.~E.}\ \bibnamefont {Schwartz}},\ and\ \bibinfo {author} {\bibfnamefont {K.~P.}\ \bibnamefont {O’Brien}},\ }\bibfield  {title} {\bibinfo {title} {Interferometric {Purcell} suppression of spontaneous emission in a superconducting qubit},\ }\href {https://doi.org/10.1103/PhysRevApplied.23.024068} {\bibfield  {journal} {\bibinfo  {journal} {Physical Review Applied}\ }\textbf {\bibinfo {volume}
  {23}},\ \bibinfo {pages} {024068} (\bibinfo {year} {2025})}\BibitemShut {NoStop}%
\bibitem [{\citenamefont {Gambetta}\ \emph {et~al.}(2007)\citenamefont {Gambetta}, \citenamefont {Braff}, \citenamefont {Wallraff}, \citenamefont {Girvin},\ and\ \citenamefont {Schoelkopf}}]{gambetta_protocols_2007}%
  \BibitemOpen
  \bibfield  {author} {\bibinfo {author} {\bibfnamefont {J.}~\bibnamefont {Gambetta}}, \bibinfo {author} {\bibfnamefont {W.~A.}\ \bibnamefont {Braff}}, \bibinfo {author} {\bibfnamefont {A.}~\bibnamefont {Wallraff}}, \bibinfo {author} {\bibfnamefont {S.~M.}\ \bibnamefont {Girvin}},\ and\ \bibinfo {author} {\bibfnamefont {R.~J.}\ \bibnamefont {Schoelkopf}},\ }\bibfield  {title} {\bibinfo {title} {Protocols for optimal readout of qubits using a continuous quantum nondemolition measurement},\ }\href {https://doi.org/10.1103/PhysRevA.76.012325} {\bibfield  {journal} {\bibinfo  {journal} {Physical Review A}\ }\textbf {\bibinfo {volume} {76}},\ \bibinfo {pages} {012325} (\bibinfo {year} {2007})}\BibitemShut {NoStop}%
\bibitem [{\citenamefont {Touzard}\ \emph {et~al.}(2019)\citenamefont {Touzard}, \citenamefont {Kou}, \citenamefont {Frattini}, \citenamefont {Sivak}, \citenamefont {Puri}, \citenamefont {Grimm}, \citenamefont {Frunzio}, \citenamefont {Shankar},\ and\ \citenamefont {Devoret}}]{touzard_gated_2019}%
  \BibitemOpen
  \bibfield  {author} {\bibinfo {author} {\bibfnamefont {S.}~\bibnamefont {Touzard}}, \bibinfo {author} {\bibfnamefont {A.}~\bibnamefont {Kou}}, \bibinfo {author} {\bibfnamefont {N.}~\bibnamefont {Frattini}}, \bibinfo {author} {\bibfnamefont {V.}~\bibnamefont {Sivak}}, \bibinfo {author} {\bibfnamefont {S.}~\bibnamefont {Puri}}, \bibinfo {author} {\bibfnamefont {A.}~\bibnamefont {Grimm}}, \bibinfo {author} {\bibfnamefont {L.}~\bibnamefont {Frunzio}}, \bibinfo {author} {\bibfnamefont {S.}~\bibnamefont {Shankar}},\ and\ \bibinfo {author} {\bibfnamefont {M.}~\bibnamefont {Devoret}},\ }\bibfield  {title} {\bibinfo {title} {Gated {Conditional} {Displacement} {Readout} of {Superconducting} {Qubits}},\ }\href {https://doi.org/10.1103/PhysRevLett.122.080502} {\bibfield  {journal} {\bibinfo  {journal} {Physical Review Letters}\ }\textbf {\bibinfo {volume} {122}},\ \bibinfo {pages} {080502} (\bibinfo {year} {2019})}\BibitemShut {NoStop}%
\bibitem [{\citenamefont {McClure}\ \emph {et~al.}(2016)\citenamefont {McClure}, \citenamefont {Paik}, \citenamefont {Bishop}, \citenamefont {Steffen}, \citenamefont {Chow},\ and\ \citenamefont {Gambetta}}]{mcclure_rapid_2016}%
  \BibitemOpen
  \bibfield  {author} {\bibinfo {author} {\bibfnamefont {D.}~\bibnamefont {McClure}}, \bibinfo {author} {\bibfnamefont {H.}~\bibnamefont {Paik}}, \bibinfo {author} {\bibfnamefont {L.}~\bibnamefont {Bishop}}, \bibinfo {author} {\bibfnamefont {M.}~\bibnamefont {Steffen}}, \bibinfo {author} {\bibfnamefont {J.~M.}\ \bibnamefont {Chow}},\ and\ \bibinfo {author} {\bibfnamefont {J.~M.}\ \bibnamefont {Gambetta}},\ }\bibfield  {title} {\bibinfo {title} {Rapid {Driven} {Reset} of a {Qubit} {Readout} {Resonator}},\ }\href {https://doi.org/10.1103/PhysRevApplied.5.011001} {\bibfield  {journal} {\bibinfo  {journal} {Physical Review Applied}\ }\textbf {\bibinfo {volume} {5}},\ \bibinfo {pages} {011001} (\bibinfo {year} {2016})}\BibitemShut {NoStop}%
\bibitem [{\citenamefont {Hazra}\ \emph {et~al.}(2025)\citenamefont {Hazra}, \citenamefont {Dai}, \citenamefont {Connolly}, \citenamefont {Kurilovich}, \citenamefont {Wang}, \citenamefont {Frunzio},\ and\ \citenamefont {Devoret}}]{hazra_benchmarking_2025}%
  \BibitemOpen
  \bibfield  {author} {\bibinfo {author} {\bibfnamefont {S.}~\bibnamefont {Hazra}}, \bibinfo {author} {\bibfnamefont {W.}~\bibnamefont {Dai}}, \bibinfo {author} {\bibfnamefont {T.}~\bibnamefont {Connolly}}, \bibinfo {author} {\bibfnamefont {P.}~\bibnamefont {Kurilovich}}, \bibinfo {author} {\bibfnamefont {Z.}~\bibnamefont {Wang}}, \bibinfo {author} {\bibfnamefont {L.}~\bibnamefont {Frunzio}},\ and\ \bibinfo {author} {\bibfnamefont {M.}~\bibnamefont {Devoret}},\ }\bibfield  {title} {\bibinfo {title} {Benchmarking the {Readout} of a {Superconducting} {Qubit} for {Repeated} {Measurements}},\ }\href {https://doi.org/10.1103/PhysRevLett.134.100601} {\bibfield  {journal} {\bibinfo  {journal} {Physical Review Letters}\ }\textbf {\bibinfo {volume} {134}},\ \bibinfo {pages} {100601} (\bibinfo {year} {2025})},\ \bibinfo {note} {publisher: American Physical Society}\BibitemShut {NoStop}%
\bibitem [{\citenamefont {Swiadek}\ \emph {et~al.}(2024)\citenamefont {Swiadek}, \citenamefont {Shillito}, \citenamefont {Magnard}, \citenamefont {Remm}, \citenamefont {Hellings}, \citenamefont {Lacroix}, \citenamefont {Ficheux}, \citenamefont {Zanuz}, \citenamefont {Norris}, \citenamefont {Blais}, \citenamefont {Krinner},\ and\ \citenamefont {Wallraff}}]{swiadek_enhancing_2024}%
  \BibitemOpen
  \bibfield  {author} {\bibinfo {author} {\bibfnamefont {F.}~\bibnamefont {Swiadek}}, \bibinfo {author} {\bibfnamefont {R.}~\bibnamefont {Shillito}}, \bibinfo {author} {\bibfnamefont {P.}~\bibnamefont {Magnard}}, \bibinfo {author} {\bibfnamefont {A.}~\bibnamefont {Remm}}, \bibinfo {author} {\bibfnamefont {C.}~\bibnamefont {Hellings}}, \bibinfo {author} {\bibfnamefont {N.}~\bibnamefont {Lacroix}}, \bibinfo {author} {\bibfnamefont {Q.}~\bibnamefont {Ficheux}}, \bibinfo {author} {\bibfnamefont {D.~C.}\ \bibnamefont {Zanuz}}, \bibinfo {author} {\bibfnamefont {G.~J.}\ \bibnamefont {Norris}}, \bibinfo {author} {\bibfnamefont {A.}~\bibnamefont {Blais}}, \bibinfo {author} {\bibfnamefont {S.}~\bibnamefont {Krinner}},\ and\ \bibinfo {author} {\bibfnamefont {A.}~\bibnamefont {Wallraff}},\ }\bibfield  {title} {\bibinfo {title} {Enhancing {Dispersive} {Readout} of {Superconducting} {Qubits} through {Dynamic} {Control} of the {Dispersive} {Shift}: {Experiment} and {Theory}},\ }\href
  {https://doi.org/10.1103/PRXQuantum.5.040326} {\bibfield  {journal} {\bibinfo  {journal} {PRX Quantum}\ }\textbf {\bibinfo {volume} {5}},\ \bibinfo {pages} {040326} (\bibinfo {year} {2024})}\BibitemShut {NoStop}%
\bibitem [{\citenamefont {Spring}\ \emph {et~al.}(2025)\citenamefont {Spring}, \citenamefont {Milanovic}, \citenamefont {Sunada}, \citenamefont {Wang}, \citenamefont {Van~Loo}, \citenamefont {Tamate},\ and\ \citenamefont {Nakamura}}]{spring_fast_2025}%
  \BibitemOpen
  \bibfield  {author} {\bibinfo {author} {\bibfnamefont {P.~A.}\ \bibnamefont {Spring}}, \bibinfo {author} {\bibfnamefont {L.}~\bibnamefont {Milanovic}}, \bibinfo {author} {\bibfnamefont {Y.}~\bibnamefont {Sunada}}, \bibinfo {author} {\bibfnamefont {S.}~\bibnamefont {Wang}}, \bibinfo {author} {\bibfnamefont {A.~F.}\ \bibnamefont {Van~Loo}}, \bibinfo {author} {\bibfnamefont {S.}~\bibnamefont {Tamate}},\ and\ \bibinfo {author} {\bibfnamefont {Y.}~\bibnamefont {Nakamura}},\ }\bibfield  {title} {\bibinfo {title} {Fast {Multiplexed} {Superconducting}-{Qubit} {Readout} with {Intrinsic} {Purcell} {Filtering} {Using} a {Multiconductor} {Transmission} {Line}},\ }\href {https://doi.org/10.1103/PRXQuantum.6.020345} {\bibfield  {journal} {\bibinfo  {journal} {PRX Quantum}\ }\textbf {\bibinfo {volume} {6}},\ \bibinfo {pages} {020345} (\bibinfo {year} {2025})}\BibitemShut {NoStop}%
\bibitem [{\citenamefont {Connolly}\ \emph {et~al.}(2025)\citenamefont {Connolly}, \citenamefont {Kurilovich}, \citenamefont {Kurilovich}, \citenamefont {Bøttcher}, \citenamefont {Hazra}, \citenamefont {Dai}, \citenamefont {Ding}, \citenamefont {Joshi}, \citenamefont {Nho}, \citenamefont {Diamond}, \citenamefont {Weiss}, \citenamefont {Fatemi}, \citenamefont {Frunzio}, \citenamefont {Glazman},\ and\ \citenamefont {Devoret}}]{connolly_full_2025}%
  \BibitemOpen
  \bibfield  {author} {\bibinfo {author} {\bibfnamefont {T.}~\bibnamefont {Connolly}}, \bibinfo {author} {\bibfnamefont {P.~D.}\ \bibnamefont {Kurilovich}}, \bibinfo {author} {\bibfnamefont {V.~D.}\ \bibnamefont {Kurilovich}}, \bibinfo {author} {\bibfnamefont {C.~G.~L.}\ \bibnamefont {Bøttcher}}, \bibinfo {author} {\bibfnamefont {S.}~\bibnamefont {Hazra}}, \bibinfo {author} {\bibfnamefont {W.}~\bibnamefont {Dai}}, \bibinfo {author} {\bibfnamefont {A.~Z.}\ \bibnamefont {Ding}}, \bibinfo {author} {\bibfnamefont {V.~R.}\ \bibnamefont {Joshi}}, \bibinfo {author} {\bibfnamefont {H.}~\bibnamefont {Nho}}, \bibinfo {author} {\bibfnamefont {S.}~\bibnamefont {Diamond}}, \bibinfo {author} {\bibfnamefont {D.~K.}\ \bibnamefont {Weiss}}, \bibinfo {author} {\bibfnamefont {V.}~\bibnamefont {Fatemi}}, \bibinfo {author} {\bibfnamefont {L.}~\bibnamefont {Frunzio}}, \bibinfo {author} {\bibfnamefont {L.~I.}\ \bibnamefont {Glazman}},\ and\ \bibinfo {author} {\bibfnamefont {M.~H.}\ \bibnamefont {Devoret}},\ }\href
  {https://doi.org/10.48550/arXiv.2506.05306} {\bibinfo {title} {Full characterization of measurement-induced transitions of a superconducting qubit}} (\bibinfo {year} {2025}),\ \bibinfo {note} {arXiv:2506.05306 [quant-ph]}\BibitemShut {NoStop}%
\bibitem [{\citenamefont {Kitaev}(2003)}]{kitaev_fault-tolerant_2003}%
  \BibitemOpen
  \bibfield  {author} {\bibinfo {author} {\bibfnamefont {A.~Y.}\ \bibnamefont {Kitaev}},\ }\bibfield  {title} {\bibinfo {title} {Fault-tolerant quantum computation by anyons},\ }\href {https://doi.org/10.1016/S0003-4916(02)00018-0} {\bibfield  {journal} {\bibinfo  {journal} {Annals of Physics}\ }\textbf {\bibinfo {volume} {303}},\ \bibinfo {pages} {2} (\bibinfo {year} {2003})}\BibitemShut {NoStop}%
\bibitem [{\citenamefont {Aliferis}\ and\ \citenamefont {Terhal}(2007)}]{aliferis_fault-tolerant_2007}%
  \BibitemOpen
  \bibfield  {author} {\bibinfo {author} {\bibfnamefont {P.}~\bibnamefont {Aliferis}}\ and\ \bibinfo {author} {\bibfnamefont {B.~M.}\ \bibnamefont {Terhal}},\ }\bibfield  {title} {\bibinfo {title} {Fault-tolerant quantum computation for local leakage faults},\ }\href@noop {} {\bibfield  {journal} {\bibinfo  {journal} {Quantum Info. Comput.}\ }\textbf {\bibinfo {volume} {7}},\ \bibinfo {pages} {139} (\bibinfo {year} {2007})}\BibitemShut {NoStop}%
\bibitem [{\citenamefont {McEwen~et al.}(2021)}]{mcewen_et_al_removing_2021}%
  \BibitemOpen
  \bibfield  {author} {\bibinfo {author} {\bibfnamefont {M.}~\bibnamefont {McEwen~et al.}},\ }\bibfield  {title} {\bibinfo {title} {Removing leakage-induced correlated errors in superconducting quantum error correction},\ }\href {https://doi.org/10.1038/s41467-021-21982-y} {\bibfield  {journal} {\bibinfo  {journal} {Nature Communications}\ }\textbf {\bibinfo {volume} {12}},\ \bibinfo {pages} {1761} (\bibinfo {year} {2021})}\BibitemShut {NoStop}%
\bibitem [{\citenamefont {Miao}\ and\ \citenamefont {McEwen~et al.}(2023)}]{miao_overcoming_2023}%
  \BibitemOpen
  \bibfield  {author} {\bibinfo {author} {\bibfnamefont {K.~C.}\ \bibnamefont {Miao}}\ and\ \bibinfo {author} {\bibfnamefont {M.}~\bibnamefont {McEwen~et al.}},\ }\bibfield  {title} {\bibinfo {title} {Overcoming leakage in quantum error correction},\ }\href {https://doi.org/10.1038/s41567-023-02226-w} {\bibfield  {journal} {\bibinfo  {journal} {Nature Physics}\ }\textbf {\bibinfo {volume} {19}},\ \bibinfo {pages} {1780} (\bibinfo {year} {2023})}\BibitemShut {NoStop}%
\bibitem [{\citenamefont {Hays}\ \emph {et~al.}(2021)\citenamefont {Hays}, \citenamefont {Fatemi}, \citenamefont {Bouman}, \citenamefont {Cerrillo}, \citenamefont {Diamond}, \citenamefont {Serniak}, \citenamefont {Connolly}, \citenamefont {Krogstrup}, \citenamefont {Nygård}, \citenamefont {Levy~Yeyati}, \citenamefont {Geresdi},\ and\ \citenamefont {Devoret}}]{hays_coherent_2021}%
  \BibitemOpen
  \bibfield  {author} {\bibinfo {author} {\bibfnamefont {M.}~\bibnamefont {Hays}}, \bibinfo {author} {\bibfnamefont {V.}~\bibnamefont {Fatemi}}, \bibinfo {author} {\bibfnamefont {D.}~\bibnamefont {Bouman}}, \bibinfo {author} {\bibfnamefont {J.}~\bibnamefont {Cerrillo}}, \bibinfo {author} {\bibfnamefont {S.}~\bibnamefont {Diamond}}, \bibinfo {author} {\bibfnamefont {K.}~\bibnamefont {Serniak}}, \bibinfo {author} {\bibfnamefont {T.}~\bibnamefont {Connolly}}, \bibinfo {author} {\bibfnamefont {P.}~\bibnamefont {Krogstrup}}, \bibinfo {author} {\bibfnamefont {J.}~\bibnamefont {Nygård}}, \bibinfo {author} {\bibfnamefont {A.}~\bibnamefont {Levy~Yeyati}}, \bibinfo {author} {\bibfnamefont {A.}~\bibnamefont {Geresdi}},\ and\ \bibinfo {author} {\bibfnamefont {M.~H.}\ \bibnamefont {Devoret}},\ }\bibfield  {title} {\bibinfo {title} {Coherent manipulation of an {Andreev} spin qubit},\ }\href {https://doi.org/10.1126/science.abf0345} {\bibfield  {journal} {\bibinfo  {journal} {Science}\ }\textbf {\bibinfo {volume} {373}},\
  \bibinfo {pages} {430} (\bibinfo {year} {2021})}\BibitemShut {NoStop}%
\bibitem [{\citenamefont {Magnard}\ \emph {et~al.}(2018)\citenamefont {Magnard}, \citenamefont {Kurpiers}, \citenamefont {Royer}, \citenamefont {Walter}, \citenamefont {Besse}, \citenamefont {Gasparinetti}, \citenamefont {Pechal}, \citenamefont {Heinsoo}, \citenamefont {Storz}, \citenamefont {Blais},\ and\ \citenamefont {Wallraff}}]{magnard_fast_2018}%
  \BibitemOpen
  \bibfield  {author} {\bibinfo {author} {\bibfnamefont {P.}~\bibnamefont {Magnard}}, \bibinfo {author} {\bibfnamefont {P.}~\bibnamefont {Kurpiers}}, \bibinfo {author} {\bibfnamefont {B.}~\bibnamefont {Royer}}, \bibinfo {author} {\bibfnamefont {T.}~\bibnamefont {Walter}}, \bibinfo {author} {\bibfnamefont {J.-C.}\ \bibnamefont {Besse}}, \bibinfo {author} {\bibfnamefont {S.}~\bibnamefont {Gasparinetti}}, \bibinfo {author} {\bibfnamefont {M.}~\bibnamefont {Pechal}}, \bibinfo {author} {\bibfnamefont {J.}~\bibnamefont {Heinsoo}}, \bibinfo {author} {\bibfnamefont {S.}~\bibnamefont {Storz}}, \bibinfo {author} {\bibfnamefont {A.}~\bibnamefont {Blais}},\ and\ \bibinfo {author} {\bibfnamefont {A.}~\bibnamefont {Wallraff}},\ }\bibfield  {title} {\bibinfo {title} {Fast and {Unconditional} {All}-{Microwave} {Reset} of a {Superconducting} {Qubit}},\ }\href {https://doi.org/10.1103/PhysRevLett.121.060502} {\bibfield  {journal} {\bibinfo  {journal} {Physical Review Letters}\ }\textbf {\bibinfo {volume} {121}},\ \bibinfo
  {pages} {060502} (\bibinfo {year} {2018})}\BibitemShut {NoStop}%
\bibitem [{\citenamefont {Sunada}\ \emph {et~al.}(2022)\citenamefont {Sunada}, \citenamefont {Kono}, \citenamefont {Ilves}, \citenamefont {Tamate}, \citenamefont {Sugiyama}, \citenamefont {Tabuchi},\ and\ \citenamefont {Nakamura}}]{sunada_fast_2022}%
  \BibitemOpen
  \bibfield  {author} {\bibinfo {author} {\bibfnamefont {Y.}~\bibnamefont {Sunada}}, \bibinfo {author} {\bibfnamefont {S.}~\bibnamefont {Kono}}, \bibinfo {author} {\bibfnamefont {J.}~\bibnamefont {Ilves}}, \bibinfo {author} {\bibfnamefont {S.}~\bibnamefont {Tamate}}, \bibinfo {author} {\bibfnamefont {T.}~\bibnamefont {Sugiyama}}, \bibinfo {author} {\bibfnamefont {Y.}~\bibnamefont {Tabuchi}},\ and\ \bibinfo {author} {\bibfnamefont {Y.}~\bibnamefont {Nakamura}},\ }\bibfield  {title} {\bibinfo {title} {Fast {Readout} and {Reset} of a {Superconducting} {Qubit} {Coupled} to a {Resonator} with an {Intrinsic} {Purcell} {Filter}},\ }\href {https://doi.org/10.1103/PhysRevApplied.17.044016} {\bibfield  {journal} {\bibinfo  {journal} {Physical Review Applied}\ }\textbf {\bibinfo {volume} {17}},\ \bibinfo {pages} {044016} (\bibinfo {year} {2022})}\BibitemShut {NoStop}%
\bibitem [{\citenamefont {Houzet}\ \emph {et~al.}(2019)\citenamefont {Houzet}, \citenamefont {Serniak}, \citenamefont {Catelani}, \citenamefont {Devoret},\ and\ \citenamefont {Glazman}}]{houzet_photon-assisted_2019}%
  \BibitemOpen
  \bibfield  {author} {\bibinfo {author} {\bibfnamefont {M.}~\bibnamefont {Houzet}}, \bibinfo {author} {\bibfnamefont {K.}~\bibnamefont {Serniak}}, \bibinfo {author} {\bibfnamefont {G.}~\bibnamefont {Catelani}}, \bibinfo {author} {\bibfnamefont {M.}~\bibnamefont {Devoret}},\ and\ \bibinfo {author} {\bibfnamefont {L.}~\bibnamefont {Glazman}},\ }\bibfield  {title} {\bibinfo {title} {Photon-assisted charge-parity jumps in a superconducting qubit},\ }\href {https://doi.org/10.1103/PhysRevLett.123.107704} {\bibfield  {journal} {\bibinfo  {journal} {Physical Review Letters}\ }\textbf {\bibinfo {volume} {123}},\ \bibinfo {pages} {107704} (\bibinfo {year} {2019})}\BibitemShut {NoStop}%
\bibitem [{\citenamefont {Kishmar}\ \emph {et~al.}(2026)\citenamefont {Kishmar}, \citenamefont {Kurilovich}, \citenamefont {Klots}, \citenamefont {Connolly}, \citenamefont {Aleiner},\ and\ \citenamefont {Kurilovich}}]{kishmar_quasiparticle-induced_2026}%
  \BibitemOpen
  \bibfield  {author} {\bibinfo {author} {\bibfnamefont {M.}~\bibnamefont {Kishmar}}, \bibinfo {author} {\bibfnamefont {P.~D.}\ \bibnamefont {Kurilovich}}, \bibinfo {author} {\bibfnamefont {A.}~\bibnamefont {Klots}}, \bibinfo {author} {\bibfnamefont {T.}~\bibnamefont {Connolly}}, \bibinfo {author} {\bibfnamefont {I.~L.}\ \bibnamefont {Aleiner}},\ and\ \bibinfo {author} {\bibfnamefont {V.~D.}\ \bibnamefont {Kurilovich}},\ }\bibfield  {title} {\bibinfo {title} {Quasiparticle-induced decoherence of a driven superconducting qubit},\ }\href {https://doi.org/10.1103/3q5z-tk7c} {\bibfield  {journal} {\bibinfo  {journal} {Physical Review Applied}\ }\textbf {\bibinfo {volume} {26}},\ \bibinfo {pages} {014091} (\bibinfo {year} {2026})},\ \bibinfo {note} {publisher: American Physical Society}\BibitemShut {NoStop}%
\bibitem [{\citenamefont {Chowdhury}\ \emph {et~al.}(2026)\citenamefont {Chowdhury}, \citenamefont {Hays}, \citenamefont {Jha}, \citenamefont {Serniak}, \citenamefont {Orlando}, \citenamefont {Grover},\ and\ \citenamefont {Oliver}}]{chowdhury_theory_2026}%
  \BibitemOpen
  \bibfield  {author} {\bibinfo {author} {\bibfnamefont {S.}~\bibnamefont {Chowdhury}}, \bibinfo {author} {\bibfnamefont {M.}~\bibnamefont {Hays}}, \bibinfo {author} {\bibfnamefont {S.~R.}\ \bibnamefont {Jha}}, \bibinfo {author} {\bibfnamefont {K.}~\bibnamefont {Serniak}}, \bibinfo {author} {\bibfnamefont {T.~P.}\ \bibnamefont {Orlando}}, \bibinfo {author} {\bibfnamefont {J.~A.}\ \bibnamefont {Grover}},\ and\ \bibinfo {author} {\bibfnamefont {W.~D.}\ \bibnamefont {Oliver}},\ }\bibfield  {title} {\bibinfo {title} {Theory of quasiparticle generation by microwave drives in superconducting qubits},\ }\href {https://doi.org/10.1103/xpxb-d9xd} {\bibfield  {journal} {\bibinfo  {journal} {Physical Review Applied}\ }\textbf {\bibinfo {volume} {25}},\ \bibinfo {pages} {014042} (\bibinfo {year} {2026})},\ \bibinfo {note} {publisher: American Physical Society}\BibitemShut {NoStop}%
\bibitem [{\citenamefont {Anferov}\ \emph {et~al.}(2024)\citenamefont {Anferov}, \citenamefont {Lee}, \citenamefont {Zhao}, \citenamefont {Simon},\ and\ \citenamefont {Schuster}}]{anferov_improved_2024}%
  \BibitemOpen
  \bibfield  {author} {\bibinfo {author} {\bibfnamefont {A.}~\bibnamefont {Anferov}}, \bibinfo {author} {\bibfnamefont {K.-H.}\ \bibnamefont {Lee}}, \bibinfo {author} {\bibfnamefont {F.}~\bibnamefont {Zhao}}, \bibinfo {author} {\bibfnamefont {J.}~\bibnamefont {Simon}},\ and\ \bibinfo {author} {\bibfnamefont {D.~I.}\ \bibnamefont {Schuster}},\ }\bibfield  {title} {\bibinfo {title} {Improved coherence in optically defined niobium trilayer-junction qubits},\ }\href {https://doi.org/10.1103/PhysRevApplied.21.024047} {\bibfield  {journal} {\bibinfo  {journal} {Physical Review Applied}\ }\textbf {\bibinfo {volume} {21}},\ \bibinfo {pages} {024047} (\bibinfo {year} {2024})}\BibitemShut {NoStop}%
\bibitem [{\citenamefont {Paik}\ \emph {et~al.}(2016)\citenamefont {Paik}, \citenamefont {Mezzacapo}, \citenamefont {Sandberg}, \citenamefont {McClure}, \citenamefont {Abdo}, \citenamefont {Córcoles}, \citenamefont {Dial}, \citenamefont {Bogorin}, \citenamefont {Plourde}, \citenamefont {Steffen}, \citenamefont {Cross}, \citenamefont {Gambetta},\ and\ \citenamefont {Chow}}]{paik_experimental_2016}%
  \BibitemOpen
  \bibfield  {author} {\bibinfo {author} {\bibfnamefont {H.}~\bibnamefont {Paik}}, \bibinfo {author} {\bibfnamefont {A.}~\bibnamefont {Mezzacapo}}, \bibinfo {author} {\bibfnamefont {M.}~\bibnamefont {Sandberg}}, \bibinfo {author} {\bibfnamefont {D.}~\bibnamefont {McClure}}, \bibinfo {author} {\bibfnamefont {B.}~\bibnamefont {Abdo}}, \bibinfo {author} {\bibfnamefont {A.}~\bibnamefont {Córcoles}}, \bibinfo {author} {\bibfnamefont {O.}~\bibnamefont {Dial}}, \bibinfo {author} {\bibfnamefont {D.}~\bibnamefont {Bogorin}}, \bibinfo {author} {\bibfnamefont {B.}~\bibnamefont {Plourde}}, \bibinfo {author} {\bibfnamefont {M.}~\bibnamefont {Steffen}}, \bibinfo {author} {\bibfnamefont {A.}~\bibnamefont {Cross}}, \bibinfo {author} {\bibfnamefont {J.}~\bibnamefont {Gambetta}},\ and\ \bibinfo {author} {\bibfnamefont {J.~M.}\ \bibnamefont {Chow}},\ }\bibfield  {title} {\bibinfo {title} {Experimental {Demonstration} of a {Resonator}-{Induced} {Phase} {Gate} in a {Multiqubit} {Circuit}-{QED} {System}},\ }\href
  {https://doi.org/10.1103/PhysRevLett.117.250502} {\bibfield  {journal} {\bibinfo  {journal} {Physical Review Letters}\ }\textbf {\bibinfo {volume} {117}},\ \bibinfo {pages} {250502} (\bibinfo {year} {2016})}\BibitemShut {NoStop}%
\bibitem [{\citenamefont {Eickbusch}\ \emph {et~al.}(2022)\citenamefont {Eickbusch}, \citenamefont {Sivak}, \citenamefont {Ding}, \citenamefont {Elder}, \citenamefont {Jha}, \citenamefont {Venkatraman}, \citenamefont {Royer}, \citenamefont {Girvin}, \citenamefont {Schoelkopf},\ and\ \citenamefont {Devoret}}]{eickbusch_fast_2022}%
  \BibitemOpen
  \bibfield  {author} {\bibinfo {author} {\bibfnamefont {A.}~\bibnamefont {Eickbusch}}, \bibinfo {author} {\bibfnamefont {V.}~\bibnamefont {Sivak}}, \bibinfo {author} {\bibfnamefont {A.~Z.}\ \bibnamefont {Ding}}, \bibinfo {author} {\bibfnamefont {S.~S.}\ \bibnamefont {Elder}}, \bibinfo {author} {\bibfnamefont {S.~R.}\ \bibnamefont {Jha}}, \bibinfo {author} {\bibfnamefont {J.}~\bibnamefont {Venkatraman}}, \bibinfo {author} {\bibfnamefont {B.}~\bibnamefont {Royer}}, \bibinfo {author} {\bibfnamefont {S.~M.}\ \bibnamefont {Girvin}}, \bibinfo {author} {\bibfnamefont {R.~J.}\ \bibnamefont {Schoelkopf}},\ and\ \bibinfo {author} {\bibfnamefont {M.~H.}\ \bibnamefont {Devoret}},\ }\bibfield  {title} {\bibinfo {title} {Fast universal control of an oscillator with weak dispersive coupling to a qubit},\ }\href {https://doi.org/10.1038/s41567-022-01776-9} {\bibfield  {journal} {\bibinfo  {journal} {Nature Physics}\ }\textbf {\bibinfo {volume} {18}},\ \bibinfo {pages} {1464} (\bibinfo {year} {2022})}\BibitemShut {NoStop}%
\bibitem [{\citenamefont {Campagne-Ibarcq}\ \emph {et~al.}(2020)\citenamefont {Campagne-Ibarcq}, \citenamefont {Eickbusch}, \citenamefont {Touzard}, \citenamefont {Zalys-Geller}, \citenamefont {Frattini}, \citenamefont {Sivak}, \citenamefont {Reinhold}, \citenamefont {Puri}, \citenamefont {Shankar}, \citenamefont {Schoelkopf}, \citenamefont {Frunzio}, \citenamefont {Mirrahimi},\ and\ \citenamefont {Devoret}}]{campagne-ibarcq_quantum_2020}%
  \BibitemOpen
  \bibfield  {author} {\bibinfo {author} {\bibfnamefont {P.}~\bibnamefont {Campagne-Ibarcq}}, \bibinfo {author} {\bibfnamefont {A.}~\bibnamefont {Eickbusch}}, \bibinfo {author} {\bibfnamefont {S.}~\bibnamefont {Touzard}}, \bibinfo {author} {\bibfnamefont {E.}~\bibnamefont {Zalys-Geller}}, \bibinfo {author} {\bibfnamefont {N.~E.}\ \bibnamefont {Frattini}}, \bibinfo {author} {\bibfnamefont {V.~V.}\ \bibnamefont {Sivak}}, \bibinfo {author} {\bibfnamefont {P.}~\bibnamefont {Reinhold}}, \bibinfo {author} {\bibfnamefont {S.}~\bibnamefont {Puri}}, \bibinfo {author} {\bibfnamefont {S.}~\bibnamefont {Shankar}}, \bibinfo {author} {\bibfnamefont {R.~J.}\ \bibnamefont {Schoelkopf}}, \bibinfo {author} {\bibfnamefont {L.}~\bibnamefont {Frunzio}}, \bibinfo {author} {\bibfnamefont {M.}~\bibnamefont {Mirrahimi}},\ and\ \bibinfo {author} {\bibfnamefont {M.~H.}\ \bibnamefont {Devoret}},\ }\bibfield  {title} {\bibinfo {title} {Quantum error correction of a qubit encoded in grid states of an oscillator},\ }\href
  {https://doi.org/10.1038/s41586-020-2603-3} {\bibfield  {journal} {\bibinfo  {journal} {Nature}\ }\textbf {\bibinfo {volume} {584}},\ \bibinfo {pages} {368} (\bibinfo {year} {2020})}\BibitemShut {NoStop}%
\bibitem [{\citenamefont {Sivak}\ \emph {et~al.}(2023)\citenamefont {Sivak}, \citenamefont {Eickbusch}, \citenamefont {Royer}, \citenamefont {Singh}, \citenamefont {Tsioutsios}, \citenamefont {Ganjam}, \citenamefont {Miano}, \citenamefont {Brock}, \citenamefont {Ding}, \citenamefont {Frunzio}, \citenamefont {Girvin}, \citenamefont {Schoelkopf},\ and\ \citenamefont {Devoret}}]{sivak_real-time_2023}%
  \BibitemOpen
  \bibfield  {author} {\bibinfo {author} {\bibfnamefont {V.~V.}\ \bibnamefont {Sivak}}, \bibinfo {author} {\bibfnamefont {A.}~\bibnamefont {Eickbusch}}, \bibinfo {author} {\bibfnamefont {B.}~\bibnamefont {Royer}}, \bibinfo {author} {\bibfnamefont {S.}~\bibnamefont {Singh}}, \bibinfo {author} {\bibfnamefont {I.}~\bibnamefont {Tsioutsios}}, \bibinfo {author} {\bibfnamefont {S.}~\bibnamefont {Ganjam}}, \bibinfo {author} {\bibfnamefont {A.}~\bibnamefont {Miano}}, \bibinfo {author} {\bibfnamefont {B.~L.}\ \bibnamefont {Brock}}, \bibinfo {author} {\bibfnamefont {A.~Z.}\ \bibnamefont {Ding}}, \bibinfo {author} {\bibfnamefont {L.}~\bibnamefont {Frunzio}}, \bibinfo {author} {\bibfnamefont {S.~M.}\ \bibnamefont {Girvin}}, \bibinfo {author} {\bibfnamefont {R.~J.}\ \bibnamefont {Schoelkopf}},\ and\ \bibinfo {author} {\bibfnamefont {M.~H.}\ \bibnamefont {Devoret}},\ }\bibfield  {title} {\bibinfo {title} {Real-time quantum error correction beyond break-even},\ }\href {https://doi.org/10.1038/s41586-023-05782-6} {\bibfield
  {journal} {\bibinfo  {journal} {Nature}\ }\textbf {\bibinfo {volume} {616}},\ \bibinfo {pages} {50} (\bibinfo {year} {2023})}\BibitemShut {NoStop}%
\bibitem [{\citenamefont {Place}\ \emph {et~al.}(2021)\citenamefont {Place}, \citenamefont {Rodgers}, \citenamefont {Mundada}, \citenamefont {Smitham}, \citenamefont {Fitzpatrick}, \citenamefont {Leng}, \citenamefont {Premkumar}, \citenamefont {Bryon}, \citenamefont {Vrajitoarea}, \citenamefont {Sussman}, \citenamefont {Cheng}, \citenamefont {Madhavan}, \citenamefont {Babla}, \citenamefont {Le}, \citenamefont {Gang}, \citenamefont {Jäck}, \citenamefont {Gyenis}, \citenamefont {Yao}, \citenamefont {Cava}, \citenamefont {de~Leon},\ and\ \citenamefont {Houck}}]{place_new_2021}%
  \BibitemOpen
  \bibfield  {author} {\bibinfo {author} {\bibfnamefont {A.~P.~M.}\ \bibnamefont {Place}}, \bibinfo {author} {\bibfnamefont {L.~V.~H.}\ \bibnamefont {Rodgers}}, \bibinfo {author} {\bibfnamefont {P.}~\bibnamefont {Mundada}}, \bibinfo {author} {\bibfnamefont {B.~M.}\ \bibnamefont {Smitham}}, \bibinfo {author} {\bibfnamefont {M.}~\bibnamefont {Fitzpatrick}}, \bibinfo {author} {\bibfnamefont {Z.}~\bibnamefont {Leng}}, \bibinfo {author} {\bibfnamefont {A.}~\bibnamefont {Premkumar}}, \bibinfo {author} {\bibfnamefont {J.}~\bibnamefont {Bryon}}, \bibinfo {author} {\bibfnamefont {A.}~\bibnamefont {Vrajitoarea}}, \bibinfo {author} {\bibfnamefont {S.}~\bibnamefont {Sussman}}, \bibinfo {author} {\bibfnamefont {G.}~\bibnamefont {Cheng}}, \bibinfo {author} {\bibfnamefont {T.}~\bibnamefont {Madhavan}}, \bibinfo {author} {\bibfnamefont {H.~K.}\ \bibnamefont {Babla}}, \bibinfo {author} {\bibfnamefont {X.~H.}\ \bibnamefont {Le}}, \bibinfo {author} {\bibfnamefont {Y.}~\bibnamefont {Gang}}, \bibinfo {author} {\bibfnamefont
  {B.}~\bibnamefont {Jäck}}, \bibinfo {author} {\bibfnamefont {A.}~\bibnamefont {Gyenis}}, \bibinfo {author} {\bibfnamefont {N.}~\bibnamefont {Yao}}, \bibinfo {author} {\bibfnamefont {R.~J.}\ \bibnamefont {Cava}}, \bibinfo {author} {\bibfnamefont {N.~P.}\ \bibnamefont {de~Leon}},\ and\ \bibinfo {author} {\bibfnamefont {A.~A.}\ \bibnamefont {Houck}},\ }\bibfield  {title} {\bibinfo {title} {New material platform for superconducting transmon qubits with coherence times exceeding 0.3 milliseconds},\ }\href {https://doi.org/10.1038/s41467-021-22030-5} {\bibfield  {journal} {\bibinfo  {journal} {Nature Communications}\ }\textbf {\bibinfo {volume} {12}},\ \bibinfo {pages} {1779} (\bibinfo {year} {2021})}\BibitemShut {NoStop}%
\bibitem [{\citenamefont {Connolly}\ \emph {et~al.}(2024)\citenamefont {Connolly}, \citenamefont {Kurilovich}, \citenamefont {Diamond}, \citenamefont {Nho}, \citenamefont {Bøttcher}, \citenamefont {Glazman}, \citenamefont {Fatemi},\ and\ \citenamefont {Devoret}}]{connolly_coexistence_2024}%
  \BibitemOpen
  \bibfield  {author} {\bibinfo {author} {\bibfnamefont {T.}~\bibnamefont {Connolly}}, \bibinfo {author} {\bibfnamefont {P.~D.}\ \bibnamefont {Kurilovich}}, \bibinfo {author} {\bibfnamefont {S.}~\bibnamefont {Diamond}}, \bibinfo {author} {\bibfnamefont {H.}~\bibnamefont {Nho}}, \bibinfo {author} {\bibfnamefont {C.~G.}\ \bibnamefont {Bøttcher}}, \bibinfo {author} {\bibfnamefont {L.~I.}\ \bibnamefont {Glazman}}, \bibinfo {author} {\bibfnamefont {V.}~\bibnamefont {Fatemi}},\ and\ \bibinfo {author} {\bibfnamefont {M.~H.}\ \bibnamefont {Devoret}},\ }\bibfield  {title} {\bibinfo {title} {Coexistence of {Nonequilibrium} {Density} and {Equilibrium} {Energy} {Distribution} of {Quasiparticles} in a {Superconducting} {Qubit}},\ }\href {https://doi.org/10.1103/PhysRevLett.132.217001} {\bibfield  {journal} {\bibinfo  {journal} {Physical Review Letters}\ }\textbf {\bibinfo {volume} {132}},\ \bibinfo {pages} {217001} (\bibinfo {year} {2024})}\BibitemShut {NoStop}%
\bibitem [{\citenamefont {Frattini}\ \emph {et~al.}(2018)\citenamefont {Frattini}, \citenamefont {Sivak}, \citenamefont {Lingenfelter}, \citenamefont {Shankar},\ and\ \citenamefont {Devoret}}]{frattini_optimizing_2018}%
  \BibitemOpen
  \bibfield  {author} {\bibinfo {author} {\bibfnamefont {N.~E.}\ \bibnamefont {Frattini}}, \bibinfo {author} {\bibfnamefont {V.~V.}\ \bibnamefont {Sivak}}, \bibinfo {author} {\bibfnamefont {A.}~\bibnamefont {Lingenfelter}}, \bibinfo {author} {\bibfnamefont {S.}~\bibnamefont {Shankar}},\ and\ \bibinfo {author} {\bibfnamefont {M.~H.}\ \bibnamefont {Devoret}},\ }\bibfield  {title} {\bibinfo {title} {Optimizing the nonlinearity and dissipation of a {SNAIL} parametric amplifier for dynamic range},\ }\href {https://doi.org/10.1103/PhysRevApplied.10.054020} {\bibfield  {journal} {\bibinfo  {journal} {Physical Review Applied}\ }\textbf {\bibinfo {volume} {10}},\ \bibinfo {pages} {054020} (\bibinfo {year} {2018})}\BibitemShut {NoStop}%
\bibitem [{\citenamefont {Joshi}\ \emph {et~al.}(2025)\citenamefont {Joshi}, \citenamefont {Hazra}, \citenamefont {Ding}, \citenamefont {Miano}, \citenamefont {Dai}, \citenamefont {Umasankar}, \citenamefont {Kottandavida}, \citenamefont {Liu}, \citenamefont {Frunzio},\ and\ \citenamefont {Devoret}}]{joshi_lumped-element_2025}%
  \BibitemOpen
  \bibfield  {author} {\bibinfo {author} {\bibfnamefont {V.}~\bibnamefont {Joshi}}, \bibinfo {author} {\bibfnamefont {S.}~\bibnamefont {Hazra}}, \bibinfo {author} {\bibfnamefont {A.}~\bibnamefont {Ding}}, \bibinfo {author} {\bibfnamefont {A.}~\bibnamefont {Miano}}, \bibinfo {author} {\bibfnamefont {W.}~\bibnamefont {Dai}}, \bibinfo {author} {\bibfnamefont {G.}~\bibnamefont {Umasankar}}, \bibinfo {author} {\bibfnamefont {A.}~\bibnamefont {Kottandavida}}, \bibinfo {author} {\bibfnamefont {G.}~\bibnamefont {Liu}}, \bibinfo {author} {\bibfnamefont {L.}~\bibnamefont {Frunzio}},\ and\ \bibinfo {author} {\bibfnamefont {M.}~\bibnamefont {Devoret}},\ }\bibfield  {title} {\bibinfo {title} {Lumped-element broadband {SNAIL} parametric amplifier with on-chip pump filter for multiplexed readout},\ }\href {https://doi.org/10.1103/5l86-pw21} {\bibfield  {journal} {\bibinfo  {journal} {Physical Review Applied}\ }\textbf {\bibinfo {volume} {24}},\ \bibinfo {pages} {044003} (\bibinfo {year} {2025})}\BibitemShut {NoStop}%
\bibitem [{\citenamefont {Weiss}()}]{weiss_floquet_nodate}%
  \BibitemOpen
  \bibfield  {author} {\bibinfo {author} {\bibfnamefont {D.~K.}\ \bibnamefont {Weiss}},\ }\href {https://dkweiss.net/floquet/} {\bibinfo {title} {floquet}}\BibitemShut {NoStop}%
\end{thebibliography}%


\begin{thebibliography}{34}%
\makeatletter
\providecommand \@ifxundefined [1]{%
 \@ifx{#1\undefined}
}%
\providecommand \@ifnum [1]{%
 \ifnum #1\expandafter \@firstoftwo
 \else \expandafter \@secondoftwo
 \fi
}%
\providecommand \@ifx [1]{%
 \ifx #1\expandafter \@firstoftwo
 \else \expandafter \@secondoftwo
 \fi
}%
\providecommand \natexlab [1]{#1}%
\providecommand \enquote  [1]{``#1''}%
\providecommand \bibnamefont  [1]{#1}%
\providecommand \bibfnamefont [1]{#1}%
\providecommand \citenamefont [1]{#1}%
\providecommand \href@noop [0]{\@secondoftwo}%
\providecommand \href [0]{\begingroup \@sanitize@url \@href}%
\providecommand \@href[1]{\@@startlink{#1}\@@href}%
\providecommand \@@href[1]{\endgroup#1\@@endlink}%
\providecommand \@sanitize@url [0]{\catcode `\\12\catcode `\$12\catcode `\&12\catcode `\#12\catcode `\^12\catcode `\_12\catcode `\%12\relax}%
\providecommand \@@startlink[1]{}%
\providecommand \@@endlink[0]{}%
\providecommand \url  [0]{\begingroup\@sanitize@url \@url }%
\providecommand \@url [1]{\endgroup\@href {#1}{\urlprefix }}%
\providecommand \urlprefix  [0]{URL }%
\providecommand \Eprint [0]{\href }%
\providecommand \doibase [0]{https://doi.org/}%
\providecommand \selectlanguage [0]{\@gobble}%
\providecommand \bibinfo  [0]{\@secondoftwo}%
\providecommand \bibfield  [0]{\@secondoftwo}%
\providecommand \translation [1]{[#1]}%
\providecommand \BibitemOpen [0]{}%
\providecommand \bibitemStop [0]{}%
\providecommand \bibitemNoStop [0]{.\EOS\space}%
\providecommand \EOS [0]{\spacefactor3000\relax}%
\providecommand \BibitemShut  [1]{\csname bibitem#1\endcsname}%
\let\auto@bib@innerbib\@empty
\bibitem [{\citenamefont {Place}\ \emph {et~al.}(2021)\citenamefont {Place}, \citenamefont {Rodgers}, \citenamefont {Mundada}, \citenamefont {Smitham}, \citenamefont {Fitzpatrick}, \citenamefont {Leng}, \citenamefont {Premkumar}, \citenamefont {Bryon}, \citenamefont {Vrajitoarea}, \citenamefont {Sussman}, \citenamefont {Cheng}, \citenamefont {Madhavan}, \citenamefont {Babla}, \citenamefont {Le}, \citenamefont {Gang}, \citenamefont {Jäck}, \citenamefont {Gyenis}, \citenamefont {Yao}, \citenamefont {Cava}, \citenamefont {de~Leon},\ and\ \citenamefont {Houck}}]{place_new_2021}%
  \BibitemOpen
  \bibfield  {author} {\bibinfo {author} {\bibfnamefont {A.~P.~M.}\ \bibnamefont {Place}}, \bibinfo {author} {\bibfnamefont {L.~V.~H.}\ \bibnamefont {Rodgers}}, \bibinfo {author} {\bibfnamefont {P.}~\bibnamefont {Mundada}}, \bibinfo {author} {\bibfnamefont {B.~M.}\ \bibnamefont {Smitham}}, \bibinfo {author} {\bibfnamefont {M.}~\bibnamefont {Fitzpatrick}}, \bibinfo {author} {\bibfnamefont {Z.}~\bibnamefont {Leng}}, \bibinfo {author} {\bibfnamefont {A.}~\bibnamefont {Premkumar}}, \bibinfo {author} {\bibfnamefont {J.}~\bibnamefont {Bryon}}, \bibinfo {author} {\bibfnamefont {A.}~\bibnamefont {Vrajitoarea}}, \bibinfo {author} {\bibfnamefont {S.}~\bibnamefont {Sussman}}, \bibinfo {author} {\bibfnamefont {G.}~\bibnamefont {Cheng}}, \bibinfo {author} {\bibfnamefont {T.}~\bibnamefont {Madhavan}}, \bibinfo {author} {\bibfnamefont {H.~K.}\ \bibnamefont {Babla}}, \bibinfo {author} {\bibfnamefont {X.~H.}\ \bibnamefont {Le}}, \bibinfo {author} {\bibfnamefont {Y.}~\bibnamefont {Gang}}, \bibinfo {author} {\bibfnamefont
  {B.}~\bibnamefont {Jäck}}, \bibinfo {author} {\bibfnamefont {A.}~\bibnamefont {Gyenis}}, \bibinfo {author} {\bibfnamefont {N.}~\bibnamefont {Yao}}, \bibinfo {author} {\bibfnamefont {R.~J.}\ \bibnamefont {Cava}}, \bibinfo {author} {\bibfnamefont {N.~P.}\ \bibnamefont {de~Leon}},\ and\ \bibinfo {author} {\bibfnamefont {A.~A.}\ \bibnamefont {Houck}},\ }\bibfield  {title} {\bibinfo {title} {New material platform for superconducting transmon qubits with coherence times exceeding 0.3 milliseconds},\ }\href {https://doi.org/10.1038/s41467-021-22030-5} {\bibfield  {journal} {\bibinfo  {journal} {Nature Communications}\ }\textbf {\bibinfo {volume} {12}},\ \bibinfo {pages} {1779} (\bibinfo {year} {2021})}\BibitemShut {NoStop}%
\bibitem [{\citenamefont {Ganjam}\ \emph {et~al.}(2024)\citenamefont {Ganjam}, \citenamefont {Wang}, \citenamefont {Lu}, \citenamefont {Banerjee}, \citenamefont {Lei}, \citenamefont {Krayzman}, \citenamefont {Kisslinger}, \citenamefont {Zhou}, \citenamefont {Li}, \citenamefont {Jia}, \citenamefont {Liu}, \citenamefont {Frunzio},\ and\ \citenamefont {Schoelkopf}}]{ganjam_surpassing_2024}%
  \BibitemOpen
  \bibfield  {author} {\bibinfo {author} {\bibfnamefont {S.}~\bibnamefont {Ganjam}}, \bibinfo {author} {\bibfnamefont {Y.}~\bibnamefont {Wang}}, \bibinfo {author} {\bibfnamefont {Y.}~\bibnamefont {Lu}}, \bibinfo {author} {\bibfnamefont {A.}~\bibnamefont {Banerjee}}, \bibinfo {author} {\bibfnamefont {C.~U.}\ \bibnamefont {Lei}}, \bibinfo {author} {\bibfnamefont {L.}~\bibnamefont {Krayzman}}, \bibinfo {author} {\bibfnamefont {K.}~\bibnamefont {Kisslinger}}, \bibinfo {author} {\bibfnamefont {C.}~\bibnamefont {Zhou}}, \bibinfo {author} {\bibfnamefont {R.}~\bibnamefont {Li}}, \bibinfo {author} {\bibfnamefont {Y.}~\bibnamefont {Jia}}, \bibinfo {author} {\bibfnamefont {M.}~\bibnamefont {Liu}}, \bibinfo {author} {\bibfnamefont {L.}~\bibnamefont {Frunzio}},\ and\ \bibinfo {author} {\bibfnamefont {R.~J.}\ \bibnamefont {Schoelkopf}},\ }\bibfield  {title} {\bibinfo {title} {Surpassing millisecond coherence in on chip superconducting quantum memories by optimizing materials and circuit design},\ }\href
  {https://doi.org/10.1038/s41467-024-47857-6} {\bibfield  {journal} {\bibinfo  {journal} {Nature Communications}\ }\textbf {\bibinfo {volume} {15}},\ \bibinfo {pages} {3687} (\bibinfo {year} {2024})}\BibitemShut {NoStop}%
\bibitem [{\citenamefont {Serniak}\ \emph {et~al.}(2019)\citenamefont {Serniak}, \citenamefont {Diamond}, \citenamefont {Hays}, \citenamefont {Fatemi}, \citenamefont {Shankar}, \citenamefont {Frunzio}, \citenamefont {Schoelkopf},\ and\ \citenamefont {Devoret}}]{serniak_direct_2019}%
  \BibitemOpen
  \bibfield  {author} {\bibinfo {author} {\bibfnamefont {K.}~\bibnamefont {Serniak}}, \bibinfo {author} {\bibfnamefont {S.}~\bibnamefont {Diamond}}, \bibinfo {author} {\bibfnamefont {M.}~\bibnamefont {Hays}}, \bibinfo {author} {\bibfnamefont {V.}~\bibnamefont {Fatemi}}, \bibinfo {author} {\bibfnamefont {S.}~\bibnamefont {Shankar}}, \bibinfo {author} {\bibfnamefont {L.}~\bibnamefont {Frunzio}}, \bibinfo {author} {\bibfnamefont {R.}~\bibnamefont {Schoelkopf}},\ and\ \bibinfo {author} {\bibfnamefont {M.}~\bibnamefont {Devoret}},\ }\bibfield  {title} {\bibinfo {title} {Direct dispersive monitoring of charge parity in offset-charge-sensitive transmons},\ }\href {https://doi.org/10.1103/PhysRevApplied.12.014052} {\bibfield  {journal} {\bibinfo  {journal} {Physical Review Applied}\ }\textbf {\bibinfo {volume} {12}},\ \bibinfo {pages} {014052} (\bibinfo {year} {2019})}\BibitemShut {NoStop}%
\bibitem [{\citenamefont {Connolly}\ \emph {et~al.}(2024)\citenamefont {Connolly}, \citenamefont {Kurilovich}, \citenamefont {Diamond}, \citenamefont {Nho}, \citenamefont {Bøttcher}, \citenamefont {Glazman}, \citenamefont {Fatemi},\ and\ \citenamefont {Devoret}}]{connolly_coexistence_2024}%
  \BibitemOpen
  \bibfield  {author} {\bibinfo {author} {\bibfnamefont {T.}~\bibnamefont {Connolly}}, \bibinfo {author} {\bibfnamefont {P.~D.}\ \bibnamefont {Kurilovich}}, \bibinfo {author} {\bibfnamefont {S.}~\bibnamefont {Diamond}}, \bibinfo {author} {\bibfnamefont {H.}~\bibnamefont {Nho}}, \bibinfo {author} {\bibfnamefont {C.~G.}\ \bibnamefont {Bøttcher}}, \bibinfo {author} {\bibfnamefont {L.~I.}\ \bibnamefont {Glazman}}, \bibinfo {author} {\bibfnamefont {V.}~\bibnamefont {Fatemi}},\ and\ \bibinfo {author} {\bibfnamefont {M.~H.}\ \bibnamefont {Devoret}},\ }\bibfield  {title} {\bibinfo {title} {Coexistence of {Nonequilibrium} {Density} and {Equilibrium} {Energy} {Distribution} of {Quasiparticles} in a {Superconducting} {Qubit}},\ }\href {https://doi.org/10.1103/PhysRevLett.132.217001} {\bibfield  {journal} {\bibinfo  {journal} {Physical Review Letters}\ }\textbf {\bibinfo {volume} {132}},\ \bibinfo {pages} {217001} (\bibinfo {year} {2024})}\BibitemShut {NoStop}%
\bibitem [{\citenamefont {Rehammar}\ and\ \citenamefont {Gasparinetti}(2023)}]{rehammar_low-pass_2023}%
  \BibitemOpen
  \bibfield  {author} {\bibinfo {author} {\bibfnamefont {R.}~\bibnamefont {Rehammar}}\ and\ \bibinfo {author} {\bibfnamefont {S.}~\bibnamefont {Gasparinetti}},\ }\bibfield  {title} {\bibinfo {title} {Low-{Pass} {Filter} {With} {Ultrawide} {Stopband} for {Quantum} {Computing} {Applications}},\ }\href {https://doi.org/10.1109/TMTT.2023.3238543} {\bibfield  {journal} {\bibinfo  {journal} {IEEE Transactions on Microwave Theory and Techniques}\ }\textbf {\bibinfo {volume} {71}},\ \bibinfo {pages} {3075} (\bibinfo {year} {2023})}\BibitemShut {NoStop}%
\bibitem [{\citenamefont {Koch}\ \emph {et~al.}(2007)\citenamefont {Koch}, \citenamefont {Yu}, \citenamefont {Gambetta}, \citenamefont {Houck}, \citenamefont {Schuster}, \citenamefont {Majer}, \citenamefont {Blais}, \citenamefont {Devoret}, \citenamefont {Girvin},\ and\ \citenamefont {Schoelkopf}}]{koch_charge-insensitive_2007}%
  \BibitemOpen
  \bibfield  {author} {\bibinfo {author} {\bibfnamefont {J.}~\bibnamefont {Koch}}, \bibinfo {author} {\bibfnamefont {T.~M.}\ \bibnamefont {Yu}}, \bibinfo {author} {\bibfnamefont {J.}~\bibnamefont {Gambetta}}, \bibinfo {author} {\bibfnamefont {A.~A.}\ \bibnamefont {Houck}}, \bibinfo {author} {\bibfnamefont {D.~I.}\ \bibnamefont {Schuster}}, \bibinfo {author} {\bibfnamefont {J.}~\bibnamefont {Majer}}, \bibinfo {author} {\bibfnamefont {A.}~\bibnamefont {Blais}}, \bibinfo {author} {\bibfnamefont {M.~H.}\ \bibnamefont {Devoret}}, \bibinfo {author} {\bibfnamefont {S.~M.}\ \bibnamefont {Girvin}},\ and\ \bibinfo {author} {\bibfnamefont {R.~J.}\ \bibnamefont {Schoelkopf}},\ }\bibfield  {title} {\bibinfo {title} {Charge-insensitive qubit design derived from the {Cooper} pair box},\ }\href {https://doi.org/10.1103/PhysRevA.76.042319} {\bibfield  {journal} {\bibinfo  {journal} {Physical Review A}\ }\textbf {\bibinfo {volume} {76}},\ \bibinfo {pages} {042319} (\bibinfo {year} {2007})}\BibitemShut {NoStop}%
\bibitem [{\citenamefont {Ristè}\ \emph {et~al.}(2013)\citenamefont {Ristè}, \citenamefont {Bultink}, \citenamefont {Tiggelman}, \citenamefont {Schouten}, \citenamefont {Lehnert},\ and\ \citenamefont {DiCarlo}}]{riste_millisecond_2013}%
  \BibitemOpen
  \bibfield  {author} {\bibinfo {author} {\bibfnamefont {D.}~\bibnamefont {Ristè}}, \bibinfo {author} {\bibfnamefont {C.~C.}\ \bibnamefont {Bultink}}, \bibinfo {author} {\bibfnamefont {M.~J.}\ \bibnamefont {Tiggelman}}, \bibinfo {author} {\bibfnamefont {R.~N.}\ \bibnamefont {Schouten}}, \bibinfo {author} {\bibfnamefont {K.~W.}\ \bibnamefont {Lehnert}},\ and\ \bibinfo {author} {\bibfnamefont {L.}~\bibnamefont {DiCarlo}},\ }\bibfield  {title} {\bibinfo {title} {Millisecond charge-parity fluctuations and induced decoherence in a superconducting transmon qubit},\ }\href {https://doi.org/10.1038/ncomms2936} {\bibfield  {journal} {\bibinfo  {journal} {Nature Communications}\ }\textbf {\bibinfo {volume} {4}},\ \bibinfo {pages} {1913} (\bibinfo {year} {2013})}\BibitemShut {NoStop}%
\bibitem [{\citenamefont {Serniak}\ \emph {et~al.}(2018)\citenamefont {Serniak}, \citenamefont {Hays}, \citenamefont {de~Lange}, \citenamefont {Diamond}, \citenamefont {Shankar}, \citenamefont {Burkhart}, \citenamefont {Frunzio}, \citenamefont {Houzet},\ and\ \citenamefont {Devoret}}]{serniak_hot_2018}%
  \BibitemOpen
  \bibfield  {author} {\bibinfo {author} {\bibfnamefont {K.}~\bibnamefont {Serniak}}, \bibinfo {author} {\bibfnamefont {M.}~\bibnamefont {Hays}}, \bibinfo {author} {\bibfnamefont {G.}~\bibnamefont {de~Lange}}, \bibinfo {author} {\bibfnamefont {S.}~\bibnamefont {Diamond}}, \bibinfo {author} {\bibfnamefont {S.}~\bibnamefont {Shankar}}, \bibinfo {author} {\bibfnamefont {L.}~\bibnamefont {Burkhart}}, \bibinfo {author} {\bibfnamefont {L.}~\bibnamefont {Frunzio}}, \bibinfo {author} {\bibfnamefont {M.}~\bibnamefont {Houzet}},\ and\ \bibinfo {author} {\bibfnamefont {M.}~\bibnamefont {Devoret}},\ }\bibfield  {title} {\bibinfo {title} {Hot nonequilibrium quasiparticles in transmon qubits},\ }\href {https://doi.org/10.1103/PhysRevLett.121.157701} {\bibfield  {journal} {\bibinfo  {journal} {Physical Review Letters}\ }\textbf {\bibinfo {volume} {121}},\ \bibinfo {pages} {157701} (\bibinfo {year} {2018})}\BibitemShut {NoStop}%
\bibitem [{\citenamefont {Purcell}\ \emph {et~al.}(1946)\citenamefont {Purcell}, \citenamefont {Torrey},\ and\ \citenamefont {Pound}}]{purcell_resonance_1946}%
  \BibitemOpen
  \bibfield  {author} {\bibinfo {author} {\bibfnamefont {E.~M.}\ \bibnamefont {Purcell}}, \bibinfo {author} {\bibfnamefont {H.~C.}\ \bibnamefont {Torrey}},\ and\ \bibinfo {author} {\bibfnamefont {R.~V.}\ \bibnamefont {Pound}},\ }\bibfield  {title} {\bibinfo {title} {Resonance {Absorption} by {Nuclear} {Magnetic} {Moments} in a {Solid}},\ }\href {https://doi.org/10.1103/PhysRev.69.37} {\bibfield  {journal} {\bibinfo  {journal} {Physical Review}\ }\textbf {\bibinfo {volume} {69}},\ \bibinfo {pages} {37} (\bibinfo {year} {1946})}\BibitemShut {NoStop}%
\bibitem [{\citenamefont {Kleppner}(1981)}]{kleppner_inhibited_1981}%
  \BibitemOpen
  \bibfield  {author} {\bibinfo {author} {\bibfnamefont {D.}~\bibnamefont {Kleppner}},\ }\bibfield  {title} {\bibinfo {title} {Inhibited {Spontaneous} {Emission}},\ }\href {https://doi.org/10.1103/PhysRevLett.47.233} {\bibfield  {journal} {\bibinfo  {journal} {Physical Review Letters}\ }\textbf {\bibinfo {volume} {47}},\ \bibinfo {pages} {233} (\bibinfo {year} {1981})}\BibitemShut {NoStop}%
\bibitem [{\citenamefont {Goy}\ \emph {et~al.}(1983)\citenamefont {Goy}, \citenamefont {Raimond}, \citenamefont {Gross},\ and\ \citenamefont {Haroche}}]{goy_observation_1983}%
  \BibitemOpen
  \bibfield  {author} {\bibinfo {author} {\bibfnamefont {P.}~\bibnamefont {Goy}}, \bibinfo {author} {\bibfnamefont {J.~M.}\ \bibnamefont {Raimond}}, \bibinfo {author} {\bibfnamefont {M.}~\bibnamefont {Gross}},\ and\ \bibinfo {author} {\bibfnamefont {S.}~\bibnamefont {Haroche}},\ }\bibfield  {title} {\bibinfo {title} {Observation of {Cavity}-{Enhanced} {Single}-{Atom} {Spontaneous} {Emission}},\ }\href {https://doi.org/10.1103/PhysRevLett.50.1903} {\bibfield  {journal} {\bibinfo  {journal} {Physical Review Letters}\ }\textbf {\bibinfo {volume} {50}},\ \bibinfo {pages} {1903} (\bibinfo {year} {1983})}\BibitemShut {NoStop}%
\bibitem [{\citenamefont {Blais}\ \emph {et~al.}(2004)\citenamefont {Blais}, \citenamefont {Huang}, \citenamefont {Wallraff}, \citenamefont {Girvin},\ and\ \citenamefont {Schoelkopf}}]{blais_cavity_2004}%
  \BibitemOpen
  \bibfield  {author} {\bibinfo {author} {\bibfnamefont {A.}~\bibnamefont {Blais}}, \bibinfo {author} {\bibfnamefont {R.-S.}\ \bibnamefont {Huang}}, \bibinfo {author} {\bibfnamefont {A.}~\bibnamefont {Wallraff}}, \bibinfo {author} {\bibfnamefont {S.~M.}\ \bibnamefont {Girvin}},\ and\ \bibinfo {author} {\bibfnamefont {R.~J.}\ \bibnamefont {Schoelkopf}},\ }\bibfield  {title} {\bibinfo {title} {Cavity quantum electrodynamics for superconducting electrical circuits: {An} architecture for quantum computation},\ }\href {https://doi.org/10.1103/PhysRevA.69.062320} {\bibfield  {journal} {\bibinfo  {journal} {Physical Review A}\ }\textbf {\bibinfo {volume} {69}},\ \bibinfo {pages} {062320} (\bibinfo {year} {2004})}\BibitemShut {NoStop}%
\bibitem [{\citenamefont {Houck}\ \emph {et~al.}(2008)\citenamefont {Houck}, \citenamefont {Schreier}, \citenamefont {Johnson}, \citenamefont {Chow}, \citenamefont {Koch}, \citenamefont {Gambetta}, \citenamefont {Schuster}, \citenamefont {Frunzio}, \citenamefont {Devoret}, \citenamefont {Girvin},\ and\ \citenamefont {Schoelkopf}}]{houck_controlling_2008}%
  \BibitemOpen
  \bibfield  {author} {\bibinfo {author} {\bibfnamefont {A.~A.}\ \bibnamefont {Houck}}, \bibinfo {author} {\bibfnamefont {J.~A.}\ \bibnamefont {Schreier}}, \bibinfo {author} {\bibfnamefont {B.~R.}\ \bibnamefont {Johnson}}, \bibinfo {author} {\bibfnamefont {J.~M.}\ \bibnamefont {Chow}}, \bibinfo {author} {\bibfnamefont {J.}~\bibnamefont {Koch}}, \bibinfo {author} {\bibfnamefont {J.~M.}\ \bibnamefont {Gambetta}}, \bibinfo {author} {\bibfnamefont {D.~I.}\ \bibnamefont {Schuster}}, \bibinfo {author} {\bibfnamefont {L.}~\bibnamefont {Frunzio}}, \bibinfo {author} {\bibfnamefont {M.~H.}\ \bibnamefont {Devoret}}, \bibinfo {author} {\bibfnamefont {S.~M.}\ \bibnamefont {Girvin}},\ and\ \bibinfo {author} {\bibfnamefont {R.~J.}\ \bibnamefont {Schoelkopf}},\ }\bibfield  {title} {\bibinfo {title} {Controlling the {Spontaneous} {Emission} of a {Superconducting} {Transmon} {Qubit}},\ }\href {https://doi.org/10.1103/PhysRevLett.101.080502} {\bibfield  {journal} {\bibinfo  {journal} {Physical Review Letters}\ }\textbf
  {\bibinfo {volume} {101}},\ \bibinfo {pages} {080502} (\bibinfo {year} {2008})}\BibitemShut {NoStop}%
\bibitem [{\citenamefont {Yen}\ \emph {et~al.}(2025)\citenamefont {Yen}, \citenamefont {Ye}, \citenamefont {Peng}, \citenamefont {Wang}, \citenamefont {Cunningham}, \citenamefont {Gingras}, \citenamefont {Niedzielski}, \citenamefont {Stickler}, \citenamefont {Serniak}, \citenamefont {Schwartz},\ and\ \citenamefont {O’Brien}}]{yen_interferometric_2025}%
  \BibitemOpen
  \bibfield  {author} {\bibinfo {author} {\bibfnamefont {A.}~\bibnamefont {Yen}}, \bibinfo {author} {\bibfnamefont {Y.}~\bibnamefont {Ye}}, \bibinfo {author} {\bibfnamefont {K.}~\bibnamefont {Peng}}, \bibinfo {author} {\bibfnamefont {J.}~\bibnamefont {Wang}}, \bibinfo {author} {\bibfnamefont {G.}~\bibnamefont {Cunningham}}, \bibinfo {author} {\bibfnamefont {M.}~\bibnamefont {Gingras}}, \bibinfo {author} {\bibfnamefont {B.~M.}\ \bibnamefont {Niedzielski}}, \bibinfo {author} {\bibfnamefont {H.}~\bibnamefont {Stickler}}, \bibinfo {author} {\bibfnamefont {K.}~\bibnamefont {Serniak}}, \bibinfo {author} {\bibfnamefont {M.~E.}\ \bibnamefont {Schwartz}},\ and\ \bibinfo {author} {\bibfnamefont {K.~P.}\ \bibnamefont {O’Brien}},\ }\bibfield  {title} {\bibinfo {title} {Interferometric {Purcell} suppression of spontaneous emission in a superconducting qubit},\ }\href {https://doi.org/10.1103/PhysRevApplied.23.024068} {\bibfield  {journal} {\bibinfo  {journal} {Physical Review Applied}\ }\textbf {\bibinfo {volume}
  {23}},\ \bibinfo {pages} {024068} (\bibinfo {year} {2025})}\BibitemShut {NoStop}%
\bibitem [{\citenamefont {Manucharyan}\ \emph {et~al.}(2012)\citenamefont {Manucharyan}, \citenamefont {Masluk}, \citenamefont {Kamal}, \citenamefont {Koch}, \citenamefont {Glazman},\ and\ \citenamefont {Devoret}}]{manucharyan_evidence_2012}%
  \BibitemOpen
  \bibfield  {author} {\bibinfo {author} {\bibfnamefont {V.~E.}\ \bibnamefont {Manucharyan}}, \bibinfo {author} {\bibfnamefont {N.~A.}\ \bibnamefont {Masluk}}, \bibinfo {author} {\bibfnamefont {A.}~\bibnamefont {Kamal}}, \bibinfo {author} {\bibfnamefont {J.}~\bibnamefont {Koch}}, \bibinfo {author} {\bibfnamefont {L.~I.}\ \bibnamefont {Glazman}},\ and\ \bibinfo {author} {\bibfnamefont {M.~H.}\ \bibnamefont {Devoret}},\ }\bibfield  {title} {\bibinfo {title} {Evidence for coherent quantum phase slips across a {Josephson} junction array},\ }\href {https://doi.org/10.1103/PhysRevB.85.024521} {\bibfield  {journal} {\bibinfo  {journal} {Physical Review B}\ }\textbf {\bibinfo {volume} {85}},\ \bibinfo {pages} {024521} (\bibinfo {year} {2012})}\BibitemShut {NoStop}%
\bibitem [{\citenamefont {Manucharyan}(2012)}]{manucharyan_phd_2012}%
  \BibitemOpen
  \bibfield  {author} {\bibinfo {author} {\bibfnamefont {V.~E.}\ \bibnamefont {Manucharyan}},\ }\bibfield  {title} {\bibinfo {title} {Ph.{D}. {Thesis}},\ }\href@noop {} {\bibfield  {journal} {\bibinfo  {journal} {Yale University}\ } (\bibinfo {year} {2012})}\BibitemShut {NoStop}%
\bibitem [{\citenamefont {Zhu}\ \emph {et~al.}(2013)\citenamefont {Zhu}, \citenamefont {Ferguson}, \citenamefont {Manucharyan},\ and\ \citenamefont {Koch}}]{zhu_circuit_2013}%
  \BibitemOpen
  \bibfield  {author} {\bibinfo {author} {\bibfnamefont {G.}~\bibnamefont {Zhu}}, \bibinfo {author} {\bibfnamefont {D.~G.}\ \bibnamefont {Ferguson}}, \bibinfo {author} {\bibfnamefont {V.~E.}\ \bibnamefont {Manucharyan}},\ and\ \bibinfo {author} {\bibfnamefont {J.}~\bibnamefont {Koch}},\ }\bibfield  {title} {\bibinfo {title} {Circuit {QED} with fluxonium qubits: {Theory} of the dispersive regime},\ }\href {https://doi.org/10.1103/PhysRevB.87.024510} {\bibfield  {journal} {\bibinfo  {journal} {Physical Review B}\ }\textbf {\bibinfo {volume} {87}},\ \bibinfo {pages} {024510} (\bibinfo {year} {2013})}\BibitemShut {NoStop}%
\bibitem [{\citenamefont {Ruskov}\ and\ \citenamefont {Tahan}(2024)}]{ruskov_longitudinal_2024}%
  \BibitemOpen
  \bibfield  {author} {\bibinfo {author} {\bibfnamefont {R.}~\bibnamefont {Ruskov}}\ and\ \bibinfo {author} {\bibfnamefont {C.}~\bibnamefont {Tahan}},\ }\bibfield  {title} {\bibinfo {title} {Longitudinal (curvature) couplings of an {N} -level qudit to a superconducting resonator at the adiabatic limit and beyond},\ }\href {https://doi.org/10.1103/PhysRevB.109.245303} {\bibfield  {journal} {\bibinfo  {journal} {Physical Review B}\ }\textbf {\bibinfo {volume} {109}},\ \bibinfo {pages} {245303} (\bibinfo {year} {2024})}\BibitemShut {NoStop}%
\bibitem [{\citenamefont {Gely}\ \emph {et~al.}(2018)\citenamefont {Gely}, \citenamefont {Steele},\ and\ \citenamefont {Bothner}}]{gely_nature_2018}%
  \BibitemOpen
  \bibfield  {author} {\bibinfo {author} {\bibfnamefont {M.~F.}\ \bibnamefont {Gely}}, \bibinfo {author} {\bibfnamefont {G.~A.}\ \bibnamefont {Steele}},\ and\ \bibinfo {author} {\bibfnamefont {D.}~\bibnamefont {Bothner}},\ }\bibfield  {title} {\bibinfo {title} {Nature of the {Lamb} shift in weakly anharmonic atoms: {From} normal-mode splitting to quantum fluctuations},\ }\href {https://doi.org/10.1103/PhysRevA.98.053808} {\bibfield  {journal} {\bibinfo  {journal} {Physical Review A}\ }\textbf {\bibinfo {volume} {98}},\ \bibinfo {pages} {053808} (\bibinfo {year} {2018})}\BibitemShut {NoStop}%
\bibitem [{\citenamefont {Hazra}\ \emph {et~al.}(2025)\citenamefont {Hazra}, \citenamefont {Dai}, \citenamefont {Connolly}, \citenamefont {Kurilovich}, \citenamefont {Wang}, \citenamefont {Frunzio},\ and\ \citenamefont {Devoret}}]{hazra_benchmarking_2025}%
  \BibitemOpen
  \bibfield  {author} {\bibinfo {author} {\bibfnamefont {S.}~\bibnamefont {Hazra}}, \bibinfo {author} {\bibfnamefont {W.}~\bibnamefont {Dai}}, \bibinfo {author} {\bibfnamefont {T.}~\bibnamefont {Connolly}}, \bibinfo {author} {\bibfnamefont {P.}~\bibnamefont {Kurilovich}}, \bibinfo {author} {\bibfnamefont {Z.}~\bibnamefont {Wang}}, \bibinfo {author} {\bibfnamefont {L.}~\bibnamefont {Frunzio}},\ and\ \bibinfo {author} {\bibfnamefont {M.}~\bibnamefont {Devoret}},\ }\bibfield  {title} {\bibinfo {title} {Benchmarking the {Readout} of a {Superconducting} {Qubit} for {Repeated} {Measurements}},\ }\href {https://doi.org/10.1103/PhysRevLett.134.100601} {\bibfield  {journal} {\bibinfo  {journal} {Physical Review Letters}\ }\textbf {\bibinfo {volume} {134}},\ \bibinfo {pages} {100601} (\bibinfo {year} {2025})},\ \bibinfo {note} {publisher: American Physical Society}\BibitemShut {NoStop}%
\bibitem [{\citenamefont {Touzard}\ \emph {et~al.}(2019)\citenamefont {Touzard}, \citenamefont {Kou}, \citenamefont {Frattini}, \citenamefont {Sivak}, \citenamefont {Puri}, \citenamefont {Grimm}, \citenamefont {Frunzio}, \citenamefont {Shankar},\ and\ \citenamefont {Devoret}}]{touzard_gated_2019}%
  \BibitemOpen
  \bibfield  {author} {\bibinfo {author} {\bibfnamefont {S.}~\bibnamefont {Touzard}}, \bibinfo {author} {\bibfnamefont {A.}~\bibnamefont {Kou}}, \bibinfo {author} {\bibfnamefont {N.}~\bibnamefont {Frattini}}, \bibinfo {author} {\bibfnamefont {V.}~\bibnamefont {Sivak}}, \bibinfo {author} {\bibfnamefont {S.}~\bibnamefont {Puri}}, \bibinfo {author} {\bibfnamefont {A.}~\bibnamefont {Grimm}}, \bibinfo {author} {\bibfnamefont {L.}~\bibnamefont {Frunzio}}, \bibinfo {author} {\bibfnamefont {S.}~\bibnamefont {Shankar}},\ and\ \bibinfo {author} {\bibfnamefont {M.}~\bibnamefont {Devoret}},\ }\bibfield  {title} {\bibinfo {title} {Gated {Conditional} {Displacement} {Readout} of {Superconducting} {Qubits}},\ }\href {https://doi.org/10.1103/PhysRevLett.122.080502} {\bibfield  {journal} {\bibinfo  {journal} {Physical Review Letters}\ }\textbf {\bibinfo {volume} {122}},\ \bibinfo {pages} {080502} (\bibinfo {year} {2019})}\BibitemShut {NoStop}%
\bibitem [{\citenamefont {Sunada}\ \emph {et~al.}(2022)\citenamefont {Sunada}, \citenamefont {Kono}, \citenamefont {Ilves}, \citenamefont {Tamate}, \citenamefont {Sugiyama}, \citenamefont {Tabuchi},\ and\ \citenamefont {Nakamura}}]{sunada_fast_2022}%
  \BibitemOpen
  \bibfield  {author} {\bibinfo {author} {\bibfnamefont {Y.}~\bibnamefont {Sunada}}, \bibinfo {author} {\bibfnamefont {S.}~\bibnamefont {Kono}}, \bibinfo {author} {\bibfnamefont {J.}~\bibnamefont {Ilves}}, \bibinfo {author} {\bibfnamefont {S.}~\bibnamefont {Tamate}}, \bibinfo {author} {\bibfnamefont {T.}~\bibnamefont {Sugiyama}}, \bibinfo {author} {\bibfnamefont {Y.}~\bibnamefont {Tabuchi}},\ and\ \bibinfo {author} {\bibfnamefont {Y.}~\bibnamefont {Nakamura}},\ }\bibfield  {title} {\bibinfo {title} {Fast {Readout} and {Reset} of a {Superconducting} {Qubit} {Coupled} to a {Resonator} with an {Intrinsic} {Purcell} {Filter}},\ }\href {https://doi.org/10.1103/PhysRevApplied.17.044016} {\bibfield  {journal} {\bibinfo  {journal} {Physical Review Applied}\ }\textbf {\bibinfo {volume} {17}},\ \bibinfo {pages} {044016} (\bibinfo {year} {2022})}\BibitemShut {NoStop}%
\bibitem [{\citenamefont {Spring}\ \emph {et~al.}(2025)\citenamefont {Spring}, \citenamefont {Milanovic}, \citenamefont {Sunada}, \citenamefont {Wang}, \citenamefont {Van~Loo}, \citenamefont {Tamate},\ and\ \citenamefont {Nakamura}}]{spring_fast_2025}%
  \BibitemOpen
  \bibfield  {author} {\bibinfo {author} {\bibfnamefont {P.~A.}\ \bibnamefont {Spring}}, \bibinfo {author} {\bibfnamefont {L.}~\bibnamefont {Milanovic}}, \bibinfo {author} {\bibfnamefont {Y.}~\bibnamefont {Sunada}}, \bibinfo {author} {\bibfnamefont {S.}~\bibnamefont {Wang}}, \bibinfo {author} {\bibfnamefont {A.~F.}\ \bibnamefont {Van~Loo}}, \bibinfo {author} {\bibfnamefont {S.}~\bibnamefont {Tamate}},\ and\ \bibinfo {author} {\bibfnamefont {Y.}~\bibnamefont {Nakamura}},\ }\bibfield  {title} {\bibinfo {title} {Fast {Multiplexed} {Superconducting}-{Qubit} {Readout} with {Intrinsic} {Purcell} {Filtering} {Using} a {Multiconductor} {Transmission} {Line}},\ }\href {https://doi.org/10.1103/PRXQuantum.6.020345} {\bibfield  {journal} {\bibinfo  {journal} {PRX Quantum}\ }\textbf {\bibinfo {volume} {6}},\ \bibinfo {pages} {020345} (\bibinfo {year} {2025})}\BibitemShut {NoStop}%
\bibitem [{\citenamefont {Frattini}\ \emph {et~al.}(2018)\citenamefont {Frattini}, \citenamefont {Sivak}, \citenamefont {Lingenfelter}, \citenamefont {Shankar},\ and\ \citenamefont {Devoret}}]{frattini_optimizing_2018}%
  \BibitemOpen
  \bibfield  {author} {\bibinfo {author} {\bibfnamefont {N.~E.}\ \bibnamefont {Frattini}}, \bibinfo {author} {\bibfnamefont {V.~V.}\ \bibnamefont {Sivak}}, \bibinfo {author} {\bibfnamefont {A.}~\bibnamefont {Lingenfelter}}, \bibinfo {author} {\bibfnamefont {S.}~\bibnamefont {Shankar}},\ and\ \bibinfo {author} {\bibfnamefont {M.~H.}\ \bibnamefont {Devoret}},\ }\bibfield  {title} {\bibinfo {title} {Optimizing the nonlinearity and dissipation of a {SNAIL} parametric amplifier for dynamic range},\ }\href {https://doi.org/10.1103/PhysRevApplied.10.054020} {\bibfield  {journal} {\bibinfo  {journal} {Physical Review Applied}\ }\textbf {\bibinfo {volume} {10}},\ \bibinfo {pages} {054020} (\bibinfo {year} {2018})}\BibitemShut {NoStop}%
\bibitem [{\citenamefont {Joshi}\ \emph {et~al.}(2025)\citenamefont {Joshi}, \citenamefont {Hazra}, \citenamefont {Ding}, \citenamefont {Miano}, \citenamefont {Dai}, \citenamefont {Umasankar}, \citenamefont {Kottandavida}, \citenamefont {Liu}, \citenamefont {Frunzio},\ and\ \citenamefont {Devoret}}]{joshi_lumped-element_2025}%
  \BibitemOpen
  \bibfield  {author} {\bibinfo {author} {\bibfnamefont {V.}~\bibnamefont {Joshi}}, \bibinfo {author} {\bibfnamefont {S.}~\bibnamefont {Hazra}}, \bibinfo {author} {\bibfnamefont {A.}~\bibnamefont {Ding}}, \bibinfo {author} {\bibfnamefont {A.}~\bibnamefont {Miano}}, \bibinfo {author} {\bibfnamefont {W.}~\bibnamefont {Dai}}, \bibinfo {author} {\bibfnamefont {G.}~\bibnamefont {Umasankar}}, \bibinfo {author} {\bibfnamefont {A.}~\bibnamefont {Kottandavida}}, \bibinfo {author} {\bibfnamefont {G.}~\bibnamefont {Liu}}, \bibinfo {author} {\bibfnamefont {L.}~\bibnamefont {Frunzio}},\ and\ \bibinfo {author} {\bibfnamefont {M.}~\bibnamefont {Devoret}},\ }\bibfield  {title} {\bibinfo {title} {Lumped-element broadband {SNAIL} parametric amplifier with on-chip pump filter for multiplexed readout},\ }\href {https://doi.org/10.1103/5l86-pw21} {\bibfield  {journal} {\bibinfo  {journal} {Physical Review Applied}\ }\textbf {\bibinfo {volume} {24}},\ \bibinfo {pages} {044003} (\bibinfo {year} {2025})}\BibitemShut {NoStop}%
\bibitem [{\citenamefont {McClure}\ \emph {et~al.}(2016)\citenamefont {McClure}, \citenamefont {Paik}, \citenamefont {Bishop}, \citenamefont {Steffen}, \citenamefont {Chow},\ and\ \citenamefont {Gambetta}}]{mcclure_rapid_2016}%
  \BibitemOpen
  \bibfield  {author} {\bibinfo {author} {\bibfnamefont {D.}~\bibnamefont {McClure}}, \bibinfo {author} {\bibfnamefont {H.}~\bibnamefont {Paik}}, \bibinfo {author} {\bibfnamefont {L.}~\bibnamefont {Bishop}}, \bibinfo {author} {\bibfnamefont {M.}~\bibnamefont {Steffen}}, \bibinfo {author} {\bibfnamefont {J.~M.}\ \bibnamefont {Chow}},\ and\ \bibinfo {author} {\bibfnamefont {J.~M.}\ \bibnamefont {Gambetta}},\ }\bibfield  {title} {\bibinfo {title} {Rapid {Driven} {Reset} of a {Qubit} {Readout} {Resonator}},\ }\href {https://doi.org/10.1103/PhysRevApplied.5.011001} {\bibfield  {journal} {\bibinfo  {journal} {Physical Review Applied}\ }\textbf {\bibinfo {volume} {5}},\ \bibinfo {pages} {011001} (\bibinfo {year} {2016})}\BibitemShut {NoStop}%
\bibitem [{\citenamefont {Shillito}\ \emph {et~al.}(2022)\citenamefont {Shillito}, \citenamefont {Petrescu}, \citenamefont {Cohen}, \citenamefont {Beall}, \citenamefont {Hauru}, \citenamefont {Ganahl}, \citenamefont {Lewis}, \citenamefont {Vidal},\ and\ \citenamefont {Blais}}]{shillito_dynamics_2022}%
  \BibitemOpen
  \bibfield  {author} {\bibinfo {author} {\bibfnamefont {R.}~\bibnamefont {Shillito}}, \bibinfo {author} {\bibfnamefont {A.}~\bibnamefont {Petrescu}}, \bibinfo {author} {\bibfnamefont {J.}~\bibnamefont {Cohen}}, \bibinfo {author} {\bibfnamefont {J.}~\bibnamefont {Beall}}, \bibinfo {author} {\bibfnamefont {M.}~\bibnamefont {Hauru}}, \bibinfo {author} {\bibfnamefont {M.}~\bibnamefont {Ganahl}}, \bibinfo {author} {\bibfnamefont {A.~G.}\ \bibnamefont {Lewis}}, \bibinfo {author} {\bibfnamefont {G.}~\bibnamefont {Vidal}},\ and\ \bibinfo {author} {\bibfnamefont {A.}~\bibnamefont {Blais}},\ }\bibfield  {title} {\bibinfo {title} {Dynamics of {Transmon} {Ionization}},\ }\href {https://doi.org/10.1103/PhysRevApplied.18.034031} {\bibfield  {journal} {\bibinfo  {journal} {Physical Review Applied}\ }\textbf {\bibinfo {volume} {18}},\ \bibinfo {pages} {034031} (\bibinfo {year} {2022})}\BibitemShut {NoStop}%
\bibitem [{\citenamefont {Xiao}\ \emph {et~al.}(2025)\citenamefont {Xiao}, \citenamefont {Venkatraman}, \citenamefont {Cortiñas}, \citenamefont {Chowdhury},\ and\ \citenamefont {Devoret}}]{xiao_diagrammatic_2025}%
  \BibitemOpen
  \bibfield  {author} {\bibinfo {author} {\bibfnamefont {X.}~\bibnamefont {Xiao}}, \bibinfo {author} {\bibfnamefont {J.}~\bibnamefont {Venkatraman}}, \bibinfo {author} {\bibfnamefont {R.~G.}\ \bibnamefont {Cortiñas}}, \bibinfo {author} {\bibfnamefont {S.}~\bibnamefont {Chowdhury}},\ and\ \bibinfo {author} {\bibfnamefont {M.~H.}\ \bibnamefont {Devoret}},\ }\bibfield  {title} {\bibinfo {title} {Diagrammatic method to compute the effective {Hamiltonian} of a driven nonlinear oscillator},\ }\href {https://doi.org/10.1103/d6j2-lsf7} {\bibfield  {journal} {\bibinfo  {journal} {Physical Review Applied}\ }\textbf {\bibinfo {volume} {24}},\ \bibinfo {pages} {044021} (\bibinfo {year} {2025})}\BibitemShut {NoStop}%
\bibitem [{\citenamefont {Dumas}\ \emph {et~al.}(2024)\citenamefont {Dumas}, \citenamefont {Groleau-Paré}, \citenamefont {McDonald}, \citenamefont {Muñoz-Arias}, \citenamefont {Lledó}, \citenamefont {D’Anjou},\ and\ \citenamefont {Blais}}]{dumas_measurement-induced_2024}%
  \BibitemOpen
  \bibfield  {author} {\bibinfo {author} {\bibfnamefont {M.~F.}\ \bibnamefont {Dumas}}, \bibinfo {author} {\bibfnamefont {B.}~\bibnamefont {Groleau-Paré}}, \bibinfo {author} {\bibfnamefont {A.}~\bibnamefont {McDonald}}, \bibinfo {author} {\bibfnamefont {M.~H.}\ \bibnamefont {Muñoz-Arias}}, \bibinfo {author} {\bibfnamefont {C.}~\bibnamefont {Lledó}}, \bibinfo {author} {\bibfnamefont {B.}~\bibnamefont {D’Anjou}},\ and\ \bibinfo {author} {\bibfnamefont {A.}~\bibnamefont {Blais}},\ }\bibfield  {title} {\bibinfo {title} {Measurement-{Induced} {Transmon} {Ionization}},\ }\href {https://doi.org/10.1103/PhysRevX.14.041023} {\bibfield  {journal} {\bibinfo  {journal} {Physical Review X}\ }\textbf {\bibinfo {volume} {14}},\ \bibinfo {pages} {041023} (\bibinfo {year} {2024})}\BibitemShut {NoStop}%
\bibitem [{\citenamefont {Johansson}\ \emph {et~al.}(2012)\citenamefont {Johansson}, \citenamefont {Nation},\ and\ \citenamefont {Nori}}]{johansson_qutip_2012}%
  \BibitemOpen
  \bibfield  {author} {\bibinfo {author} {\bibfnamefont {J.~R.}\ \bibnamefont {Johansson}}, \bibinfo {author} {\bibfnamefont {P.~D.}\ \bibnamefont {Nation}},\ and\ \bibinfo {author} {\bibfnamefont {F.}~\bibnamefont {Nori}},\ }\bibfield  {title} {\bibinfo {title} {{QuTiP}: {An} open-source {Python} framework for the dynamics of open quantum systems},\ }\href {https://doi.org/10.1016/j.cpc.2012.02.021} {\bibfield  {journal} {\bibinfo  {journal} {Computer Physics Communications}\ }\textbf {\bibinfo {volume} {183}},\ \bibinfo {pages} {1760} (\bibinfo {year} {2012})}\BibitemShut {NoStop}%
\bibitem [{\citenamefont {Johansson}\ \emph {et~al.}(2013)\citenamefont {Johansson}, \citenamefont {Nation},\ and\ \citenamefont {Nori}}]{johansson_qutip_2013}%
  \BibitemOpen
  \bibfield  {author} {\bibinfo {author} {\bibfnamefont {J.~R.}\ \bibnamefont {Johansson}}, \bibinfo {author} {\bibfnamefont {P.~D.}\ \bibnamefont {Nation}},\ and\ \bibinfo {author} {\bibfnamefont {F.}~\bibnamefont {Nori}},\ }\bibfield  {title} {\bibinfo {title} {{QuTiP} 2: {A} {Python} framework for the dynamics of open quantum systems},\ }\href {https://doi.org/10.1016/j.cpc.2012.11.019} {\bibfield  {journal} {\bibinfo  {journal} {Computer Physics Communications}\ }\textbf {\bibinfo {volume} {184}},\ \bibinfo {pages} {1234} (\bibinfo {year} {2013})}\BibitemShut {NoStop}%
\bibitem [{\citenamefont {Weiss}()}]{weiss_floquet_nodate}%
  \BibitemOpen
  \bibfield  {author} {\bibinfo {author} {\bibfnamefont {D.~K.}\ \bibnamefont {Weiss}},\ }\href {https://dkweiss.net/floquet/} {\bibinfo {title} {floquet}}\BibitemShut {NoStop}%
\bibitem [{\citenamefont {Kitaev}(2003)}]{kitaev_fault-tolerant_2003}%
  \BibitemOpen
  \bibfield  {author} {\bibinfo {author} {\bibfnamefont {A.~Y.}\ \bibnamefont {Kitaev}},\ }\bibfield  {title} {\bibinfo {title} {Fault-tolerant quantum computation by anyons},\ }\href {https://doi.org/10.1016/S0003-4916(02)00018-0} {\bibfield  {journal} {\bibinfo  {journal} {Annals of Physics}\ }\textbf {\bibinfo {volume} {303}},\ \bibinfo {pages} {2} (\bibinfo {year} {2003})}\BibitemShut {NoStop}%
\bibitem [{\citenamefont {{Google Quantum AI}}(2025)}]{google_quantum_ai_quantum_2025}%
  \BibitemOpen
  \bibfield  {author} {\bibinfo {author} {\bibnamefont {{Google Quantum AI}}},\ }\bibfield  {title} {\bibinfo {title} {Quantum error correction below the surface code threshold},\ }\href {https://doi.org/10.1038/s41586-024-08449-y} {\bibfield  {journal} {\bibinfo  {journal} {Nature}\ }\textbf {\bibinfo {volume} {638}},\ \bibinfo {pages} {920} (\bibinfo {year} {2025})}\BibitemShut {NoStop}%
\end{thebibliography}%
 
\end{document}


\widetext
 
\title{Supplementary information for ``High-frequency dispersive readout of a superconducting qubit''}

\author{Pavel~D.~Kurilovich}
\thanks{These two authors contributed equally.\\
pkurilovich@fas.harvard.edu, tom.connolly@yale.edu}
\extrathanks{Present address: Department of Physics, Harvard University, Cambridge, Massachusetts 02138, USA}
\affiliation{Departments of Applied Physics and Physics, Yale University, New Haven, Connecticut 06520, USA}
\author{Thomas~Connolly}
\thanks{These two authors contributed equally.\\
pkurilovich@fas.harvard.edu,
tom.connolly@yale.edu}
\affiliation{Departments of Applied Physics and Physics, Yale University, New Haven, Connecticut 06520, USA}
\author{Charlotte~G.~L.~B\o ttcher}
\thanks{{Present address: Department of Applied Physics, Stanford University, Stanford, California 94305, USA}}
\affiliation{Departments of Applied Physics and Physics, Yale University, New Haven, Connecticut 06520, USA}
 
\author{Daniel~K.~Weiss}
\thanks{{Present address: Quantum Circuits, Inc., New Haven, CT, USA}}
\affiliation{Departments of Applied Physics and Physics, Yale University, New Haven, Connecticut 06520, USA}
\affiliation{Yale Quantum Institute, Yale University, New Haven, Connecticut 06511, USA}
 
\author{Sumeru~Hazra}
\affiliation{Departments of Applied Physics and Physics, Yale University, New Haven, Connecticut 06520, USA}
\author{Vidul~R.~Joshi}
\thanks{{Present address: Microsoft Quantum}}
\affiliation{Departments of Applied Physics and Physics, Yale University, New Haven, Connecticut 06520, USA}
 
\author{Andy~Z.~Ding}
\affiliation{Departments of Applied Physics and Physics, Yale University, New Haven, Connecticut 06520, USA}
\author{Heekun~Nho}
\affiliation{Departments of Applied Physics and Physics, Yale University, New Haven, Connecticut 06520, USA}
\author{Spencer~Diamond}
\affiliation{Departments of Applied Physics and Physics, Yale University, New Haven, Connecticut 06520, USA}
\author{Vladislav~D.~Kurilovich}\thanks{{Present address: Google Quantum AI, 301 Mentor Dr, Goleta, CA93111, USA}}
\affiliation{Departments of Applied Physics and Physics, Yale University, New Haven, Connecticut 06520, USA}
 
\author{Mykola~Kishmar}
\affiliation{Physics Department, Columbia University, New York, New York 10027, USA}
 
\author{Wei~Dai}
\affiliation{Departments of Applied Physics and Physics, Yale University, New Haven, Connecticut 06520, USA}
\author{Valla~Fatemi}
\affiliation{Departments of Applied Physics and Physics, Yale University, New Haven, Connecticut 06520, USA}
\affiliation{School of Applied and Engineering Physics, Cornell University, Ithaca, New York 14853, USA}
\author{Luigi Frunzio}
\affiliation{Departments of Applied Physics and Physics, Yale University, New Haven, Connecticut 06520, USA}
\author{Leonid~I.~Glazman}
\affiliation{Departments of Applied Physics and Physics, Yale University, New Haven, Connecticut 06520, USA}
\affiliation{Yale Quantum Institute, Yale University, New Haven, Connecticut 06511, USA}
\author{Michel~H.~Devoret}\thanks{michel.devoret@yale.edu\\
{Present address: Physics Dept., U.C. Santa Barbara, Santa Barbara, California 93106, USA and Google Quantum AI, 301 Mentor Dr, Goleta, California 93111, USA}}
\affiliation{Departments of Applied Physics and Physics, Yale University, New Haven, Connecticut 06520, USA}
 
\date{\today}
 
\maketitle
 
\tableofcontents
\newpage
\section{Experimental setup}
\subsection{Wiring diagram}
\begin{figure}[h!]
  \begin{center}
    \includegraphics[scale = 1.0]{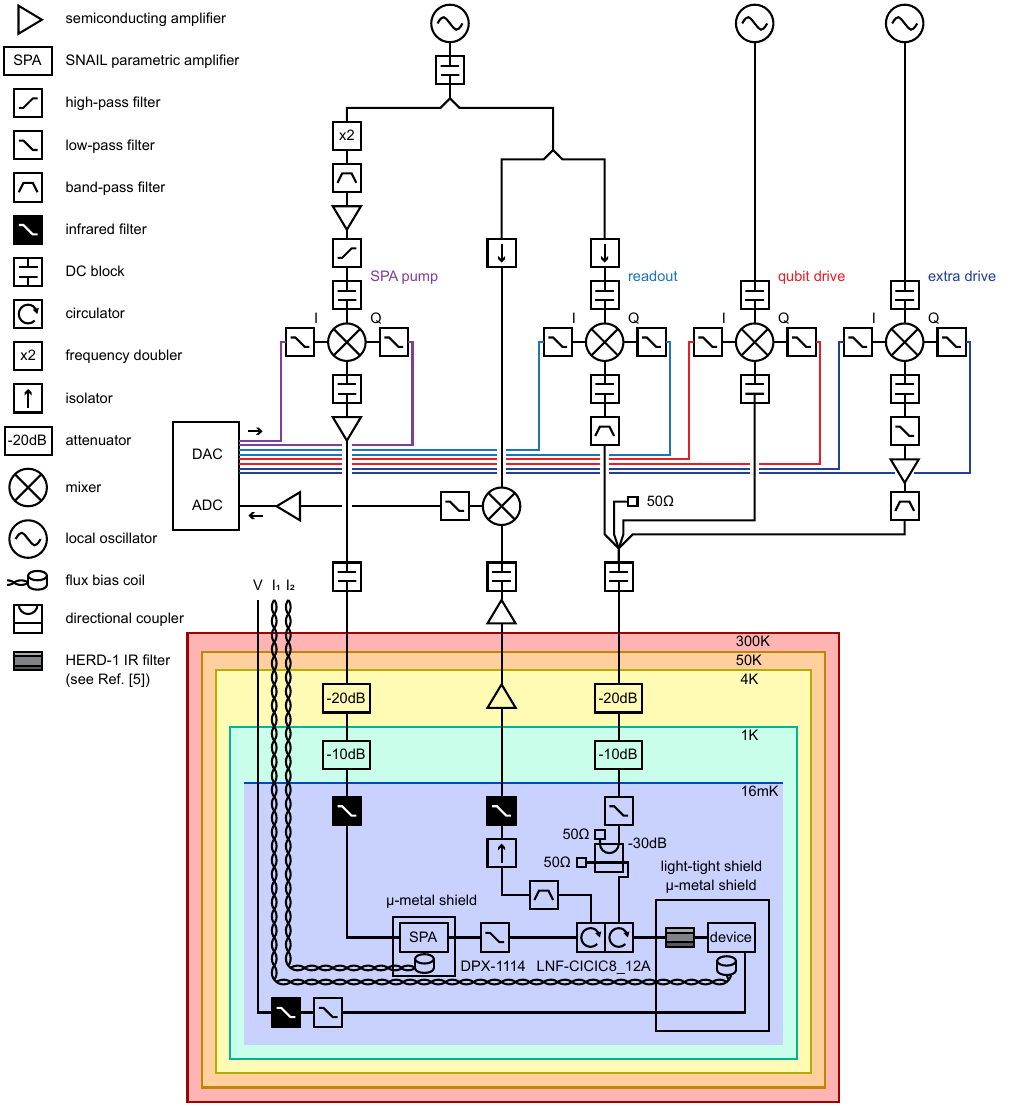}
    \caption{Wiring diagram of the setup. The RuOx thermometer on the mixing chamber reads 16 mK; the excited-state population of the qubit indicates that the actual temperature is at most 11 mK.}
    \label{fig:wiring}
  \end{center}
\end{figure}

\subsection{Fabrication details}
Our transmon is fabricated on an annealed EFG Sapphire substrate. The transmon Josephson junction is an Al/AlOx/Al trilayer made using the bridge-free technique. The transmon pads, readout resonator, and ground plane are made of tantalum. This was shown to improve the quality factor compared to an all-aluminum transmon fabrication \cite{place_new_2021, ganjam_surpassing_2024}. The Al/AlOx/Al junctions and the tantalum structures are demonstrated in Figure~\ref{fig:fab}a.
 
The fabrication of the tantalum layer follows Ref.~\cite{ganjam_surpassing_2024}. The bridge-free Josephson junction fabrication is similar to that described in Ref.~\cite{serniak_direct_2019} except for two nuances. First, when forming the Josephson junction, we oxidized the aluminum film at a higher pressure of 50 Torr for a longer duration of 30 minutes. This allowed us to substantially lower the Josephson energy compared to Ref.~\cite{serniak_direct_2019} without reducing the junction size. Second, compared to Ref.~\cite{serniak_direct_2019} we used more aggressive ion milling in order to make a good contact between tantalum and aluminum. Ion milling was similar to that employed in Ref.~\cite{ganjam_surpassing_2024}. 
 
\subsection{Packaging details}
 
\begin{figure}[h]
  \begin{center}
    \includegraphics[scale = 1]{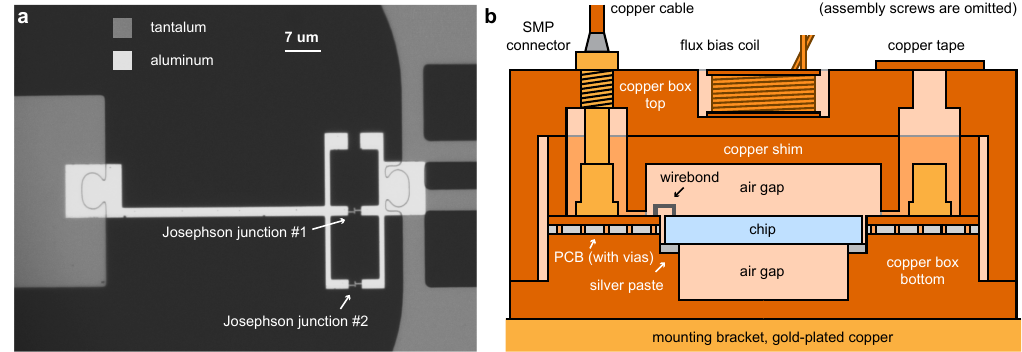}
    \caption{Chip and packaging details. \textbf{a}, Zoom in on the Josephson junction region of our device. Al/AlOx/Al superconducting quantum interference device (SQUID) is interfaced with the tantalum pads and the ground plane. The on-chip flux lines (on the right from the SQUID) and an additional SQUID arm are not used in the present experiment. The flux lines are shorted to the ground plane with wirebonds and the extra SQUID arm is left open. \textbf{b}, Schematic of the package hosting the chip (scale not preserved).}
    \label{fig:fab}
  \end{center}
\end{figure}

The 7 mm by 7 mm chip hosting the transmon was placed in a copper package. To reduce the participation of lossy copper in the device and to increase the frequency of spurious box modes, we carved an air-gap below and above the chip. The schematic of the package is shown in Figure~\ref{fig:fab}b. The package is then embedded in an additional light-tight shield which protects it from stray infrared radiation; this suppresses the qubit heating and reduces quasiparticle-related decoherence. The shield is analogous to that described in Ref.~\cite{connolly_coexistence_2024}. Inside the shield, we used a low-loss HERD-1 low-pass filter \cite{rehammar_low-pass_2023} to suppress the infrared radiation coming from the readout line. We use small dots of silver paste at the corners of the sapphire chip to thermalize it to the copper package.
 
\section{Transmon superconducting qubit and its parameters in our work}
 
In this section, we give a brief primer on transmon superconducting qubits. We highlight the analogy between the Hamiltonian of a transmon with that of a quantum mechanical pendulum placed in a gravity field. Detailed discussion of the transmon and its operation principles can be found in Ref.~\cite{koch_charge-insensitive_2007}. We also provide the list of parameters used in our experiment.
 
\subsection{Transmon Hamiltonian}
 
The transmon is a superconducting circuit in which a Josephson tunnel junction is shunted by a capacitor. The Cooper pairs forming the superconducting condensate can be transferred between the capacitor pads by tunneling through the junction. The transmon Hamiltonian is given by
\begin{equation}
    \label{eq:ham}
    H = 4 E_C (N - n_g)^2 - E_J \cos \varphi,
\end{equation}
where, $\varphi$ is the operator of superconducting phase difference across the junction and $N$ is the operator of the number of Cooper pairs transferred across it. The charge and phase operators are conjugate to each other, $[N, e^{i\varphi}]=e^{i\varphi}$; equivalently this relationship can be expressed as $N=-i\partial_\varphi$. In Eq.~\eqref{eq:ham}, $E_C$ is the transmon charging energy and $E_J$ is the Josephson energy of the junction. In a transmon qubit $E_J\gg E_C$. Parameter $n_g$ describes the so-called offset charge. Offset charge may stem from the presence of stray charges in the transmon environment; in that case $n_g$ drifts with time, typically with a $1/f$ power spectrum \cite{riste_millisecond_2013, serniak_hot_2018}.
 
Notably, the Hamiltonian of a transmon, Eq.~\eqref{eq:ham}, is formally equivalent to that of a quantum-mechanical pendulum placed in a gravity field \cite{koch_charge-insensitive_2007}. In this analogy, the charging energy plays the role of kinetic energy and Josephson energy plays the role of potential energy. The offset charge plays the role of vector potential, assuming that the pendulum is charged and placed in a uniform magnetic field perpendicular to the plane of its motion.
 
\subsection{Spectrum of a transmon}
 
The energy spectrum of the transmon can be separated into two regions. First, a set of states is localized within the Josephson potential well, $|E|<E_J$. Second, a second set of states has energies exceeding the well edge, $E > E_J$. 
 
The spectrum of states deep inside of the Josephson well resembles that of an oscillator. In the regime $E_J \gg E_C$, the frequency of this oscillator is given by $\hbar \wq = \sqrt{8E_JE_C} - E_C$.  Due to non-linearity -- which stems from the Josephson potential -- the oscillator is anharmonic. This means that the energy spacing between the neighboring levels, $\Delta E_n = E_{n+1} - E_n$, depends on $n$. In the considered regime $E_J \gg E_C$, the anharmonicity is given by $\Delta E_{n+1} - \Delta E_{n} = -E_C$ for states deep in the Josephson well. Therefore, anharmonicity is small compared to the level spacing $\sim\hbar\wq$. In the discussion so far, we omitted the dependence of energy levels on the offset charge. Inside of the well, this dependence is weak \cite{koch_charge-insensitive_2007}.
 
In contrast, states above the Josephson well, $E>E_J$, have a strong offset-charge sensitivity. This is because for these states, the $n_g$-dependent charging energy term in the Hamiltonian is dominant.
 
Within the pendulum analogy discussed above, the states with energies above the well correspond to running solutions where the pendulum skips over the upper equilibrium point $\varphi = \pi$.
 
\subsection{Parameters of our transmon}
 
For our transmon, the parameters of Hamiltonian~\eqref{eq:ham} are $E_J/h = 2.195\ghz$, $E_C/h = 0.036\ghz$, resulting in the qubit frequency $\wq/2\pi = 0.757\ghz$. The Josephson and charging energies are determined from the measured transmon frequency and anharmonicity by comparing these parameters with a numerical simulation.
 
The relaxation time of our qubit is $388\:\mu\mathrm{s}$. The coherence time at our operation point is $T_2 = 24\us$ (echo); it is set by flux noise, as we bias the SQUID away from its sweet spot. Short coherence time does not affect the readout properties of the qubit. Moreover, it can be straightforwardly extended by designing a SQUID such that the coherence sweet spot and the optimal point for readout coincide.
 
\section{Purcell effect in high-frequency readout}
\label{sec:purcell}
\begin{figure}[t]
  \begin{center}
    \includegraphics[scale = 1]{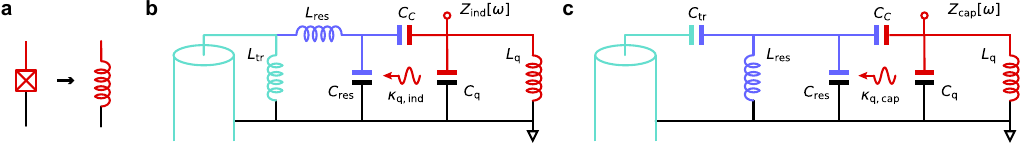}
    \caption{
    Schematics of the lumped-element circuits used to estimate the Purcell decay rate for the high-frequency readout. \textbf{a}, To estimate the Purcell decay rate, we use classical circuit analysis. To this end, we replace the Josephson junction with a linear inductor. \textbf{b}, Notations of circuit elements in the case when the readout resonator is inductively coupled to the transmission line. This lumped circuit can be used to model a $\lambda/4$ readout resonator used in our manuscript. \textbf{c}, Same for the case of a resonator capacitively coupled to the transmission line.}
    \label{fig:purcell}
  \end{center}
\end{figure}
The transmon inherits dissipation through its capacitive coupling to a lossy readout resonator. This is known as the Purcell effect \cite{purcell_resonance_1946, kleppner_inhibited_1981, goy_observation_1983}; it places a limit on the coherence time of the qubit. In this section, we analyze the Purcell effect in the limit where the frequency of the readout resonator vastly exceeds that of the transmon, $\wres\gg \wq$. We show that the Purcell effect becomes suppressed by $\wres/\wq$ raised to a high power, compared to the well-known expression \cite{blais_cavity_2004, houck_controlling_2008} derived for the case $|\wres - \wq| \ll \wq$. The degree of suppression depends on whether the coupling between the resonator and the transmission line is capacitive or inductive; we consider both cases here. Throughout the derivation we neglect direct coupling between the transmon and the transmission line. {We also assume that the impedance of a transmission line is independent of frequency.} Finally, at the end of the section we compare the two limits to a finite-element simulation of our device. 
 
\subsection{Lumped element models for the Purcell effect}
 
To describe the Purcell effect, we resort to classical circuit analysis. To this end, we replace the Josephson junction of our transmon with a linear inductance. The resulting lumped model of our device is shown in Fig.~\ref{fig:purcell}. The parameters of this model can be constrained by the measured qubit and resonator frequencies. Assuming that the coupling between the resonator and the transmission line is weak, we obtain
\begin{equation}
    \wres^2 = \frac{\cq + C_C}{\lres C_\Sigma^2},\quad \wq^2 = \frac{1}{\lqb(\cq + C_C)},\quad C_\Sigma^2 = \cres \cq + \cres C_C + \cq C_C.
\end{equation}
In deriving these equations, we assumed $\wres\gg\wq$ and neglected corrections of order $\eta^2(\wq/\wres)^2$, where coupling efficiency $0<\eta<1$ is defined\footnote{We implicitly assumed that $\eta$ is sufficiently far from unity. The case $1 - \eta \ll 1$ should be treated separately, and is outside of the scope of our work.} as
\begin{equation}
\label{eq:eta}
    \eta = \frac{C_C}{\sqrt{(\cres + C_C)(\cq+C_C)}}.    
\end{equation} 
Qualitatively, neglecting these corrections amounts to replacing the qubit inductance with an open at $\omega = \wres$ and resonator inductance with a short at $\omega =\wq$.
 
We begin by finding the resonator linewidth in terms of the circuit parameters of Fig.~\ref{fig:purcell}. For the inductive and capacitive coupling between the resonator and the transmission line we find respectively
\begin{equation}
    \kappa_\mathrm{ind} = \frac{L_\mathrm{tr}^2 \wres^2}{Z_0 \lres},\quad \kappa_\mathrm{cap} = \frac{\wres^2 Z_0 C_\mathrm{tr}^2 (\cq+C_C)}{C_\Sigma^2}.
\end{equation}
{Here, $Z_0$ is the characteristic impedance of the transmission line.} Next, we find the qubit linewidth -- which quantifies the dissipation due to the Purcell effect -- by evaluating the impedance of the island close to the qubit frequency ($|\omega-\wq|\ll \wq$). Explicitly, for the two cases we find
\begin{equation}
    Z_\mathrm{ind}[\omega] = \frac{1}{\frac{1}{i\omega \lqb} + i \omega(\cq + C_C) + \frac{\omega^4 C_C^2 L_\mathrm{tr}^2}{Z_0}},\quad Z_\mathrm{cap}[\omega] = \frac{1}{\frac{1}{i\omega \lqb} + i \omega(\cq + C_C) + \omega^6 C_C^2 C_\mathrm{tr}^2 \lres^2 Z_0}
\end{equation}
Correspondingly, we obtain the qubit linewidth in both cases
\begin{equation}
    \label{eq:kappa_eta}
    \kappa_{\mathrm{q},\mathrm{ind}} = \frac{\eta^2}{1-\eta^2}\frac{\wq^4}{\wres^4}\kappa_\mathrm{ind},\quad\kappa_{\mathrm{q},\mathrm{cap}} = \frac{\eta^2}{1-\eta^2} \frac{\wq^6}{\wres^6}\kappa_\mathrm{cap}.
\end{equation}
Alternatively, we can rewrite these equations in term of the coupling strength $g = \frac{1}{2}\eta \sqrt{\wq \wres}$:
\begin{equation}
\label{eq:purcell_correct}
    \kappa_{\mathrm{q},\mathrm{ind}} = \frac{1}{1-\eta^2}\frac{4\wq^3}{\wres^3}\frac{g^2}{\wres^2}\kappa_\mathrm{ind},\quad\kappa_{\mathrm{q},\mathrm{cap}} = \frac{1}{1-\eta^2} \frac{4\wq^5}{\wres^5}\frac{g^2}{\wres^2}\kappa_\mathrm{cap}.
\end{equation}
Notably, these equations are different from a standard expression for the Purcell effect \cite{blais_cavity_2004, houck_controlling_2008} derived under the assumptions $|\wres - \wq|\ll \wq$ and $\eta^2\ll1$,
\begin{equation}
\label{eq:stand_purcell}
    \kappa_\mathrm{q} = \frac{g^2}{(\wres - \wq)^2}\kappa.
\end{equation}
Compared to this expression, Purcell decay is suppressed by multiple powers of a large parameter $\wres/\wq$. This suppression arises from the frequency dependence of the impedance of the coupling elements. For example, capacitive coupling between the resonator and the transmission line becomes shunted by the resonator inductance at small frequencies. Additionally, linewidths in Eq.~\eqref{eq:purcell_correct} contain an extra prefactor of $1/(1-\eta^2)$ compared to Eq.~\eqref{eq:stand_purcell}. We note that deviations from the standard formula for the Purcell effect have been recently discussed in Ref.~\cite{yen_interferometric_2025}. Similar suppression is present in fluxonium qubits \cite{manucharyan_evidence_2012}.
 
\subsection{Finite-element estimate of the Purcell limit}
 
The coupling between the readout resonator and the transmission line in our device is not purely inductive, so the two limits treated above bracket the actual circuit. For the parameters of our device, Eq.~\eqref{eq:purcell_correct} gives a Purcell limit on the qubit lifetime of $11\ms$ for purely inductive coupling and $1.6\:\mathrm{s}$ for purely capacitive coupling; the two differ by the factor $(\omega_{\rm r}/\omega_{\rm q})^2 \approx 150$. To go beyond the lumped modeling, we simulated the device with a Sonnet finite-element solver. Assuming a perfect $50\:\Omega$ load, the simulation gives a Purcell limit of $13\ms$. The device therefore lies close to the inductive limit, and the inductive scaling quoted in the main text is the conservative of the two.
 
\section{Dispersive shift in high-frequency readout}
\label{sec:disp_shift}
In dispersive readout, a transmon qubit is capacitively coupled to a readout resonator, see Fig.~1 of the main text. Due to the transmon non-linearity, the frequency of the resonator shifts by $\chi$ when the transmon is excited from its ground state $|0\rangle$ to the first excited state $|1\rangle$. By measuring the phase of the signal reflected from the resonator, it is then possible to infer the transmon state. In this section, we theoretically evaluate the dispersive shift for the high-frequency readout regime, ${\omega_r \gg \wq}$. We express $\chi$ as a function of parameters characterizing the transmon, the resonator, and their coupling.
 
A potential concern in the high-frequency readout regime is the reduction of the dispersive shift $\chi$ with the increase of ${\omega_r/\wq}$. The readout signal, correspondingly, becomes diminished. However, as we explain below, this issue does not pose a significant limitation due to a favorable asymmetric scaling: unwanted resonances are exponentially suppressed as a function of the ratio ${\omega_r/\wq}$; the dispersive shift $\chi$, in contrast, reduces only as a power law in the same parameter. This disparity means that one can compensate for the power-law signal loss by increasing the qubit-resonator coupling strength, while at the same time maintaining the exponential protection from spurious resonances.
 
First, in Section \ref{sec:chi_calc}, we derive a new expression for the dispersive shift $\chi$ that is valid both in the high-frequency readout regime $\wres\gg\wq$ and in the standard regime $|\wres - \wq|\ll \wres$. This expression relates $\chi$ to the charging energy of the transmon $E_C$, the resonator and qubit frequencies $\wres$ and $\wq$, as well as to the strength of coupling between the resonator and the qubit $g$. Then, in Section \ref{sec:compare_to_exp}, we compare the experimentally measured dispersive shifts to the obtained expression. We find that theory and experiment are in an excellent agreement. In Section \ref{sec:disp_resonances}, we explain the complications in the behavior of $\chi$ between the two limiting cases, $|\wres - \wq|\ll \wres$ and $\wres\gg\wq$. We show that in this crossover regime $\chi$ goes through a sequence of resonances. Finally, in Section \ref{sec:chi_power_law} we explicitly show that as $\wres/\wq$ is increased, $\chi$ reduces as a power law of this parameter. Specifically, $\chi\propto \wq/\wres$ for capacitive coupling between the readout resonator and the transmon, assuming that the capacitances are fixed.
 
Throughout this section, we assume that the Josephson energy $E_J$ vastly exceeds the charging energy $E_C$, such that the circuit is deep in the transmon regime. For the parameters of our transmon $E_J/E_C \sim 60$ and thus this assumption is well justified.
 
\subsection{Calculation of the dispersive shift}
\label{sec:chi_calc}
 
The Hamiltonian of a transmon coupled to a readout resonator is given by
\begin{equation}
\label{eq:res_tr_ham}
    H = 4 E_C (N - n_g)^2 - E_J \cos \varphi + \hbar g \frac{N}{\nzpf} (a + a^\dagger) + \hbar\omega_\mathrm{r,bare} (a^\dagger a + 1/2).
\end{equation}
 
The parameter $\nzpf = (E_J / 32 E_C)^{1/4}$ denotes the zero point fluctuations of the transmon charge. The readout resonator has bare frequency $\omega_\mathrm{r,bare}$; the corresponding creation and annihilation operators are $a^\dagger$ and $a$, respectively. Parameter $g$ denotes the coupling strength between the readout resonator and the transmon.
 
The bare resonator frequency $\omega_\mathrm{r,bare}$ does not account for the presence of the Josephson junction; it would be the resonator frequency if the Josephson junction was replaced by an open circuit. We can compute corrections to this bare frequency due to the presence of the junction using second-order perturbation theory. When the transmon is in its $n$-th excited state, $|n\rangle$, the frequency shift of the resonator $\chi_n$ is \cite{manucharyan_phd_2012, zhu_circuit_2013, serniak_direct_2019}
\begin{equation}
\label{eq:shift_general_formula}
\hbar\chi_n	=\frac{(\hbar g)^{2}}{\nzpf^2}\sum_{m\neq n}|\langle n|N|m\rangle|^{2}\frac{2E_{m\rightarrow n}}{E_{m\rightarrow n}^{2}-(\hbar\omega_\mathrm{r,bare})^{2}}.
\end{equation}
Here, $E_{m\rightarrow n} = E_n - E_m$, with $E_n$ the energy of the $n$-th excited transmon state (for computational states $n = 0,1$). In the high-frequency readout regime, we can safely restrict the summation to $m = n\pm 1$ as long as $E_J \gg E_C$. One concern with this restriction is the contribution of terms with $|E_{m\rightarrow n}|$ close to $\hbar \omega_\mathrm{r,bare}$, corresponding to unwanted resonances. For these terms with $m\neq n\pm 1$, the denominator in Eq.~\eqref{eq:shift_general_formula} becomes small. The matrix element for the corresponding transitions is, however, exponentially small in the ratio $\omega_\mathrm{r,bare} / \wq$, as explained in the main text. This justifies restricting the summations $m = n\pm 1$ for a sufficiently large $\omega_\mathrm{r,bare} / \wq$. We analyze the case when these transitions become important in Section~\ref{sec:disp_resonances}.
 
For the transmon in the ground state, the resonator frequency shift is given by
\begin{equation}
\label{eq:chi0}
\hbar\chi_0 = \frac{(\hbar g)^{2}}{\nzpf^2}|\langle0|N|1\rangle|^{2}\frac{2E_{0\rightarrow1}}{(\hbar\omega_\mathrm{r,bare})^{2} - E_{0\rightarrow1}^{2}}.
\end{equation}
Here, the matrix element $|\langle0| N|1\rangle|^{2}$ is close to $\nzpf^2$ in the transmon regime.
As we explain below, however, it is important to go beyond this approximation when computing the difference between $\chi_1$ and $\chi_0$.
 
For the transmon in the first excited state, the resonator frequency shift is
\begin{equation}
\label{eq:chi1}
\hbar\chi_1	= - \frac{(\hbar g)^{2}}{\nzpf^2}|\langle0|N|1\rangle|^{2}\frac{2E_{0\rightarrow1}}{(\hbar\omega_\mathrm{r,bare})^{2} - E_{0\rightarrow1}^{2}}+\frac{(\hbar g)^{2}}{\nzpf^2}|\langle1|N|2\rangle|^{2}\frac{2E_{1\rightarrow2}}{(\hbar\omega_\mathrm{r,bare})^{2} - E_{1\rightarrow2}^{2}}.
\end{equation}
The dispersive shift, quantity most relevant for qubit readout, is defined as $\hbar\chi	=\hbar\chi_{1}-\hbar\chi_{0}$. We find
\begin{equation}
\label{eq:shift_intermediate_formula}
\hbar\chi = 2\frac{(\hbar g)^{2}}{\nzpf^2}\left(|\langle1|N|2\rangle|^{2}\frac{E_{1\rightarrow2}}{(\hbar\omega_\mathrm{r,bare})^{2} - E_{1\rightarrow2}^{2}} -|\langle0|N|1\rangle|^{2}\frac{2E_{0\rightarrow1}}{(\hbar\omega_\mathrm{r,bare})^{2} - E_{0\rightarrow1}^{2}} \right)
\end{equation}
\subsubsection{Corrections to the charge matrix elements}
To further simplify expression \eqref{eq:shift_intermediate_formula}, care is needed when evaluating the matrix elements of $N$. To get a valid expression for $\chi$, it is essential to account for the corrections to these matrix elements due to the transmon non-linearity. As we explain below, taking these corrections into the account results in
\begin{equation}
\label{eq:me}
|\langle1|N|2\rangle|^{2}	 = 2|\langle1|N|0\rangle|^{2}\left(1-\frac{E_{C}}{E_{0\rightarrow1}}\right) = 2|\langle1|N|0\rangle|^{2}\frac{E_{1\rightarrow2}}{E_{0\rightarrow1}}.
\end{equation}
The simplifying assumption $|\langle1|N|2\rangle|^{2} = 2 |\langle 0|N|1\rangle|^{2}$ would lead to an incorrect result for the dispersive shift. This occurs because in the leading approximation in $E_{C}$ the dispersive shift vanishes in Eq.~\eqref{eq:shift_intermediate_formula}. It is then necessary to keep track of corrections linear in $E_{C}$.
 
To derive Eq.~\eqref{eq:me}, one can approximate the transmon Hamiltonian with that of a non-linear oscillator with a quartic nonlinearity. This is achieved by replacing $N = \nzpf (b + b^\dagger)$, $\varphi = -i \pzpf (b - b^\dagger)$, and expanding the cosine potential of the transmon to the fourth order. Here, zero point fluctuations of the phase, $\pzpf$, are related to that of the charge via $\pzpf \nzpf = 1 / 2$. The resulting Hamiltonian reads
\begin{equation}
    H_\mathrm{tr} = H_0 + V,\quad H_0 = \hbar\wq (b^\dagger b + 1/2),\quad V = -\frac{E_C}{12} (b - b^\dagger)^4.
\end{equation}
Here, we treat $V$ as a perturbation. To compute the matrix element $|\langle i|N|i+1\rangle|^2$, we expand transmon eigenstates $|i\rangle$ and $|i+1\rangle$ into the eigenstates of $H_0$ up to the first order in $V$. This way, we find
\begin{gather}
    |i\rangle = |i\rangle_0 - \frac{E_C}{24\hbar\wq}(4i+6)\sqrt{i+1}\sqrt{i+2}|i+2\rangle_0 + ...\:,\label{eq:istate}\\
    |i+1\rangle = |i + 1\rangle_0 + (4i + 2)\frac{E_C}{24\hbar\wq}\sqrt{i+1}\sqrt{i}|i-1\rangle_0 + ...\:.\label{eq:iplus1state}
\end{gather}
Here, ``0'' subscript indicates the eigenstates of $H_0$ and $...$ denote terms irrelevant for the calculation of $|\langle i|N|i+1\rangle|^2$. Substituting Eq.~\eqref{eq:istate} and Eq.~\eqref{eq:iplus1state} into the matrix element in question, we find
\begin{equation}
    |\langle i|N|i + 1\rangle|^2 = (i+1) N_\mathrm{zpf}^2\left(1 - \frac{E_C}{\hbar\wq}(i+1)\right),
\end{equation}
which results in Eq.~\eqref{eq:me}.
\subsubsection{Final result for the dispersive shift}
Substituting Eq.~\eqref{eq:me} into Eq.~\eqref{eq:shift_intermediate_formula}, we obtain
\begin{equation}
\label{eq:shift_final_formula}
\hbar\chi	= -8E_{C}\frac{(\hbar g)^{2}(\hbar\omega_\mathrm{r,bare})^{2}}{\left[(\hbar\omega_\mathrm{r,bare})^{2} - E_{0\rightarrow1}^{2}\right]\left[(\hbar\omega_\mathrm{r,bare})^{2}-E_{1\rightarrow2}^{2}\right]}
\end{equation}
Neglecting the qubit energies $E_{0\rightarrow1}$ and $E_{1\rightarrow2}$ in the denominators and also neglecting the distinction between $\wres$ and $\omega_\mathrm{r,bare}$, we obtain a simplified expression for the dispersive shift in the high-frequency readout regime,
\begin{equation}
\label{eq:disp_simple}
\hbar\chi = - 8E_C \frac{g^2}{\wres^2},\quad \wres/\wq\gg1.
\end{equation}
We note that equation \eqref{eq:shift_final_formula} can be alternatively derived using the formalism of Ref.~\cite{ruskov_longitudinal_2024}. An equation analogous to Eq.~\eqref{eq:shift_final_formula} was also derived in Ref.~\cite{gely_nature_2018} via normal mode decomposition. In contrast to our approach, the approach of Ref.~\cite{gely_nature_2018} neglects the difference between $\wq = E_{0\rightarrow1}/\hbar$ and $E_{1\rightarrow2}/\hbar$.
 
As long as $\omega_\mathrm{r,bare} \gg \wq$, cavity pulls in higher transmon states can be approximated as $\chi_n = \chi_0 - n \chi$ for states within the Josephson well, $E_n < E_J$. Equation \eqref{eq:chi0} for the cavity pull $\chi_0$ also simplifies in the case of high-frequency readout. Namely, for $\omega_\mathrm{r,bare}\gg \wq$ we obtain
\begin{equation}
\label{eq:chi0simple}
    \chi_0 = 2\wq \frac{g^2}{\wres^2}.
\end{equation}
We note that in the validity regime of perturbation theory, $g\ll \wres$, cavity pull satisfies $\chi_0 \ll \wq \ll \wres$. For the parameters used in our experiment we obtain $\chi_0/2\pi = 4.7\:\mathrm{MHz}$. For this reason, in the main text and in Section~\ref{sec:purcell}, we do not draw a distinction between $\wres$ and $\omega_\mathrm{r,bare}$.
 
\subsubsection{Reproducing the result for dispersive shift in the small detuning limit}
We note that Eq.~\eqref{eq:shift_final_formula} also correctly describes the dispersive shift in the regime $|\omega_\mathrm{r,bare} - \wq|\ll \wq$ (in addition to the regime $\omega_\mathrm{r,bare}\gg \wq$). Indeed, in this regime the summation in Eq.~\eqref{eq:shift_general_formula} can also be restricted to $m = n\pm 1$ which was the only assumption in our derivation. To see that this is case, we note that in the small detuning regime we can approximate
\begin{gather}
\hbar\chi	=-8E_{C}\frac{(\hbar g)^{2}(\hbar\omega_\mathrm{r,bare})^{2}}{(E_{0\rightarrow1}-\hbar\omega_\mathrm{r,bare})(E_{1\rightarrow2}-\hbar\omega_\mathrm{r,bare})(E_{0\rightarrow1}+\hbar\omega_\mathrm{r,bare})(E_{1\rightarrow2}+\hbar\omega_\mathrm{r,bare})}\approx\notag\\\approx-8E_{C}\frac{(\hbar g)^{2}(\hbar\omega_\mathrm{r,bare})^{2}}{(E_{0\rightarrow1}-\hbar\omega_\mathrm{r,bare})(E_{1\rightarrow2}-\hbar\omega_\mathrm{r,bare})\cdot2\hbar\omega_\mathrm{r,bare}\cdot2\hbar\omega_\mathrm{r,bare}} = \notag\\
	=-2E_{C}\frac{(\hbar g)^{2}}{(\hbar\omega_\mathrm{r,bare} - E_{0\rightarrow1})(\hbar\omega_\mathrm{r,bare} - E_{1\rightarrow2})}=-2E_{C}\frac{g^{2}}{\Delta(\Delta-E_{C}/\hbar)}.\label{eq:koch-like}
\end{gather}
Here, we introduced the conventional notation $\Delta = \wq - \omega_\mathrm{r,bare}$. Eq.~\eqref{eq:koch-like} coincides with the established result of Ref.~\cite{koch_charge-insensitive_2007}.
 
\subsection{Comparison to the experiment}
\label{sec:compare_to_exp}
 
We experimentally measure the resonator frequencies for the qubit initialized in its different states. To this end, we send a short measurement pulse to the readout resonator and record the phase of the reflected signal. After collecting sufficient statistics, we change the frequency of the pulse and observe how the phase changes. As long as the transmon is kept in state $|n\rangle$, the phase winds by $2\pi$ around the shifted resonator frequency $\omega_{n} = \omega_{\mathrm{r,bare}} + \chi_n$ (where $\omega_{\mathrm{r,bare}}$ is the frequency of the resonator not dressed by its coupling to the qubit) as the probe frequency is changed. Finding the center of such a ``phase-roll'', we extract $\omega_n$. In practice, we do not actively keep the transmon in a given state. Rather, we rely on thermal fluctuations to populate the higher transmon states. As a result, with enough statistics, we can simultaneously observe phase-rolls corresponding to different transmon states and extract the resulting frequencies $\omega_n$.
\begin{figure*}
  \begin{center}
    \includegraphics[scale = 1]{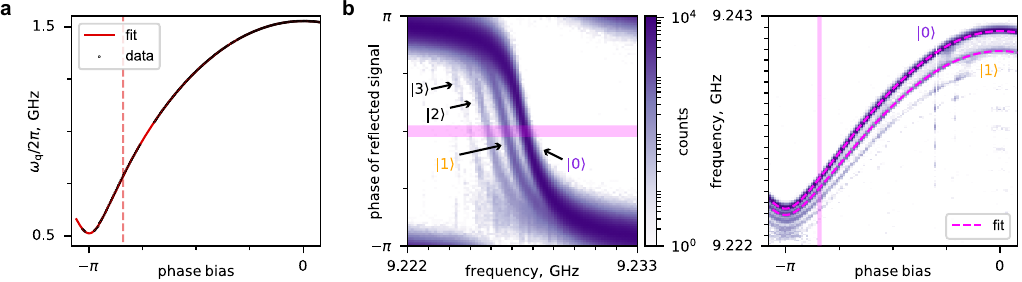}
    \caption{Qubit and resonator spectroscopy. \textbf{a}, Frequency of the $|0\rangle \rightarrow |1\rangle$ transition of our transmon as a function of SQUID phase bias. The red vertical line corresponds to the operation point considered in the main text of the manuscript. \textbf{b}, Single-tone spectroscopy of the readout resonator. Left panel: histogram of the phase of the signal reflected from the resonator as a function of signal frequency. Different phase-rolls correspond to different transmon states (measurement is performed in a single-shot readout regime). The middle of each phase-roll indicates the resonator frequency for the corresponding transmon state. Right panel: resulting dependence of dispersive shifts on the phase bias. The theory fit is produced using Eqs.~\eqref{eq:chi0} and \eqref{eq:shift_final_formula}. Purple vertical line -- operation point of the main text.}
    \label{fig:dispersive_shifts_exp}
  \end{center}
\end{figure*}
We repeat this experiment at different values of the transmon Josephson energy, which we control by threading the magnetic flux through the SQUID loop of our device. The resulting dependence of the resonator frequencies $\omega_n$ on flux bias is shown in Figure~\ref{fig:dispersive_shifts_exp}.
 
Theoretically, the dependence of frequencies $\omega_0$ and $\omega_1$ on the flux bias stems from two contributions. First, qubit frequency $\wq$ in Eqs.~\eqref{eq:shift_final_formula} and \eqref{eq:chi0simple} changes with flux. Second, the coupling strength $g$ changes with flux. Indeed, in terms of the circuit parameters of Figure~\ref{fig:purcell}, the coupling strength can be expressed as
\begin{equation}
\label{eq:g}
g = \frac{1}{2}\eta \sqrt{\wq \omega_\mathrm{r,bare}},
\end{equation}
where parameter $\eta$ is defined in Eq.~\eqref{eq:eta}. Therefore, the flux dependence of $g$ is also contained in the qubit frequency $\wq$.
We take both of the described contributions into the account when computing the frequencies $\omega_0$ and $\omega_1$.
 
We compare the observed flux bias dependence of $\omega_0$ and $\omega_1$ to that predicted by Eqs.~\eqref{eq:shift_final_formula} and \eqref{eq:chi0simple}. To this end, we simultaneously fit the measured resonator frequencies to $\omega_0 = \omega_\mathrm{r,bare} + \chi_0$ and $\omega_1 = \omega_\mathrm{r, bare} + \chi_1$ using $\omega_\mathrm{r, bare}$ and $\eta$ as fitting parameters. We find that the theory is in the excellent agreement with the data both for $\omega_0$ and for $\omega_1$. From this measurement we extract $\omega_\mathrm{r, bare}/2\pi = 9.2233\ghz$ and $\eta = 0.38$.
 
\subsection{Resonances in the dispersive shift}
\label{sec:disp_resonances}
\begin{figure*}
  \begin{center}
    \includegraphics[scale = 1]{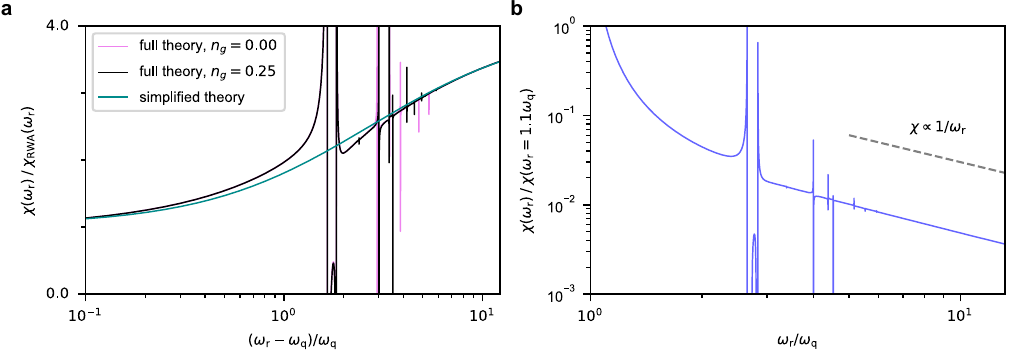}
    \caption{Dependence of the dispersive shift on the frequency of the readout resonator. Parameters used in the plots are $E_J = 2.2\ghz$, $E_C = 36\mhz$. \textbf{a}, Comparison between simplified expression for the dispersive shift, Eq.~\ref{eq:shift_final_formula}, and the full expression, Eq.~\ref{eq:shift_general_formula}. Black and magenta solid lines show the results of Eq.~\eqref{eq:shift_general_formula} for the offset charge $n_g=0.25$ and $n_g = 0.00$, respectively. The curve is normalized by the ``standard'' formula for the dispersive shift, Eq.~\eqref{eq:koch-like-2}, evaluated for the same values of parameters. Green solid line corresponds to a simplified expression for the dispersive shift, Eq.~\eqref{eq:shift_final_formula}. In the intermediate frequency range, $|\omega_\mathrm{r,bare}-\wq| \sim \wq$, the dispersive shift goes through a sequence of resonances.  More resonances will appear if we go to higher order in perturbation theory. As explained in the main text, the resonances become exponentially suppressed in the regime $\omega_\mathrm{r,bare}/\wq\gg 1$. \textbf{b}, At high frequencies of the readout resonator, dispersive shift falls off as $\chi \propto \wq / \wres$ (assuming that the coupling efficiency $\eta$ is fixed). The plot is obtained by numerically evaluating Eq.~\eqref{eq:shift_general_formula}.}
    \label{fig:death_zone}
  \end{center}
\end{figure*}

Next, we analyze how well Eq.~\eqref{eq:shift_final_formula} describes the dispersive shift for an arbitrary resonator frequency. To this end, we compare Eq.~\eqref{eq:shift_final_formula} to the full expression Eq.~\eqref{eq:shift_general_formula}, in which we do not restrict the summation to terms with $n=m\pm 1$. The result of this comparison is presented in Figure~\ref{fig:death_zone}a, where we plot the dispersive shift as a function of the detuning between the readout resonator and the transmon. For visualization purposes, we normalize the dispersive shift by the conventional formula of Ref.~\cite{koch_charge-insensitive_2007} derived for the case $|\omega_\mathrm{r,bare}-\wq|\ll\wq$,
\begin{equation}
\label{eq:koch-like-2}
\hbar\chi_\mathrm{RWA}=-2E_C \frac{g^2}{\Delta(\Delta-E_C / \hbar)},\quad \Delta = \wq - \omega_\mathrm{r,bare}.
\end{equation}
 
From the figure, it is apparent that Eq.~\eqref{eq:shift_final_formula} correctly captures the dispersive shift both in the small-detuning regime, $|\omega_\mathrm{r,bare} - \wq|\ll \wq$, and in the high-frequency readout regime\footnote{Here, we tacitly assume that the coupling between the resonator and the transmon is weak enough such that the perturbative treatment in the derivation of the dispersive shift is valid.}, $\omega_\mathrm{r,bare}/\wq\gg1$. However, in the intermediate frequency regime, $|\omega_\mathrm{r,bare} - \wq| \gtrsim \wq$, the approximate equation \eqref{eq:shift_final_formula} no longer holds. In this regime the summation in Eq.~\eqref{eq:shift_general_formula} cannot be truncated to terms with $n=m\pm1$ since transitions to higher transmon states might come into resonance with the readout frequency. The dispersive shift goes through a sequence of divergences as the readout frequency is increased. These resonances might lead to undesired readout-induced state transitions. Thus, we expect the performance of dispersive readout to be poor in the regime $|\omega_\mathrm{r,bare} - \wq| \gtrsim \wq$. Only when $\omega_\mathrm{r,bare}/\wq \gg 1$ do the unwanted resonances vanish exponentially, leading to the improved performance of the dispersive readout demonstrated in the main text.
 
As is also evident from the figure, the standard result of Ref.~\cite{koch_charge-insensitive_2007} correctly predicts the dispersive shift when $|\wq - \omega_\mathrm{r,bare}|\ll\wq$. It fails, however, to capture the behavior of the dispersive shift outside of this regime. At high readout frequencies, it is essential to take into the account the counter-rotating terms in the coupling between the transmon and the resonator; these terms are neglected in Eq.~\eqref{eq:koch-like-2}. When $\omega_\mathrm{r,bare}\gg\wq$, counter-rotating terms enhance the dispersive shift roughly by a factor of four compared to the small detuning regime (for the same ratio between the coupling strength and detuning).
 
\subsection{Power-law decay of dispersive shift with detuning}
\label{sec:chi_power_law}
 
Finally, we discuss the asymptotic behavior of the dispersive shift $\chi$ when the resonator frequency is high, $\wres \gg \wq$. To determine this behavior, one needs to substitute the equation for the coupling strength $g$, Eq.~\eqref{eq:g}, into Eq.~\eqref{eq:disp_simple} for the dispersive shift. As a result we get
\begin{equation}
    \hbar\chi = - 2 E_C \eta^2\frac{\wq}{\wres}.
\end{equation}
Therefore, as long as the coupling efficiency $\eta$ stays fixed, the dispersive shift scales as $\chi\propto \wq/\wres$. In Figure~\ref{fig:death_zone}b, we explicitly demonstrate this behavior by numerically evaluating $\chi$ using Eq.~\eqref{eq:shift_general_formula}.

\section{Matrix elements of the unwanted resonances}
\label{sec:me_analytics}
 
In this section, we give an analytical estimate for the matrix element $\langle f|N|i\rangle$ controlling the strength of the unwanted resonances in the case where the final state $|f\rangle$ lies outside of the Josephson well. Additionally, we numerically show that the matrix element remains small even when the final state lies within the Josephson well.
 
\subsection{Estimate of the matrix element}
 
In the regime relevant for our experiment, $\hbar\wres > 2E_J$, the final state $|f\rangle$ of the resonant transition lies above the Josephson well. As explained in the main text, the matrix element for the transition from either of the computational states to $|f\rangle$ is exponentially suppressed. In this section, we provide an intuitive derivation of this suppression, obtaining the exponent for the matrix element decay. We omit the offset charge as it is irrelevant for finding the exponent.
 
We begin by noting that for a state far outside of the well the charging term of Hamiltonian~\eqref{eq:ham} dominates, so that $|f\rangle$ is close to an eigenstate of the Cooper pair number operator with charge $N_\mathrm{f}$,
\begin{equation}
\label{eq:Nres}
     4E_C N_\mathrm{f}^2 \approx \hbar\wres - E_J,\qquad N_\mathrm{f} \approx \sqrt{\frac{\hbar\wres - E_J}{4E_C}}.
\end{equation}
Nevertheless, due to the Josephson coupling, state $|f\rangle$ contains charge components besides $|N_\mathrm{f}\rangle$. This includes components in the vicinity of $N=0$, where the computational states are localized. It is these small $N$ components that are responsible for the transition between computational and non-computational states. Thus, to crudely estimate the exponent for the decay of the matrix element with the increase of $f$, with logarithmic precision we can approximate
\begin{equation}
    \ln|\langle f|N|i\rangle| \sim \ln|\langle N=0|f\rangle|.
\end{equation}
We estimate the resulting overlap using perturbation theory in the Josephson coupling. In the charge basis, the Josephson term acts as a hopping between neighboring charge states. Reaching zero charge from $|f\rangle$ requires $N_\mathrm{f}$ such hops. Therefore, for the transition amplitude, we find
\begin{equation}
\label{eq:psi_charge}
    \ln|\langle N=0|f\rangle| \approx \sum_{k=0}^{N_\mathrm{f} - 1}\ln\left(\frac{E_J/2}{4E_C (N_\mathrm{f}^2 - k^2)}\right) \approx -N_\mathrm{f}\left[\ln\left(\frac{8}{e^2}\frac{4 E_C N_\mathrm{f}^2}{E_J}\right)\right].
\end{equation}
Here we used the Stirling formula for the factorials resulting from the sum over $k$. As is apparent from Eq.~\eqref{eq:psi_charge}, the matrix element decays faster than exponentially for states outside of the Josephson well. We note that while Eq.~\eqref{eq:psi_charge} correctly predicts the exponent, it is not suitable for correctly capturing the pre-factor in front of the exponent.
 
\subsection{Suppression of the matrix element inside of the Josephson well}

\begin{figure*}
  \begin{center}
    \includegraphics[scale = 1]{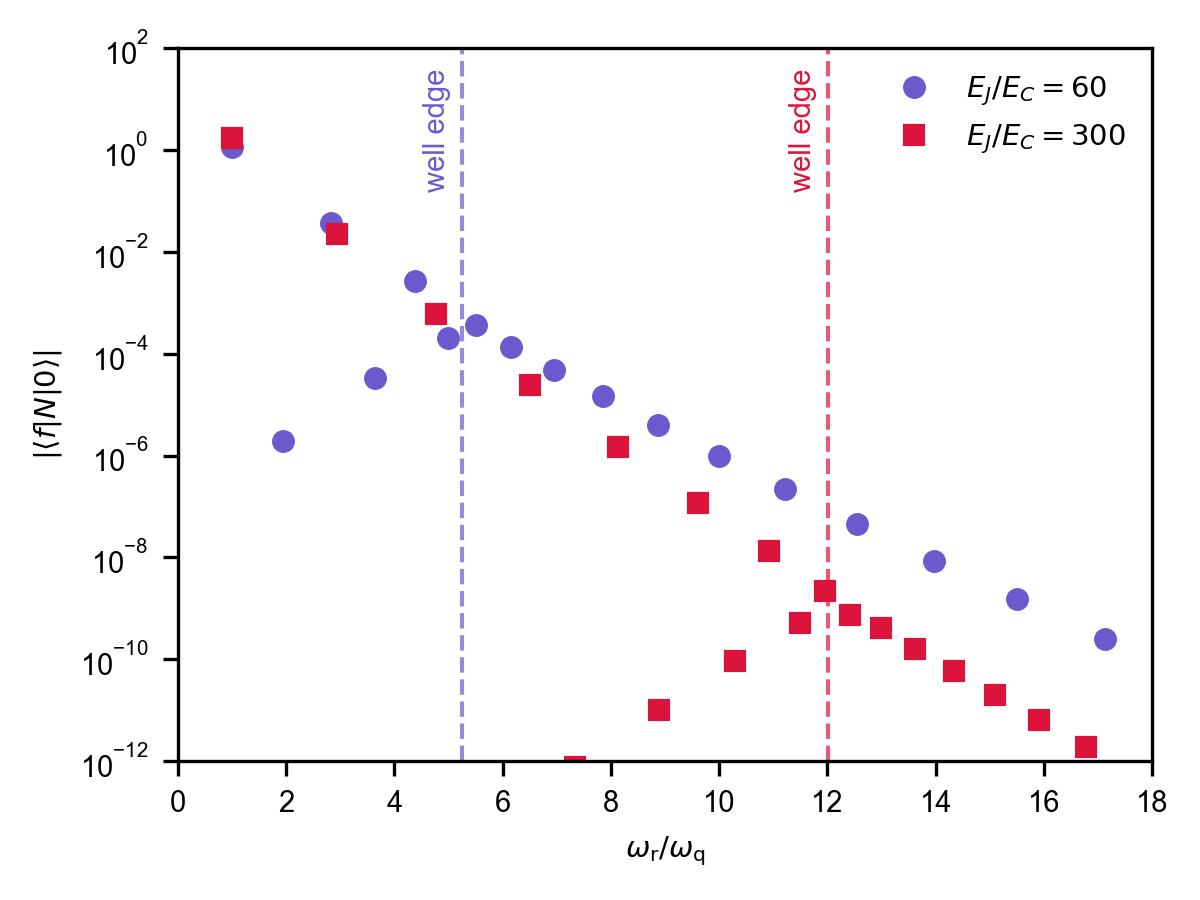}
    \caption{Matrix elements $|\langle f|N|i\rangle|$ corresponding to a single-photon unwanted resonance as a function of $\wres/\wq$, for the parameters of our device ($E_J/E_C = 60$) and for $E_J/E_C = 300$. Vertical lines mark the frequency at which the resonant final state crosses the edge of the Josephson well in each case. $n_g = 1/4$.}
    \label{fig:me_ejec}
  \end{center}
\end{figure*}

Next, we demonstrate that having the final state $|f\rangle$ lie outside of the Josephson well is not essential to achieve the matrix element suppression. To this end, we repeat the numerical evaluation of the matrix elements for a transmon with $E_J/E_C = 300$. For these parameters, the Josephson well accommodates more levels and its edge is only reached at $\wres/\wq \approx 12$. The result for the matrix elements is shown in Figure~\ref{fig:me_ejec}. At a fixed $\wres/\wq$, the matrix elements are \textit{smaller} than for the parameters of our device, even though the resonant final state remains inside the well. The price of increasing $E_J/E_C$ is the reduction of the anharmonicity at fixed $\wq$.

\section{Leakage probability due to the unwanted resonances}
\label{sec:leakage_lz}
 
In this section, we use the matrix elements for unwanted resonance to estimate the leakage probability per readout. This allows us to compare the readout performance across different readout frequencies. Our estimate of the leakage probability is derived using the Landau-Zener formula for traversing an anti-crossing.
 
To estimate the leakage probability, we note that during the readout pulse, the photon number in the resonator changes. Therefore, due to the AC Stark shift, the energies of the transmon levels sweep a range of frequencies. In an unwanted resonance, the frequency of a transition between a computational and a non-computational state dynamically passes through the resonator frequency. The detuning from the resonance $\delta$ follows
\begin{equation}
    \hbar\delta(t) = E_{i\rightarrow f}\big(n(t)\big) - \hbar\wres,\quad \frac{d\delta}{dt} = \frac{1}{\hbar}\frac{\partial E_{i\rightarrow f}}{\partial n}\,\dot{n},
\end{equation}
where the dependence of transition energy on the resonator photon number $n$ is determined by the AC Stark shift. The leakage probability per traversal is given by the Landau-Zener formula,
\begin{equation}
\label{eq:lz}
    P_{i\rightarrow f} = 1 - \exp\left(-\frac{2\pi|\Omega_{i\rightarrow f}|^2}{|d\delta/dt|}\right)\approx \frac{2\pi|\Omega_{i\rightarrow f}|^2}{|d\delta/dt|},
\end{equation}
with the transition amplitude $\hbar\Omega_{i\rightarrow f} = \hbar g \sqrt{n}\,\langle f|N|i\rangle/\nzpf$.
 
Regardless of the particular pulse shape, the pulse fills the resonator with photons on the timescale of its linewidth. Therefore, for the order-of-magnitude estimate of the leakage probability we replace $\dot{\nbar}\sim\kappa\nbar$. Both the numerator and the denominator of Eq.~\eqref{eq:lz} are then proportional to the photon number $n$, so that it drops out:
\begin{equation}
\label{eq:lz_intermediate}
    P_{i\rightarrow f} \sim 2\pi\,\frac{\hbar g^2\, |\langle f|N|i\rangle|^2}{\nzpf^2\,\kappa\,|\partial E_{i\rightarrow f}/\partial n|}.
\end{equation}
Using $\hbar\chi = -8E_C g^2/\wres^2$ together with the Stark shift of the transition ($|\partial E_{i\rightarrow f}/\partial n| \approx 2 E_J |\chi| / \omega_\mathrm{q}$), this reduces to
\begin{equation}
\label{eq:lz_final}
    P_{i\rightarrow f} \simeq \frac{2\pi\hbar\wres^2\,|\langle f|N|i\rangle|^2}{E_J\,\kappa}.
\end{equation}
The comparison across readout frequencies is thus naturally made in terms of the leakage probability rather than a maximal allowed photon number. The dephasing of the computational state does not affect this result: the transition probability integrated across the resonance is set by the swept area and is independent of the broadening rate.
 
\begin{figure*}
  \begin{center}
    \includegraphics[scale = 1]{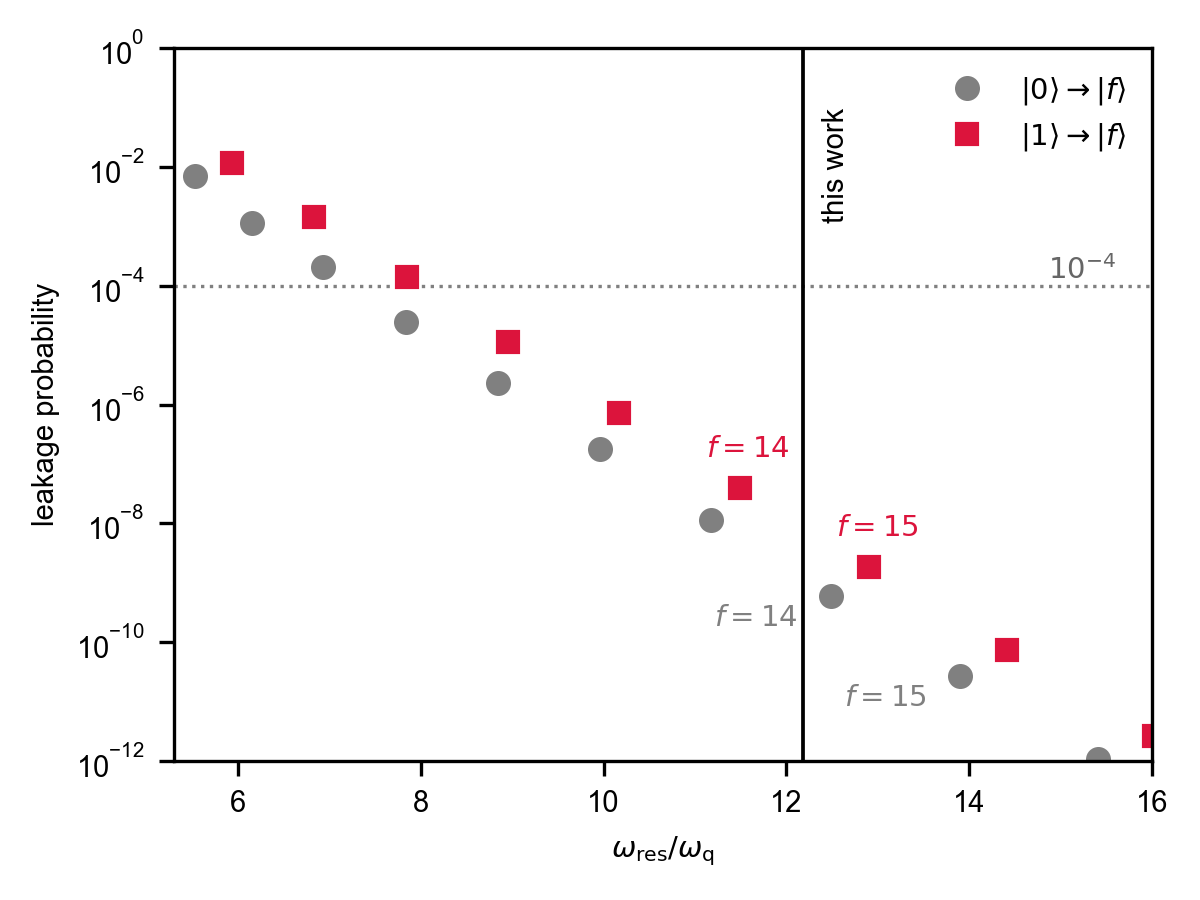}
    \caption{Leakage probability per readout due to the unwanted resonances, given by probability of two traversals evaluated in Eq.~\eqref{eq:lz_final}, as a function of $\wres/\wq$, for the transitions out of $|0\rangle$ and out of $|1\rangle$. Transmon parameters and $\kappa$ are those of our device; $n_g = 1/4$. The vertical line marks the operation point of the main text. The horizontal dotted line marks the probability of $10^{-4}$.}
    \label{fig:leakage_lz}
  \end{center}
\end{figure*}
 
In Figure~\ref{fig:leakage_lz}, we plot the probability of leakage out of both computational states, holding the transmon parameters and the resonator linewidth at the values of our device. We account for two traversals of the resonance per pulse and add the probabilities incoherently (assuming that the measurement-induced dephasing is sufficiently strong). From the figure it is apparent that a ratio $\wres/\wq \approx 9$ is sufficient to constrain the leakage probability to $\lesssim 10^{-5}$. Reducing the ratio below this value is undesirable: leakage to the states near and above the well edge is particularly damaging, since the return to the computational subspace proceeds through a long cascade of transitions with small matrix elements.

\section{Quantum non-demolition fidelity metric for readout characterization}
In this section, we elaborate on the definition of quantum non-demolition (QND) fidelity used in the main text of the manuscript. Notably, this metric of readout performance is different from other commonly used metrics: readout fidelity and ``repeatability''. As we elaborate on below, the latter metrics do not faithfully account for leakage errors, i.e., transmon excitation to non-computational states \cite{hazra_benchmarking_2025}. They are thus not suitable for describing the readout performance for applications sensitive to leakage such as quantum error correction.
 
\subsection{Definition of QND fidelity}
We begin by rigorously defining the error probabilities $\varepsilon_\mathrm{sep}$ and $\varepsilon_\mathrm{trans}$ discussed in the main text. To this end, we introduce the quantity $P(i_s, j_m | k_s)$ where $k_s$ to the right of the separating line stands for the prepared state $|k_s\rangle$; to the left of the separating line, $j_m$ stands for the measurement result and $i_s$ stands for the post-readout state $|i_s\rangle$. As such $P(i_s, j_m | k_s)$ is the probability that the measurement outcome is $j_m$ and that the post-readout state is $|i_s\rangle$ conditioned on system being prepared in state $|k_s\rangle$ before the readout.

We introduce the QND fidelity as
\begin{equation}
    \label{eq:qnd_fid}
    \mathcal{Q} = \frac{1}{2}\left[P(1_s,1_m|1_s) + P(0_s, 0_m|0_s)\right].
\end{equation}
The conditional probabilities in this definition can be decomposed as
\begin{equation}
\label{eq:two_sums}
P(i_s,i_m|i_s) = 1 - \sum_{j_s\neq i_s} P(j_s|i_s)-  \sum_{j_m\neq i_s} P(i_s, j_m|i_s).
\end{equation}
Here, $P(j_s|i_s)$ is the probability that the state after the readout is $|j_s\rangle$ given that the state before the readout is $|i_s\rangle$ (regardless of measurement outcome). Therefore, the first sum in Eq.~\eqref{eq:two_sums} describes the total probability of a \textit{transition} error. In the same equation, $P(i_s,j_m|i_s)$ with $j_m\neq i_s$ is the probability that the readout did not change the state but the state was assigned erroneously. The second sum in Eq.~\eqref{eq:two_sums} thus describes the total probability of a \textit{separation} error. We define $\varepsilon_\mathrm{trans}$ and $\varepsilon_\mathrm{sep}$ as probabilities of transition and separation errors, respectively, averaged over the computational states,
\begin{equation}
\varepsilon_\mathrm{trans} = \frac{1}{2} \left[ \sum_{j_s\neq 0_s} P(j_s|0_s) + \sum_{j_s\neq 1_s} P(j_s|1_s)\right]\quad \varepsilon_\mathrm{sep} = \frac{1}{2} \left[ \sum_{j_m \neq 0_s} P(0_s, j_m|0_s) + \sum_{j_m \neq 1_s} P(1_s,j_m|1_s)\right].
\end{equation}
We can further break down the transition error probability into the bit flip and leakage probabilities, $\varepsilon_\mathrm{trans} = \varepsilon_\mathrm{trans}^{\mathrm{bit-flip}} + \varepsilon_\mathrm{trans}^{\mathrm{leakage}}$, where
\begin{equation}
\varepsilon_\mathrm{trans}^{\mathrm{bit-flip}} = \frac{1}{2} \left[ P(1_s|0_s) + P(0_s|1_s)\right], \quad   \varepsilon_\mathrm{trans}^{\mathrm{leakage}} = \frac{1}{2} \sum_{j_s>1} \left[ P(j_s|0_s) +  P(j_s|1_s)\right].
\end{equation}
\subsection{Comparison of QND fidelity to readout fidelity and repeatability}
A more conventional definition of the readout fidelity is
\begin{equation}
    \label{eq:ro_fid}
    \mathcal{F} = \frac{1}{2}\left[P(0_m|0_s) + P(1_m|1_s)\right],
\end{equation}
where $P(i_m|i_s)$ is the probability of the readout outcome to be $i_m$ if system was prepared in state $|i_s\rangle$ before the readout. This metric, however, does not adequately account for the possibility of leakage. Assume for example, that the excited transmon state $|1\rangle$ can leak to a non-computational state $|j\rangle$ with a large probability. Assume also that the state $|j\rangle$ always mistakenly yields the readout outcome $1_m$. In this case, the readout fidelity, defined by Eq.~\eqref{eq:ro_fid} will not reflect the presence of a leakage error; the fidelity can be close to unity even when the probability of leakage is large. This situation was in fact observed in our recent experiment Ref.~\cite{hazra_benchmarking_2025}. Clearly, the QND fidelity defined in Eq.~\eqref{eq:qnd_fid} is more suitable for quantifying the measurement performance in the context of quantum error correction compared to the readout fidelity defined in Eq.~\eqref{eq:ro_fid}.
 
We note that the QND fidelity can never exceed the readout fidelity, $\mathcal{Q} \leq \mathcal{F}$. This can be seen by noting that
\begin{equation}
    P(i_m|i_s) = P(i_s,i_m|i_s) + \sum_{j_s\neq i_s} P(j_s,i_m|i_s) \geq P(i_s,i_m|i_s)
\end{equation}
and using the definitions given by Eqs.~\eqref{eq:qnd_fid} and \eqref{eq:ro_fid}.
 
To define readout repeatability -- which is a conventional measure of the QND character of readout -- 
we consider a sequence of two back-to-back measurements. We then introduce $P(i_{m2}|j_{m1},k_s)$ as the probability of the second readout yielding outcome $i_{m2}$ given that the outcome of the first readout is $j_{m1}$ and that the system is prepared in state $|k_s\rangle$. The repeatability is then defined as
\begin{equation}
    \label{eq:repeat}
    \mathcal{R} = \frac{1}{2}\left[P(0_{m2}|0_{m1},0_s) + P(1_{m2}|1_{m1},1_s)\right].
\end{equation}
Similarly to the readout fidelity, however, this metric might overlook the leakage as leakage states can be mistakenly assigned to 0 or 1. 
For applications sensitive to leakage errors, repeatability should thus be replaced with the QND fidelity defined in Eq.~\eqref{eq:qnd_fid}. Generally, the QND fidelity does not exceed repeatability, $\mathcal{Q} \leq \mathcal{R}$.
 
Several recent works \cite{touzard_gated_2019, sunada_fast_2022, spring_fast_2025} name the repeatability $\mathcal{R}$, as defined in Eq.~\eqref{eq:repeat}, the ``QNDness'' or ``QND fidelity''. We avoid this label for $\mathcal{R}$ and emphasize that it differs from $\mathcal{Q}$, which we call the QND fidelity. In the absence of leakage, $\mathcal{Q}$ and $\mathcal{R}$ coincide up to corrections quadratic in error probabilities. However, leakage, which is fatal for quantum error correction, might not be detected by $\mathcal{R}$, while it is adequately captured by $\mathcal{Q}$. Therefore, we use QND fidelity defined in Eq.~\eqref{eq:qnd_fid} in our work.
 
\section{Details of readout pulses, amplification, and QND fidelity measurement}
 
In this section, we elaborate on salient features of our measurement. First, we describe the parametric amplifier used in our experiment. Second, we describe the readout pulses that we employ. Third, we comment on important nuances of our QND fidelity measurement. In particular, we explain the technique that we employ to quantify the leakage to non-computational states caused by our pulses. We also demonstrate that QND fidelity that we measure remains stable on the timescale of at least 12 hours.
 
\subsection{Amplification}
In our measurement, we use a near-quantum-limited SNAIL parametric amplifier \cite{frattini_optimizing_2018} as a first stage of amplification, see Fig.~\ref{fig:wiring}. This allows us to make the measurement robust to thermal noise that is inevitably added to the signal as it leaves the dilution refrigerator. The operation of the particular amplifier that we use is described in detail in Ref.~\cite{joshi_lumped-element_2025}. 
 
SNAIL parametric amplifier can be operated either in the phase-preserving mode or in the phase-sensitive mode. In the phase-preserving amplification mode, both signal quadratures are amplified equally at the expense of adding at least half of a photon of noise to the measurement. The phase-sensitive mode avoids the added noise but only amplifies one of the quadratures. For our readout pulses designed to distinguish $|0\rangle$ and $|1\rangle$, suppressing the noise is paramount. For this reason, we use phase-sensitive amplification mode. As explained below, information about both quadratures is required to detect leakage to non-computational states. For this reason, we use phase-preserving amplification mode in our leakage check measurements.
 
\subsubsection{Amplifier parameters and measurement efficiency}
 
In Ref.~\cite{joshi_lumped-element_2025}, the amplifier is characterized at a gain of $20\:\mathrm{dB}$, where it has an instantaneous bandwidth of $250\mhz$ and adds $0.61$ photons of noise in the phase-preserving mode. In the present measurement we operate it at $26\:\mathrm{dB}$; the bandwidth is correspondingly reduced but remains in excess of the bandwidth of our readout pulses.
 
From the dependence of the separation error on the readout power we extract a measurement efficiency $\eta_m = 0.45$. This value is consistent with an independent estimate based on the known properties of the output chain, given in Table~\ref{tab:eff}. The estimate accounts for the insertion losses of the infrared filter, the circulator, and the diplexer; for the added noise of the parametric amplifier; for the noise of the following HEMT amplifier; and for the loss of the cabling, the connectors, and the sample-holder printed circuit board between the device and the amplifier. The insertion losses are room-temperature specifications, so we quote the estimate as a range, $\eta_m = 0.4$--$0.6$.
 
\begin{table}[h]
\begin{tabular}{l c}
\hline\hline
Contribution & Factor in $\eta_m$ \\
\hline
HERD-1 infrared filter ($0.15$ dB) & $0.97$\\
Circulator ($0.2$ dB) & $0.95$ \\
Diplexer ($1.2$ dB) & $0.76$ \\
Added noise of the SPA ($0.61$ photons) & $0.90$ \\
HEMT noise referred to the input & $0.94$ \\
Cabling, PCB, and connectors ($0.7$ dB) & $0.85$ \\
\hline
Total & $0.5$ \\
Measured & $0.45$\\
\hline\hline
\end{tabular}
\caption{{Budget of the measurement efficiency. The HEMT contribution is evaluated with the effective gain of the first stage, i.e. the amplifier gain reduced by the losses following it; the amplifier operates in reflection, so the diplexer is traversed twice and enters this reduction a second time.}}
\label{tab:eff}
\end{table}
 
\begin{figure*}
  \begin{center}
    \includegraphics[scale = 1]{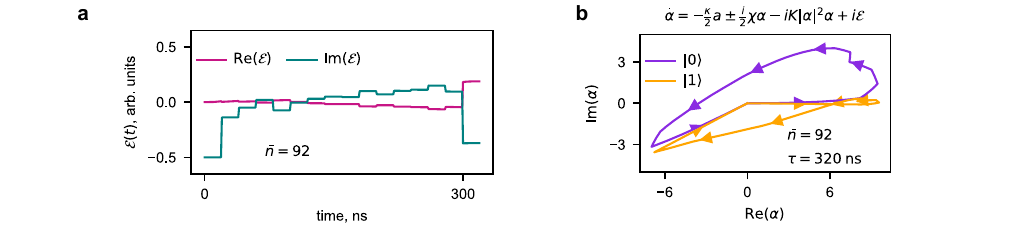}
    \caption{We optimize our shaped readout pulses to maximize signal-to-noise ratio per readout resonator photon. We require the pulses to leave no photons in the resonator after the readout. \textbf{a}, Amplitude of the drive acting on the resonator for our optimal readout pulse. The frequency of the pulse is in the middle between resonator frequencies corresponding to transmon states $|0\rangle$ and $|1\rangle$. $\bar{n}$ refers to the maximal photon number during the pulse. \textbf{b}, Numerically simulated resonator trajectories for the optimal pulse. Two curves correspond to transmon initialized in different computational states. The resonator field returns to origin at the end of the pulse for both transmon states.}
    \label{fig:pulses}
  \end{center}
\end{figure*}

\subsection{Pulse shaping}
 
Our readout pulses are numerically optimized to maximize the signal-to-noise ratio for a fixed average photon number in the resonator. When performing the optimization, we add a constraint that the pulses should leave no photons in the readout resonator when the measurement is completed. This is necessary because any remaining photons would dephase the qubit preventing its operation after the readout.

\subsubsection{Pulse optimization method}
 
Our pulse shaping technique is similar to the one previously reported in Refs.~\cite{mcclure_rapid_2016, hazra_benchmarking_2025}. We choose the frequency of our pulses to be in the middle between resonator frequencies for transmon states $|0\rangle$ and $|1\rangle$. We choose the amplitude of the drive acting on the resonator, $\mathcal{E}(t)$, to be piecewise constant. In contrast to previous works, we use a larger number of segments in our pulses; the segment duration that we choose is 20 ns. We numerically optimize the drive amplitude in each segment to improve the overall performance of the pulse while leaving the resonator empty of photons at the end of the pulse. The resulting amplitude profile is shown in Fig.~\ref{fig:pulses}.
 
We note that in our pulse optimization it is important to account for the residual non-linearity of the readout resonator. This nonlinearity comes from the coupling between the resonator and the transmon. It can be estimated as
\begin{equation}
    \hbar K = \hbar\frac{d\wres}{d\bar{n}} = -\frac{\hbar^2\chi^2}{\ 4E_C}.
\end{equation}
For the parameters used in the main text, this yields $K/h = -0.0056\:\mathrm{MHz}$. Therefore, for the photon numbers that we employ, $\bar{n} \sim 100$, the non-linear resonator frequency shift may be comparable to the resonator linewidth $\kappa$ and dispersive shift $\chi$.

\subsubsection{Measurement of residual photon number}
\begin{figure*}
  \begin{center}
    \includegraphics[scale = 1]{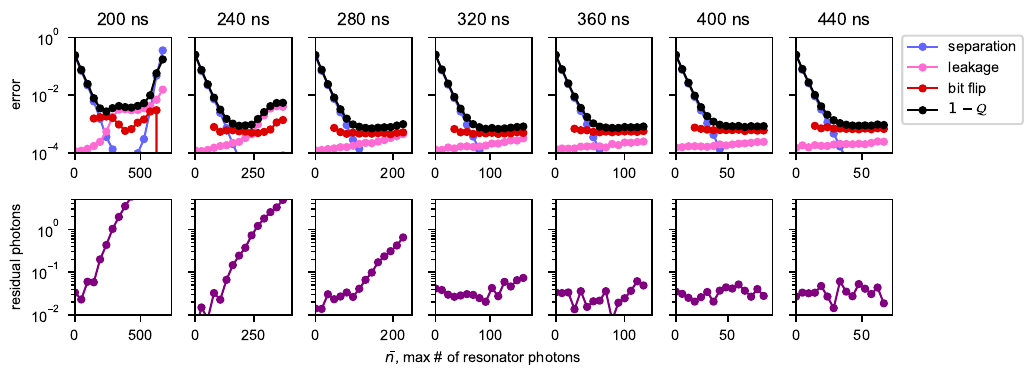}
    \caption{QND fidelity and residual photon number as a function of readout power and duration.}
    \label{fig:residual_nbar}
  \end{center}
\end{figure*}
 
To verify that our readout pulse empties the resonator after the measurement, we perform a ring-down experiment. In this experiment, we play a single readout pulse and continue to acquire the demodulated signal after the readout pulse is over. Any photons remaining in the resonator at the end of the pulse will leak out on a timescale of $1/\kappa \approx 88\:\mathrm{ns}$. We can determine the number of photons remaining in the resonator based on the amplitude of the acquired signal after the pulse ends. To do this, we must calibrate the amplitude of the amplified room-temperature signal to the amplitude of the coherent state inside the resonator. To this end, we first play a continuous-wave tone at the cavity frequency and measure the AC Stark shift $\delta\omega$ of the transmon $|0\rangle$-$|1\rangle$ transition. In conjunction with the known dispersive shift $\chi$, this tells us the amplitude of the cavity field: $|\alpha|^2 = \delta \omega / \chi$. Once the cavity has reached steady-state, we turn off the drive and acquire the resonator ring-down. We can then convert the known $|\alpha|$ to the amplitude of the signal at our room-temperature digitizer.
 
The performance of our numerically optimized pulses with different durations is demonstrated in Fig.~\ref{fig:residual_nbar}. From the figure it is apparent that the pulses are effective at emptying the resonator from photons at the end of the pulse. Only for the shortest pulses with large powers, $\bar{n} \gtrsim 170$, does the residual photon number exceed $0.1$. This error could be due to our uncertainty on the resonator parameters $\chi$, $\kappa$, and nonlinearity $K$, or it could be due to higher-order nonlinearity.

\subsection{Leakage checks}
In the main text of the manuscript, we presented a protocol for measuring QND fidelity by separately quantifying the probability of qubit transition errors and measurement separation errors. In this protocol, we repeatedly measure the qubit multiple times in a row. When the assignment changes from consistently $|0\rangle$ to consistently $|1\rangle$ or vice versa, we conclude a transition error occurred.
When one measurement in the middle of the sequence disagrees with the measurements before and after, we identify a separation error. Counting the number of errors, we estimate their probabilities.
 
There is, however, an important complication in the described protocol due to the possibility of leakage to non-computational transmon states. Our readout is only effective at distinguishing states $|0\rangle$ and $|1\rangle$ from each other; non-computational states are mistakenly assigned to either $|0\rangle$ or $|1\rangle$. Because of this, the protocol described above does not correctly quantify the probability of transition errors to leakage states. Moreover, leakage corrupts further measurements in the sequence. While the outcomes of corrupted measurements are still assigned to $|0\rangle$ or $|1\rangle$, the statistics of these outcomes do not contain information about the discrimination of computational states.
 
In order to remedy this problem, we interleave repeated qubit measurement sequences with leakage checks. These checks can distinguish the non-computational states from the computational ones, but are not effective at distinguishing $|0\rangle$ from $|1\rangle$. The leakage checks allow us to both quantify the transition probability to non-computational states and to discard the sequences where the transmon starts from a non-computational state when computing the QND fidelity.
 
The leakage checks differ from computational state measurements in three aspects. First, the leakage checks are performed in the phase-preserving mode of our quantum limited amplifier. Both quadratures of the signal contain information about leakage. In contrast, computational state measurements are performed in phase-sensitive amplification mode. Second, the leakage checks are performed with an increased measurement duration. This allows us to suppress the separation errors and also distinguish different non-computational states from each other. Finally, the leakage checks are performed at a lower readout frequency. This improves the contrast between the computational and non-computational states at the cost of sacrificing the contrast between the computational states.
 
The example histograms of the leakage check measurements are shown in Fig.~\ref{fig:leak_blobs}a,b. It is clear from the figure that the checks distinguish the computational and non-computational states. They moreover distinguish non-computational states $|2\rangle$, $|3\rangle$ and $|4+\rangle$ (where $|4+\rangle$ means $|4\rangle$ or any of the higher states). This resolution allows us to be confident that the leakage transitions that we see in our experiment are unrelated to spurious resonances in the transmon spectrum. Indeed, activation of such resonances would excite the transmon to states $|4+\rangle$. As demonstrated in Fig.~\ref{fig:leak_blobs}c, the leakage observed in our experiment predominantly leaves the transmon in states $|2\rangle$ and $|3\rangle$.
 
\subsection{Stability of the QND fidelity with time}
 
\begin{figure*}
  \begin{center}
    \includegraphics[scale = 1]{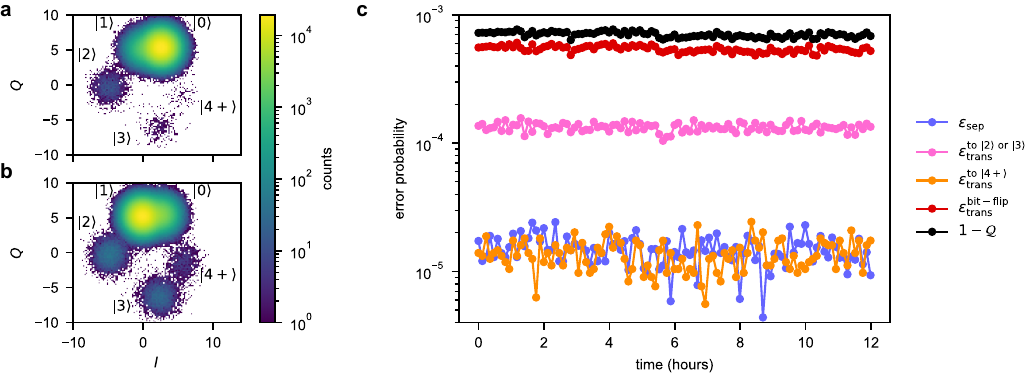}
    \caption{\textbf{a},\textbf{b}, Histogram of leakage check measurement outcomes after the sequence of 18 qubit measurements described in Fig.~3 of the main text. Only shots where qubit began in the computational subspace are shown. In panel \textbf{a}, the sequences which are not preceded by a $\pi$-pulse are shown. In panel \textbf{b}, sequences which are preceded by a $\pi$-pulse are shown. We identify blobs corresponding to transmon states $|0\rangle$ to $|3\rangle$, and a separate blob where states $|4\rangle$ and higher reside. \textbf{c}, Measurement of QND fidelity over the course of about 12 hours. The measurement is performed using the optimized $320\:\mathrm{ns}$-long pulse discussed in the main text. The QND fidelity remains stable, with most fluctuation arising from changes in the bit-flip rate. This suggests that our leakage errors are not sensitive to offset charge, which typically fluctuates on a timescale of minutes. Most leakage takes the transmon to states $|2\rangle$ and $|3\rangle$. This measurement only places an upper bound on the leakage rate to highly-excited states, since the population in $|4+\rangle$ may be due to a sequence of two excitations over the course of the 18 qubit measurements.}
    \label{fig:leak_blobs}
  \end{center}
\end{figure*}

Fig.~\ref{fig:leak_blobs}c also demonstrates that the QND fidelity that we measure is stable for at least $12$ hours. This further confirms that our readout is not limited by the unwanted resonances. Indeed, we expect the unwanted resonances to be strongly sensitive to the offset charge $n_g$, see Fig.~\ref{fig:floquet1}b. The offset charge was previously observed to fluctuate on the scale of minutes \cite{riste_millisecond_2013, serniak_hot_2018}. We would thus see significant variation of the QND fidelity in our 12-hour sweep if unwanted resonances were relevant for the readout performance. In contrast, we see minimal variation with time that can be explained by the drift of the dielectric environment of the qubit.
 
\section{Details of Floquet simulations}
 
In this section, we explain the numerical procedure we use to find drive frequencies and powers at which unwanted resonances occur. In these simulations, we model the readout signal as a classical monochromatic drive applied to the transmon, following Refs.~\cite{shillito_dynamics_2022, xiao_diagrammatic_2025, dumas_measurement-induced_2024}. The resonances can then be identified as regions in the parameter space where the considered computational state appreciably hybridizes with non-computational states due to the presence of the drive. The information about this hybridization is contained in the Floquet modes of the model. To quantify the hybridization, we introduce the quantity
\begin{equation}
\label{eq:hybridization}
    \Theta = 1 - |\langle c | \psi\rangle|^2,
\end{equation}
where $|c\rangle$ is the Stark-shifted computational state and $|\psi\rangle$ is the Floquet mode which has the highest overlap with $|c\rangle$. Away from unwanted resonances $|\psi\rangle\approx |c\rangle$ and $\Theta$ is close to zero. In contrast, exactly on resonance, the Floquet mode becomes an equal superposition between the computational state and a non-computational state leading to $\Theta = 1/2$.
 
Below, we present the full details of our numerical approach. First, we explain how we find the Floquet modes by diagonalizing the propagator over one drive period. Then, to build intuition for this approach, we find the Floquet modes in a simple driven two-level-system model. In this case we can explicitly compute the hybridization $\Theta$. We then detail how our technique applies to the case of a transmon qubit. An additional complication in this case (as compared to the two-level system model) comes from the presence of AC Stark shifts for all transmon levels. We clarify how this effect can be calibrated out. Finally, from numerical simulations, we extract transition amplitudes corresponding to unwanted resonances. We show that in our parameter range they can be well captured by a simple analytical expression.
 
\subsection{Definition of Floquet modes through the propagator}
 
Assume that the system is described by a Hamiltonian $H(t)$ periodic with period $T$, $H(t+T) = H(t)$. 
The dynamics of the system can then be decomposed in terms of the Floquet modes, defined as the eigenvectors of the propagator over one drive period,
\begin{equation}
\label{eq:propagator}
    U(0,T)|\Phi_m\rangle = e^{-i\epsilon_m T/\hbar}|\Phi_m\rangle,\quad U(0,T) = \mathcal{T}\exp\left(-\frac{i}{\hbar} \int_0^T H(t)dt\right).
\end{equation}
Here, $|\Phi_{m}\rangle$ is the Floquet mode indexed by $m$ and $\epsilon_m$ is the associated quasienergy. The symbol $\mathcal{T}$ denotes time ordering.
 
The evolution operator $U(0, T)$ can always be represented in the form $U(0,T) = \exp(-i H_F T / \hbar)$. We call the operator $H_F$ the static effective Hamiltonian. If its explicit form is known, diagonalization of the propagator reduces to that of the static effective Hamiltonian. Otherwise, the Floquet modes can be found by diagonalizing the propagator numerically. For the case of a transmon, we rely on the latter approach.
 
\subsection{Simple example of identifying resonances via Floquet modes}
\label{sec:simple_model}
Here, we present a simple two-level model of a resonance in which the hybridization induced by the drive can be computed analytically. This model demonstrates how the resonances can be identified by finding the Floquet modes and evaluating the hybridization $\Theta$ defined in Eq.~\eqref{eq:hybridization}.
The intuition developed here is directly applicable to the case of a driven transmon.
 
\begin{figure*}
  \begin{center}
    \includegraphics[scale = 1]{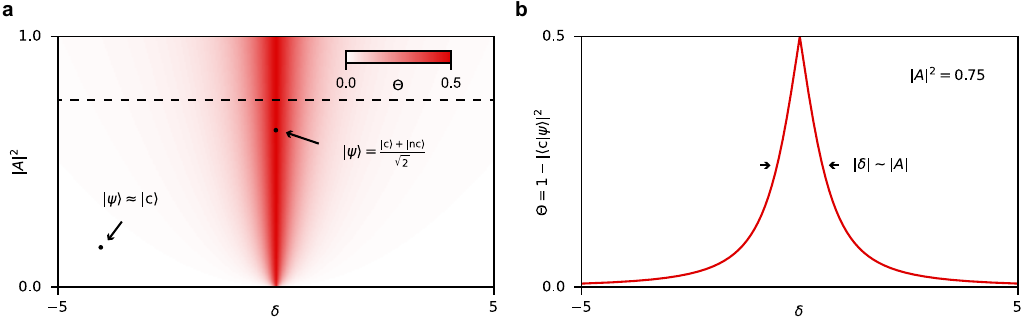}
    \caption{\label{fig:simple_model} A drive can induce hybridization between one of the computational states, $|\mathrm{c}\rangle$, and a non-computational state $|\mathrm{nc}\rangle$. We quantify the degree of hybridization $\Theta$ by computing the squared overlap between the Floquet mode $|\psi\rangle$ closest to $|\mathrm{c}\rangle$ and the non-computational state $|\mathrm{nc}\rangle$. \textbf{a}, Hybridization $\Theta$ as a function of drive power $|A|^2$ and detuning of the drive $\delta$ from the transition between $|\mathrm{c}\rangle$ and $|\mathrm{nc}\rangle$ computed within a two-level toy model, see Eq.~\eqref{eq:simplemodel}. Away from resonance, we have $|\delta|\gg |A|$, thus the Floquet mode is close to $|c\rangle$ resulting in $\Theta\ll 1$. Close to resonance, $|\delta|\ll |A|$, the Floquet mode is an equal superposition of $|\mathrm{c}\rangle$ and $|\mathrm{nc}\rangle$ resulting in $\Theta = 1/2$. \textbf{b}, A cut of panel \textbf{a} at $|A|^2 = 0.75$. The width of the resonant feature in $\Theta(\delta)$ is determined by the drive matrix element $|A|$.}
  \end{center}
\end{figure*}

In our two-level model, we take one state to play the role of a computational state $|\rm{c}\rangle$ and the other that of a non-computational state $|\rm{nc}\rangle$. The Hamiltonian of the model in the basis $\{|\rm{nc}\rangle,\:|\rm{c}\rangle\}$ is given by
\begin{align}
\label{eq:simplemodel}
\frac{H}{\hbar}= \frac{\omega}{2}(1 + \sigma_{z}) +\frac{A}{2}(\sigma_{+}e^{-i\omega_{d}t} + \sigma_{-}e^{i\omega_{d}t}),
\end{align}
where the $\sigma_-,\:\sigma_+,\:\sigma_z$ are the standard Pauli operators.
Within this model, static effective Hamiltonian $H_F$ can be found explicitly by going to the rotating frame, 
\begin{align}
H_F = \frac{\delta}{2}(\sigma_{z}+1)+\frac{A}{2}\sigma_{x},\quad \delta = \omega - \omega_d.
\end{align}
Here, we omitted an unimportant constant term. Diagonalizing the static effective Hamiltonian, we find the quasienergies; they are given by $\epsilon_{\pm}/\hbar=\pm \Omega / 2$ with $\Omega = \sqrt{A^2 + \delta^2}$. The associated Floquet modes are given by
\begin{align}
|\epsilon_{+}\rangle &= \cos(\theta/2)|\rm{nc}\rangle + \sin(\theta/2)|\rm{c}\rangle, \\ \nonumber 
|\epsilon_{-}\rangle &= \sin(\theta/2)|\rm{nc}\rangle -\cos(\theta/2)|\rm{c}\rangle,
\end{align}
where $\theta=\arctan(A/\delta)$. Given these Floquet modes, we can now proceed to compute the hybridization
\begin{align}
\Theta = 1 - \mathrm{max}\left(|\langle c|\epsilon_{-}\rangle|^2, |\langle c|\epsilon_{+}\rangle|^2\right)=\frac{(A/\delta)^2}{2[1+(A/\delta)^2+\sqrt{1+(A/\delta)^2}]}.
\label{eq:Theta}
\end{align}
This function peaks at $1/2$ for $\delta=0$, indicating a resonance, see Fig.~\ref{fig:simple_model}.
 
\subsection{Using Floquet modes to identify resonances in a transmon qubit}
In this section, we elaborate on the numerical procedure for finding resonances in the spectrum of a driven transmon. As explained above, we determine the resonant conditions by computing Floquet modes and using them to compute the hybridization $\Theta$ between the computational and non-computational states. Throughout this section, we use AC Stark shift of the $|0\rangle$ to $|1\rangle$ transition to quantify drive power. This parameter is related to the photon number as $\delta\omega = \chi \bar{n}$.
 
\begin{figure*}
  \begin{center}
    \includegraphics[scale = 1]{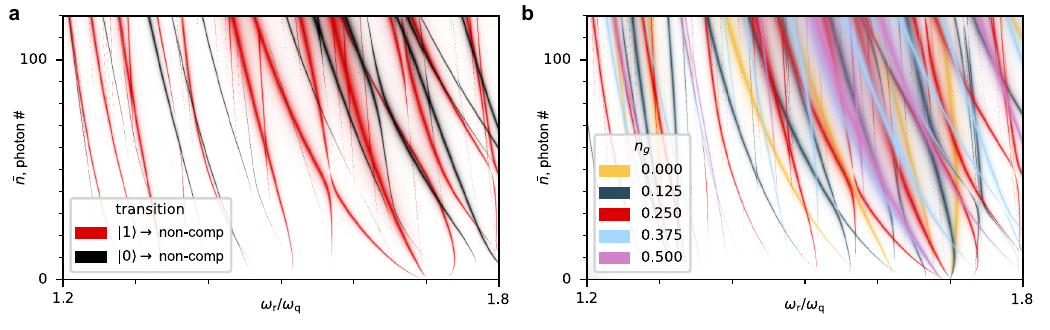}
    \caption{\label{fig:floquet1} Unwanted resonances for $E_J = 2.2\ghz$, $E_C = 36\mhz$, $\chi = -0.9\mhz$. The resonances are plotted as a function of drive frequency and power. \textbf{a},  Unwanted resonances for transmon initialized in the ground state (black) and the first excited state (red). Offset charge is set to $n_g = 0.25$. \textbf{b}, Spectrum of unwanted resonances for transmon prepared in the state $|1\rangle$ as a  function of the offset charge $n_g$. The spectrum is highly $n_g$-dependent.}
  \end{center}
\end{figure*}

The Hamiltonian of a driven transmon is given by
\begin{align}
\label{eq:semi-classical}
H = 4 E_{C}(N-n_{g})^2 - E_{J}\cos\varphi + \hbar\zeta N \cos(\wres t),
\end{align}
where $\wres$ is the readout resonator frequency and $\zeta$ is the amplitude of the drive acting on the transmon. The drive amplitude can be related to the induced AC-stark shift $\delta\omega$ by
\begin{align}
\label{eq:zeta_stark}
 \frac{\delta\omega}{\wq} = -\frac{1}{8} \frac{\zeta^2\wres^2}{(\wres^2-\wq^2)^2},
\end{align}
where we assumed $E_J/E_C \gg 1$ and $|\wres - \wq|\gg E_C/\hbar$. Together with $\delta\omega = \chi\nbar$, Eq.~\eqref{eq:zeta_stark} provides the correspondence between the drive amplitude of the semi-classical model and the photon number of the quantum description. We use it to express the drive power in terms of $\nbar$ in Fig.~2 of the main text and in Figs.~\ref{fig:floquet1} and \ref{fig:floquet2}.
To pinpoint the combinations of $\wres$ and $\delta\omega$ that activate the resonances, we compute the Floquet modes $|\Phi_{m}\rangle$ associated with Eq.~\eqref{eq:semi-classical}. This can no longer be done analytically: unlike Section~\ref{sec:simple_model}, there is no rotating frame eliminating all time dependence in Eq.~\eqref{eq:semi-classical}. For this reason, we resort to computing the Floquet modes numerically using the open-source package \texttt{QuTiP} \cite{johansson_qutip_2012, johansson_qutip_2013}. This is achieved by diagonalizing the propagator \eqref{eq:propagator} with $H(t)$ given in Eq.~\eqref{eq:semi-classical}.

There is, however, an important complication not captured by the simple model considered in Section~\ref{sec:simple_model}. Namely, under the applied drive, the Floquet modes are AC Stark shifted from the ``bare'' undriven states. This effect results from the combination of the transmon nonlinearity and the drive; it is present even away from any spurious resonances. In fact, the Stark-shifted Floquet modes can differ substantially from their undriven counterparts. When computing the hybridization $\Theta$, we should specifically consider the Stark-shifted computational and non-computational states.
 
To determine the structure of the Stark-shifted states, following Ref.~\cite{xiao_diagrammatic_2025}, we first identify for each ($\zeta,\wres$) the Floquet mode $|\ell(\zeta, \wres)\rangle$ with maximum overlap with the bare undriven transmon state $|\ell\rangle$. We then decompose this Floquet mode in the undriven basis. The decomposition coefficients can be expressed as a power series of $\zeta$ and $\wres$:
\begin{align}
\langle k | \ell(\zeta, \wres)\rangle = \sum_{ij}C_{ijk\ell} \zeta^i\wres^j.
\end{align}
The decomposition coefficients are determined by fitting the numerically extracted coefficients to a polynomial truncated to $i,j\leq 4$. Importantly, we exclude from this fit any points $(\zeta, \wres)$ where the overlap between the Floquet mode and any bare ``undriven" state is below a threshold (typically taken to be 0.8). This is needed to avoid fitting any spurious resonances.
 
Because the overlap with the bare state eventually drops below the threshold for a sufficiently strong drive, we perform the above procedure iteratively~\cite{xiao_diagrammatic_2025}. We split the drive amplitudes under consideration into multiple windows. In the first window (smallest drive amplitudes), we proceed as above. In the next window, we take as the ``bare state'' the fitted state from the previous window, and so on.
Once all Floquet modes have been extracted and labeled in this way, we perform a final fit over the whole range of drive frequencies and amplitudes to yield a result free from numerical artifacts at the window boundaries.
 
Having now obtained both the Floquet modes as well as the numerically-fit AC-Stark-shifted computational states, we are now in a position to calculate the hybridization $\Theta$. 
In this way we obtain Figure 2 of the main text of the manuscript. The code utilized to perform these numerical simulations is available as an open-source Python package \texttt{floquet} \cite{weiss_floquet_nodate}.
 
\subsection{Additional Floquet simulations}
Judging from Figure 2 of the main text it may seem that undesired resonances can be avoided in the standard readout regime $\wres\sim\wq$ by correctly choosing the readout frequency. This strategy is invalidated, however, by two facts. First, the landscape of resonances is different for transmon state $|1\rangle$ -- shown in Figure 2 -- and for state $|0\rangle$. The landscape of undesired resonances is much more dense if both states are considered simultaneously, see Figure~\ref{fig:floquet1}a. Second, the resonances have a pronounced dependence on the offset charge $n_g$, even when $E_J / E_C = 60$, see Figure~\ref{fig:floquet1}b. This is because the majority of resonances hybridize computational states to charge-sensitive states near the top or outside of the Josephson well. Random drifts of $n_g$ with time \cite{riste_millisecond_2013, serniak_hot_2018} make it hard to avoid the resonances without a dedicated offset-charge control.
 
\subsection{Comparison of transition amplitudes between simulation and simple analytical theory}
In our regime, the strongest undesired resonances employ one drive photon and excite the qubit to a non-computational state. This happens when the drive frequency $\wres$ satisfies $\hbar\wres = E_{j} - E_{i}$, where $E_{i}$ is the energy of the initial computational transmon state, and $E_{j}$ is the energy of the final non-computational state. For $\wres$ close to the resonance, the hybridization between the computational state and the non-computational state can be described by a $2 \times 2$ static effective Hamiltonian 
\begin{equation}
\label{eq:2x2}
H_F = 
    \begin{pmatrix}
    E_i & \hbar\Omega_{i\rightarrow j}\\
     \hbar\Omega_{i\rightarrow j}^\star & E_j - \hbar\wres
    \end{pmatrix}.
\end{equation}
Transition amplitude $\Omega_{i\rightarrow j}$ depends on $\zeta$. At small drive powers this dependence is given by
\begin{equation}
    \label{eq:low-power}
    \Omega_{i\rightarrow j}^\mathrm{low-power} = \frac{1}{2}\zeta \langle j|N|i\rangle,
\end{equation}
At higher powers, $\Omega_{i\rightarrow j}$ deviates from the low-power expression \eqref{eq:low-power} due to the possibility of reaching the final state through a sequence of intermediate virtual states. To evaluate the importance of these corrections, we extract the transition amplitude $\Omega_{i\rightarrow j}$ from our Floquet numerical simulation. This can equivalently be done either by evaluating the anti-crossing of quasienergies of the Floquet modes corresponding to the considered states or by fitting the hybridization $\Theta$ to Eq.~\eqref{eq:Theta}. Notably, to numerically find the transition amplitude, it is important to account for the AC Stark shift of energies $E_i$ and $E_j$.
 
\begin{figure*}
  \begin{center}
    \includegraphics[scale = 1]{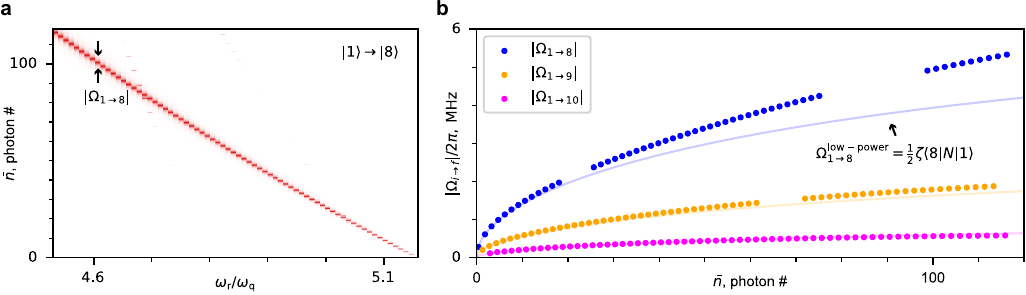}
    \caption{\label{fig:floquet2} Transition amplitudes corresponding to the unwanted resonances. Parameters used for the plot are same as in Figure~\ref{fig:floquet1} and in Figure 2 of the main text. \textbf{a}, Hybridization between the computational state $|i\rangle = |1\rangle$ and a non-computational state $|j\rangle = |8\rangle$ as a function of drive power and frequency. The resonant feature occurs when $\hbar\wres = E_8 - E_1$; the width of this feature is determined by the transition amplitude $\Omega_{1\rightarrow 8}$. This allows us to numerically find $\Omega_{1\rightarrow 8}$ as a function of power. \textbf{b}, Comparison between the low-power expression \eqref{eq:low-power} for the transition amplitude (solid lines) and the numerically extracted transition amplitudes (dots). Gaps in the numerical data occur where several resonances cross; there Eq.~\eqref{eq:2x2} does not correctly describe the behavior of the system and $\Omega_{i\rightarrow j}$ is ill-defined.}
  \end{center}
\end{figure*}

The result of this procedure is shown in Figure~\ref{fig:floquet2}. As is apparent from the figure, the corrections to the low-power expression \eqref{eq:low-power} are small at powers used in our experiment. This justifies the use of Eq.~\eqref{eq:low-power} in the main text when estimating the transition amplitudes. We note that the transitions relevant for our experiment are $|0\rangle \rightarrow |14\rangle$ and $|1\rangle \rightarrow |15\rangle$.
 
\section{Role of improved readout performance for quantum error correction}
 
In quantum error correction, the state of a logical qubit is redundantly encoded into an entangled state of multiple physical qubits. By measuring syndromes of errors happening on the physical qubits, the errors can be detected and reverted before the logical state is corrupted. The prototypical example of a quantum error correction code is a surface code \cite{kitaev_fault-tolerant_2003}. In its simplest form, qubits are arranged in a square $d \times d$ lattice; the error syndromes correspond to products of four qubit operators at each plaquette of the lattice. A recent experiment demonstrated the feasibility of quantum error correction with a surface code formed out of superconducting transmon qubits \cite{google_quantum_ai_quantum_2025}. The authors of Ref.~\cite{google_quantum_ai_quantum_2025} found that as the number of qubits in the code was increased, the probability of a logical error during a fixed time interval decreased exponentially. Specifically, they found $p_L \propto \Lambda^{-(d+1) / 2}$, where parameter $p_L$ is the logical error probability and $\Lambda$ is the parameter characterizing the effectiveness of quantum error correction. In the experiment of Ref.~\cite{google_quantum_ai_quantum_2025}, the latter parameter was limited to $\Lambda \approx 2.14$. As we now explain, better readout performance -- such as the one achieved in our work -- can boost this number up by about $35\%$.
 
In Ref.~\cite{google_quantum_ai_quantum_2025}, the measurement fidelity was limited to $99.3\%$. According to the error budget of Ref.~\cite{google_quantum_ai_quantum_2025}, measurement accounts for about $29\%$ of the total $1/\Lambda$, see Fig.~2c in Ref.~\cite{google_quantum_ai_quantum_2025}. This contribution comes from two factors. First of all, mistakes in error syndrome measurement confound the error decoding. This accounts for about $11\%$ of the error budget, see blue ``Meas.'' bar in Fig.~2c of Ref.~\cite{google_quantum_ai_quantum_2025}. Second, while error syndromes are being measured, the qubits comprising the surface code may decay. Such errors account for additional $18\%$ of the error budget, see yellow ``Idle'' bar in Fig.~2c of Ref.~\cite{google_quantum_ai_quantum_2025}.
 
As we demonstrated in our work, high-frequency dispersive readout can reduce the measurement infidelity of a transmon qubit by about an order-of-magnitude compared to the standard dispersive readout. We achieved QND measurement fidelity of $99.93\%$ appreciably higher than $99.3\%$ fidelity reported in Ref.~\cite{google_quantum_ai_quantum_2025}. We expect this improvement to yield a tenfold suppression of the contributions to $1/\Lambda$ associated with the measurements\footnote{For this estimate, it is important to account for both the increase of fidelity and the reduction of the readout duration relative to the qubit lifetime $T_1$, which is about a factor of 8 in our device compared to Ref.~\cite{google_quantum_ai_quantum_2025}.}. These would thus go from $29\%$ to about $3\%$. This would yield an overall increase of $\Lambda$ of about $35\%$. This improvement would thus significantly cut the overhead associated with quantum error correction.
 
In addition to the improvement explained above, our new readout scheme also alleviates the need for Purcell electromagnetic filters. These filters significantly complicate the chip layout and potentially lead to unwanted crosstalk. Thus, we expect that eliminating Purcell filters would further increase $\Lambda$, although this effect is hard to quantify without a detailed simulation.
 
\section{Perspective on qubit measurement without an amplifier}
 
In our work, we relied on a near-quantum limited amplifier to achieve state-of-the-art quantum non-demolition fidelity of $\mathcal{Q} = 99.93\%$. The amplifier dramatically suppresses the separation error by making the readout robust to noise in the output chain. In this section, we provide estimates that show that high-frequency readout scheme \textit{without an amplifier} should at least be on-par with conventional dispersive readout \textit{with an amplifier} in terms of $\mathcal{Q}$. We leave detailed optimization of readout parameters without the amplifier, as well as the experimental characterization of such a readout, to future work.
 
We begin by reviewing the parameters of our readout in the presence of the amplifier. In the presence of the amplifier, our best readout uses $\nbar = 92$ photons (maximal number during the pulse); the pulse duration is 320 $\mathrm{ns}$.  The error budget can be broken down as $1 - \mathcal{Q} = \varepsilon_\mathrm{sep} + \varepsilon_\mathrm{trans}^{\mathrm{bit-flip}} + \varepsilon_\mathrm{trans}^{\mathrm{leakage}}$, where $\varepsilon_\mathrm{sep} = 3 \cdot 10^{-5}$, $\varepsilon_\mathrm{trans}^{\mathrm{bit-flip}} = 5 \cdot 10^{-4}$, $\varepsilon_\mathrm{trans}^{\mathrm{leakage}} = 2 \cdot 10^{-4}$, see Fig.~3d of the main text.
 
Now, as an example, we provide an estimate which shows that \textit{without the amplifier} high-frequency readout should be able to achieve $\mathcal{Q} \sim 99.3\%$ reading out with 180 photons in $2\:\mu\mathrm{s}$. First, note that the bit-flip error probability scales as $\varepsilon_\mathrm{trans}^{\mathrm{bit-flip}} \propto t / T_1$, where $t$ is the readout duration and $T_1$ is the qubit lifetime. Therefore, we can estimate $\varepsilon_\mathrm{trans}^{\mathrm{bit-flip}} \sim 30 \cdot 10^{-4}$ for $t = 2\:\mu\mathrm{s}$. Second, the probability of the separation error can be expressed as \cite{touzard_gated_2019}
\begin{equation}
    {\varepsilon_\mathrm{sep} = 0.5 \cdot\mathrm{erfc}\left(\sqrt{\frac{\mathrm{SNR}^2}{4}}\right),\quad \mathrm{SNR}^2 \propto \eta_m \bar{n} t,}
\end{equation}
where $\eta_m$ is the measurement efficiency. In the absence of an amplifier, we can \textit{pessimistically} take the efficiency to be a factor of $20$ smaller than that of our readout with phase-sensitive amplification. For parameters $\bar{n} = 180$ and $t= 2\:\mu\mathrm{s}$, we can thus estimate ${\varepsilon_\mathrm{sep}} \sim 10 \cdot 10^{-4}$. Third, the leakage probability grows with both the photon number and the readout duration. Scaling the leakage measured in Fig.~\ref{fig:residual_nbar} with $\bar{n} t$, we estimate $\varepsilon_\mathrm{trans}^{\mathrm{leakage}} \sim 30 \cdot 10^{-4}$. As a result, we find $\mathcal{Q} \sim 99.3\%$, on par with the state of the art of $99.3\%$ reported in the literature \cite{spring_fast_2025} \textit{with the amplifier}.
 
\section{Generalization of high-frequency readout beyond transmon superconducting qubit}
In the main text of the manuscript, we have demonstrated high-frequency dispersive readout of a transmon superconducting qubit. The suppression of unwanted resonances in the qubit spectrum allowed us to use higher power for the readout signal, leading to an improved readout performance. The fact that the resonances are suppressed \textit{exponentially} makes high-frequency readout advantageous in spite of the milder \textit{power-law} suppression of the readout resonator dispersive shift.
 
However, high-frequency readout is not specific to transmons. By the same token, this technique can be used to improve the measurement of \textit{any} superconducting qubit, provided that the readout frequency sufficiently exceeds the characteristic frequency of the circuit. The core argument in favor of using high readout frequency -- exponential suppression of unwanted resonances winning over the power-law suppression of dispersive shift -- is valid for arbitrary circuit. Indeed, regardless of the circuit, unwanted resonances in the high-frequency regime necessarily involve highly-excited non-computational states. The wavefunctions for these states are quickly oscillating, in contrast to the one for the computational states. This makes transition matrix elements between the computational and non-computational states exponentially small. The power-law suppression of dispersive shift can be overcome by increasing the coupling strength between the qubit and the readout resonator. Protection from radiative decay -- another advantage of the high-frequency readout regime -- similarly applies to other superconducting circuits, as long as their characteristic frequency is small compared to that of the readout resonator.
 
To give an example, in what follows, we explicitly consider the measurement of a fluxonium superconducting qubit. We numerically evaluate the matrix elements for the unwanted resonances and demonstrate that they are exponentially suppressed in the high-frequency readout regime. As we explain below, to gain an advantage from the increased readout frequency, the latter has to vastly exceed the characteristic plasma frequency of the fluxonium circuit (as opposed to just the qubit frequency). Additionally, we calculate the dispersive shift of the readout resonator caused by its coupling to the fluxonium circuit and explicitly demonstrate that dispersive shift only decreases as a mild power-law of the readout frequency.
 
\subsection{Hamiltonian of a fluxonium qubit coupled to a readout resonator}
 
\begin{figure*}
\centering
  \begin{center}
    \includegraphics[scale = 1]{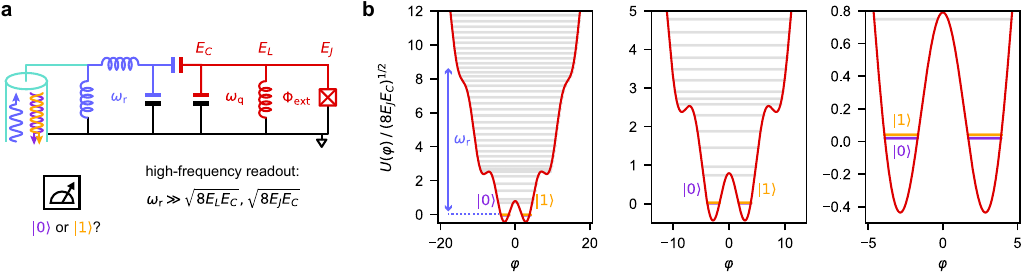} \caption{High-frequency readout of a fluxonium qubit. We use a typical set of qubit parameters $E_J / h = 5~\text{GHz}$, $E_L / h = 0.5~\text{GHz}$, $E_C / h = 1~\text{GHz}$, and $\Phi_{\rm ext}/ \Phi_0 = 0.5$, where $\Phi_0$ is the flux quantum.
    {\bf a}, Schematic of the fluxonium circuit. 
    {\bf b}, Fluxonium potential and energy levels. Three panels correspond to three resolution scales. For high-frequency readout to be effective, the readout frequency has to satisfy $\hbar\wres \gg \sqrt{8E_J E_C},\sqrt{8E_L E_C}$.}
    \label{fig:HF_readout_fluxonium}
  \end{center}
\end{figure*}
 
The Hamiltonian of a fluxonium qubit coupled to a readout resonator with frequency $\omega_{\rm r}$ is given by
\begin{align}
    H = 4 E_C N^2 + U(\varphi) + \lambda N (a + a^\dagger) + \hbar \omega_{\rm r}(a^\dagger a + 1/2), \label{eq:ham_arbitrary}
\end{align}
where $E_C = e^2/2C_{\rm q}$ is the charging energy, $\lambda$ determines the coupling strength between the qubit and the resonator (note the distinction from the definition of $g$ in Eq.~\eqref{eq:res_tr_ham}; we choose a different notation because parameter $\nzpf$ is not defined for fluxonium). The potential energy is given by
\begin{equation}
        U(\varphi) = E_L \frac{\varphi^2}{2} - E_J \cos \left(\varphi - 2 \pi \frac{\Phi_{\rm ext}}{\Phi_0} \right), \label{eq:Fluxonium_Potential}
\end{equation}
where $E_J$ is the Josephson energy of the junction and $E_L$ is the inductive energy of an inductor shunting the qubit. The schematic depiction of the considered circuit is shown in Fig.~\ref{fig:HF_readout_fluxonium}a. The structure of the low-lying energy levels is determined by the external flux bias, $\Phi_\text{ext}$. Fluxonium is often operated at $\Phi_\text{ext}/ \Phi_0 = 0.5$ in the parameter regime where $U(\varphi)$ has a double well shape at this $\Phi_\text{ext}$. The tunnel split ground states in the wells define the qubit basis. We focus on this regime in our example. The spectrum of the fluxonium qubit is shown in Fig.~\ref{fig:HF_readout_fluxonium}b.
 
\subsection{Suppression of unwanted resonances in the high-frequency readout of a fluxonium qubit}
To take advantage of the increased readout frequency when measuring the fluxonium, this frequency has to vastly exceed the characteristic frequencies of the circuit, $\hbar\wres\gg \sqrt{8E_J E_C},\sqrt{8E_L E_C}$. When this condition is fulfilled, the strongest unwanted resonance can only couple the computational states to highly excited non-computational states $|f\rangle$ with $f\gg 1$. In this case, the transition matrix element becomes exponentially suppressed due to quick oscillation of the excited state wavefunction. Notably, it is \textit{insufficient} to require $\wres\gg \wq$ for high-frequency readout to be advantageous for an arbitrary circuit such as fluxonium.
 
\begin{figure*}
\centering
  \begin{center}
    \includegraphics[scale = 1]{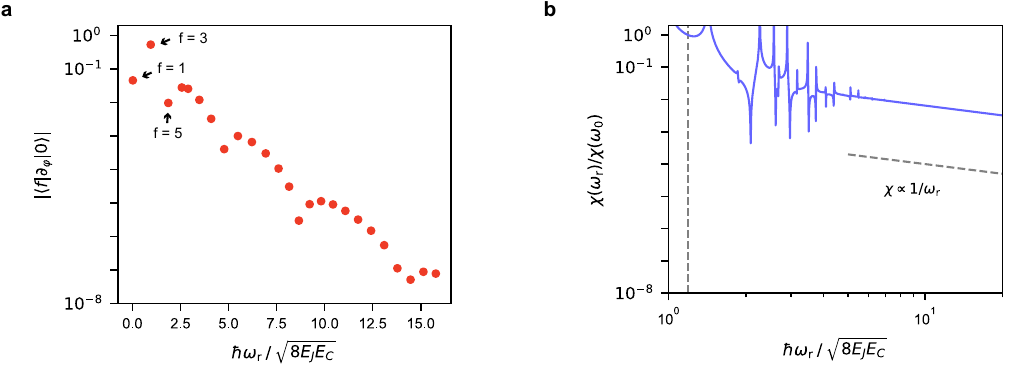}
    \caption{({\bf A}) Matrix elements for transitions caused by the unwanted resonances with different final states. The matrix elements decay exponentially as the readout frequency is increased. This behaviour is analogous to that observed for the transmon qubit, see Fig.~1 of the main text.
    ({\bf B}) The dispersive shift decays as $\chi(\wres) \propto 1/\wres$ at high frequencies. The dispersive shift is normalized by its value at a reference point $\omega_0$ indicated with a vertical dashed line. Resonances in the dispersive shift have the same origin as in the case of a transmon qubit, see discussion in Section \ref{sec:disp_resonances}.}
    \label{fig:HF_readout_fluxonium_2}
  \end{center}
\end{figure*}

Next, we explicitly demonstrate the exponential suppression of unwanted resonances via a numerical simulation. In the high-frequency readout regime, the strongest allowed resonance happens when $\hbar\wres= E_{i\rightarrow f} = E_f - E_i$. The strength of this resonance is quantified by the associated transition amplitude, $\Omega_{i\rightarrow f}$. The latter can be expressed through the matrix element of the operator coupling the resonator and the qubit,
\begin{equation}
    \hbar|\Omega_{i\rightarrow f}| = \lambda \sqrt{\bar{n}} |\langle f|N|i\rangle|= \lambda \sqrt{\bar{n}} |\langle f|\partial_\varphi|i\rangle|.
\end{equation}
In Figure~\ref{fig:HF_readout_fluxonium_2}a, we plot matrix elements $|\langle f|\partial_\varphi|i\rangle|$ for different final states $|f\rangle$. By symmetry of the fluxonium Hamiltonian at $\Phi_\text{ext}=0.5\Phi_0$, the matrix element is non-zero only when $i$ and $f$ have different parity. From the figure it is apparent that the matrix elements indeed decay exponentially as the readout frequency -- and thus the number of the relevant final state -- is increased.
 
\subsection{Dispersive shift in high-frequency readout of a fluxonium qubit}
 
Now we evaluate the dispersive shift $\chi$ of the readout resonator caused by its coupling to the fluxonium qubit. We demonstrate that as the resonator frequency is increased, the dispersive shift decays as a mild power law, in contrast to the exponential suppression of the unwanted resonances. Suppression of the dispersive shift can thus be overcome by increasing the qubit-resonator coupling strength without compromising the protection from unwanted resonances.
 
Similarly to discussion in Section~\ref{sec:disp_shift}, we can express the dispersive shift as $\chi = \chi_1 - \chi_0$. Here, $\chi_i$ is the pull experienced by the readout resonator when the qubit is placed in state $|i\rangle$. The cavity pulls are given by
\begin{equation}
\label{eq:chi_i}
    \hbar\chi_i = 2 \lambda^2 \sum_{f \neq i} \frac{E_i - E_f}{(E_i - E_f)^2 - \hbar^2\omega_{\rm r}^2} |\langle f| N |i\rangle|^2.
\end{equation}
 
Next, using Eq.~\eqref{eq:chi_i}, we evaluate the dependence of the dispersive shift $\chi = \chi_1 - \chi_0$ on the readout resonator frequency. For simplicity, we assume that the resonator frequency is tuned by changing the resonator inductance; all capacitances are held fixed. Notably, the coupling constant $\lambda$ depends on the frequency of the readout resonator. In the regime when the coupling capacitance is small compared to the qubit and the resonator capacitances, this dependence is given by
\begin{align}
    \lambda = \sqrt{4 E_C \hbar \omega_{\rm r}} \frac{C_{C}}{\sqrt{C_{\rm q} C_{\rm res}}}.  \label{eq:lambda_fluxonium}
\end{align}
Then, as is apparent from Eq.~\eqref{eq:chi_i}, in the regime $\hbar\wres\gg \sqrt{8E_J E_C},\sqrt{8 E_L E_C}$, the dispersive shift scales as $\propto 1 / \wres$. This can be seen by neglecting the energy difference $E_i - E_f$ in the denominator of Eq.~\eqref{eq:chi_i} and substituting the expression for $\lambda$, Eq.~\eqref{eq:lambda_fluxonium}, into Eq.~\eqref{eq:chi_i}. 
 
In Figure~\ref{fig:HF_readout_fluxonium_2}b, we plot dependence of the dispersive shift on the readout resonator frequency. This function consists of an overall $\propto 1/\wres$ decline and sharp resonances on top of it. The resonances occur when $\hbar\wres = E_f - E_i$ for $i=0,1$ and certain $f$. As the frequency is increased, the resonances become exponentially suppressed.
 
\bibliography{references}